\documentclass{jfm}
\usepackage{graphicx}
\usepackage{epstopdf, epsfig}

\usepackage{add-instructions} 
\usepackage{amsmath} 
\usepackage{mathtools} 
\usepackage{nicematrix} 
\usepackage{subcaption}
\usepackage{blkarray}
\usepackage{todonotes}
\usepackage[numbers]{natbib}
\usepackage{multirow}
\usepackage{booktabs}
\usepackage[pdfpagemode={UseOutlines},bookmarks=true,bookmarksopen=true,bookmarksopenlevel=0,bookmarksnumbered=true,hypertexnames=false,colorlinks,linkcolor={blue},citecolor={blue},urlcolor={red},pdfstartview={FitV},unicode,breaklinks=true]{hyperref}
\definecolor{myblack}{rgb}{0,0,0}
\definecolor{myblue}{rgb}{0.0, 0.0, 1.0}
\definecolor{myred}{rgb}{1.0, 0.0, 0.0}
\definecolor{mycyan}{rgb}{0.0, 1.0, 1.0}
\definecolor{mymagenta}{rgb}{1.0, 0.0, 1.0}
\definecolor{mygreen}{rgb}{0.0, 0.5019607843137255, 0.0}
\definecolor{myorange}{rgb}{1.0, 0.6470588235294118,0.0}
\usepackage{tikz}
\shorttitle{Shear-induced migration in pressure-driven suspension flows}
\shortauthor{M. Noori, J.D. Berry, and D.J.E. Harvie}
\title{Multifluid simulation of shear-induced migration in pressure-driven suspension flows}
\author{Mohammad Noori\aff{1}, Joseph D. Berry\aff{1}, \and Dalton J.E. Harvie\aff{1} \corresp{\email{daltonh@unimelb.edu.au}}}
\affiliation{\aff{1}Department of Chemical Engineering, The University of Melbourne, Parkville, Victoria 3010, Australia}
\begin{document}
\maketitle
\begin{abstract}
The present study simulates shear-induced migration (SIM) in semi-dilute pressure-driven Stokes suspension flows using a multi-fluid (MF) model. Building on analysis from a companion paper \citep{harvie2021tensorial}, the specific formulation uses volume-averaged phase stresses that are linked to the binary hydrodynamic interaction of spheres and suspension microstructure as represented by an anisotropic, piece-wise constant pair-distribution function (PDF). The form of the PDF is chosen to capture observations regarding the microstructure in sheared suspensions of rough particles, as reported in the literature. Specifically, a hydrodynamic roughness value is used to represent the width of the anisotropic region, and within this region the concentration of particles is higher in the compression zone than expansion zone. By numerically evaluating the hydrodynamic particle interactions and calculating the various shear and normal viscosities, the stress closure is incorporated into \citeauthor{harvie2021tensorial}'s volume-averaged MF framework, referred to as the MF-roughness model. Using multi-dimensional simulations the roughness and compression zone PDF concentration are then globally optimised to reproduce benchmark solid and velocity distributions reported in the literature for a variety of semi-dilute monodisperse suspension flows occurring within rectangular channels. For comparison, two different versions of the phenomenological stress closure by \citet{morris1999curvilinear} are additionally proposed as fully tensorial frame-invariant alternatives to the MF-roughness model. Referred to as MF-MB99-A and MF-MB99-B, these models use alternative assumptions for partitioning of the mixture normal stress between the solid and fluid phases. The optimised solid and velocity distributions from all three stress closures are similar and correlate well with the experimental data. The optimised MF-roughness viscosities correspond reasonably well with those from semi-dilute direct numerical simulations (DNS) and experimental correlations reported in the literature. The results of the two alternative MF models indicate that partitioning the mixture normal stress unequally between the solid and fluid phases (the MF-MB99-B model) causes the optimised normal viscosity of the mixture to be nearly five times larger than if the normal stresses are assumed to act only on the solid phase (the MF-MB99-A model).
\end{abstract}
\tableofcontents  
\section{Introduction}
Shear-induced migration (SIM) refers to the tendency of particles to aggregate in low-shear regions in suspension flows. A slowly sheared pressure-driven flow can alter the uniform distribution of micron-sized particles in a suspension, resulting in their migration from high shear near wall regions to the central regions characterised by the lowest shear rates. Studying this migration effect can help understand suspension behaviour across diverse biological, medical, and chemical applications. For instance, SIM can be studied in the context of analysing cell distribution in blood flow and coagulation mechanisms, as well as evaluating the performance of ultrafiltration technologies in purifying dairy products and addressing membrane fouling. Moreover, the phenomenon's relevance extends to assessing the performance of flotation fluidised beds and hydrocyclones in mineral classification or studying the resuspension of sediment beds in rivers.\\
\\
The direct measurement of steady-state SIM behaviour in pressure-driven flows has been documented in various studies. The experimental measurements conducted in two-dimensional wide rectangular channels \citep{lyon1998experimental1,lyon1998experimental2,frank2003particle,semwogerere2007development,semwogerere2008shear}, three-dimensional channels \citep{gao2009mixing,gao2010direct}, or pipes \citep{hampton1997migration,brown2009nuclear,oh2015pressure}. These suspension flows can be categorised into flows of weakly Brownian $1-3\;\mu\mbox{m}$ diameter particles and flows of larger non-Brownian particles. They are nearly unconfined and neutrally buoyant in Stokes regimes with low $Re$ numbers. The solids distribution in these flows is influenced by shear and thermal fluctuation forces at the particle scale. While the shear forces cause the migration toward the centerline of the flows, the thermal forces of Brownian particles are expansive and limit the migration toward the centerline. The importance of local particle shear forces relative to the thermal forces in the SIM of Brownian particles is characterised by the non-dimensional $Pe$ number.\\
\\
Some literature studies have measured the flows with different average $\PeB$ numbers within $(10-1000)$ to investigate the significance of thermal forces in the SIM of weakly Brownian particles \citep{semwogerere2007development,frank2003particle,brown2009nuclear,fridjonsson2016colloid}. The channel flow measurements by \citet{semwogerere2007development} show that the development length of the Brownian particles increases with $Pe$ numbers and reaches asymptotic values at around $\PeB=100$. They also found that the development length decreases as solid concentrations in the flow increase. The shorter development length for the flows of lower $\PeB$ numbers and higher concentrations is attributed to the expansive effect of the thermal interaction of the particles in limiting their shear-induced central enrichment. The development length of the non-Brownian suspension flows in a pipe is also found to decrease with solid concentration \citep{hampton1997migration}. For these flows, the confinement also affects the development length, where larger particle-to-pipe radio ratios are found to have smaller development lengths. The bidisperse flow measurements can also show the impact of the Brownian motion on SIM behaviour. In the experiments by \citet{lyon1998experimental2}, adding a smaller non-Brownian particle did not change a larger particle distribution. However, the addition of a smaller Brownian particle to the flows studied by \citet{semwogerere2008shear} significantly changed the central enrichment of a larger Brownian particle and caused a uniform distribution of them across the channel height.\\
\\
Direct numerical simulations (DNS) \citep{sierou2002rheology,morris2002microstructure,yeo2010simulation,gallier2014rheology,mari2014shear,sunol2023confined} along with the experimental optical techniques \citep{blanc2011experimental,blanc2013microstructure,xu2014microstructure,deboeuf2018imaging} are found helpful in understanding the microscopic origins of the SIM. The literature studies using these techniques reveal that the permanent shift in the trajectories of interacting particles results in an anisotropic microstructure and the SIM behaviour in the flow scale. The near-field hydrodynamic and contact interactions between particles, the flow confinement and particle-wall interactions, thermal motion, and particles' surface asperities are identified to impact the bulk flow behaviours. The DNS techniques solve the equations of motion for each individual particle. The review studies by \citet{maxey2017simulation} for the non-Brownian and \citet{larson2021methods} for the Brownian suspension flows mainly categorize these techniques based on how the fluid medium is meshed and how the flow within this medium is simulated. While these methods are robust for studying particle pair interactions, suspension microstructure, or bulk rheological properties such as shear viscosity, the high computational demand limits their application to realistic suspension flows. The number of particles considered in different DNS studies in the literature is usually on the order of $10^{2}-10^{4}$, which can be significantly lower than the numbers in real-world suspension flow applications. For instance, the number of red blood cells in an artery measuring $8\;\mbox{cm}$ in length and $4\;\mbox{mm}$ in diameter (typical dimensions of the main branches of the human heart coronaries) is within $10^{9}-10^{10}$. Furthermore, because of the computational cost and limitations in the mesh refinement, the DNS techniques require a minimum separation on the order of $10^{-4}-10^{-3}$, larger than the minimum separations achievable by smooth hard spheres. As a result, they may introduce inaccuracies in predicting the near-field lubrication interaction of particles. These demands have caused the application of the DNS techniques in the SIM behaviour of pressure-driven suspension flows to be limited to the studying fully developed solids and velocity distributions in periodic domains of sheared flows \citep{yeo2011numerical,gallier2016effect,chun2017lattice,chun2019shear,howard2022bidisperse}.\\
\\
The development of SIM in pressure-driven Stokes flows is primarily studied using continuum modelling approaches, including diffusive flux (DF), suspension balance (SB), or multifluid (MF) models. In the DF model, the flux of particles relative to the mixture (migration flux) is determined using phenomenological expressions that account for migration due to the changes in two-body interaction frequency, suspension shear viscosity, and a flux term resulting from the particles' Brownian diffusion \citep{phillips1992constitutive}. The free parameters of these flux expressions are specified based on the Couette flow experimental measurements. The model's prediction of the SIM in some pressure-driven flows did not correspond very well with the experimental measurements \citep{koh1994experimental,lyon1998experimental1,fang2002flow}. It also failed to accurately predict SIM direction in certain flow geometries. This is because the isotropic flux expressions are not linked to the suspension rheology and only account for the cross-stream migration in unidirectional flows \citep{phillips1992constitutive,fang2002flow}. The MF model is a general model that considers the phase-specific mass and momentum balances derived from averaging the local transport equations \citep{jackson1997locally,jackson1998locally,zhang1997momentum}. The closure of the model depends on the averaged expressions for the phase-specific stresses and interphase forces, whose mathematical formulations are derived only for the dilute suspensions. While the SB models are also developed based on the averaging of local transport equations, it is only consistent with the MF model when assuming that the interaction stresses in the solid and mixture phases are equal \citep{nott1994pressure,nott2011suspension,lhuillier2009migration}.\\
\\
The stress closure by \citet{morris1999curvilinear} is frequently used to simulate SIM in semi-dilute to concentrated suspension flows. This SB closure assumes equal interaction viscosities in the solid and mixture phases. The free parameters of this model are determined based on experimental measurements in some rheometer flows. Different values of these free parameters are used in the literature to simulate pressure-driven flows \citep{miller2006normal,dbouk2013shear,lecampion2014confined,oh2015pressure,guazzelli2018rheology}. However, these parameters have not been linked to the particle properties and microstructure. Furthermore, the closure's shear and normal viscosities can differ from direct rheometric measurements \citep{boyer2011unifying,zarraga2000characterization}.\\
\\
The mathematical formulation of the phase-averaged stresses that depend on suspension microstructure can be found in some literature studies using volume or ensemble averaging techniques. However, their functional forms have been derived for the dilute suspension flows. \citet{harvie2021tensorial} has proposed a MF framework based on the volume averaging technique. The interphase force in this framework causes the SIM migration flux to result from the competition between the solid and mixture phase stresses. The phase-specific stresses are linked to the pairwise hydrodynamic force, and a stresslet between the particles is averaged based on an anisotropic form of the pair distribution function (PDF) suggested for the microstructure of the sheared suspensions. These Newtonian interaction stresses are fully tensorial and linear in shear rate with some interaction viscosity coefficients. The interaction viscosities are expressed as a function of the pairwise mobility and resistance functions integrated into the PDF model. The functional form of these viscosities needs to be determined to apply the MF framework for the simulation of SIM in pressure-driven semi-dilute suspension flows.\\
\\
The present study primarily aims to determine the functional forms of the interaction viscosities within \citeauthor{harvie2021tensorial}'s MF framework for the semi-dilute sheared flows. First, the mobility and resistance functions are evaluated numerically using the semi-analytical formulation in \citet{kim2013microhydrodynamics}. Then, a piece-wise constant form of the PDF model is chosen for the sheared suspensions to calculate the volume integral of these functions. Finally, the functional form of these integrals defining the interaction viscosities is determined by curve-fitting the numerical data. This incorporation of interaction viscosities into the MF framework is referred to as the MF-roughness model. A hydrodynamic roughness will be used to define the width of the anisotropic region in the PDF model consistent with the direct microstructure measurements of the sheared suspensions. The hydrodynamic roughness value and the parameters distinguishing the PDF values in the anisotropic region are then optimised based on the experimental semi-dilute channel flows reported by \citet{semwogerere2007development} and \citet{semwogerere2008shear}. These microstructure parameters are adjusted in a global optimisation procedure to minimise the deviation of solids and velocity distribution results from the experimental data across the different suspension flows. Unlike the phenomenological SIM flux expressions in the DF or SB approaches, the MF-roughness model links it to the phase-specific stresses, suspension microstructure, and particle roughness.\\
\\
The current study also implements two modified forms of stress closure by \citet{morris1999curvilinear} into the MF model. The free parameters in each closure are then optimised based on the same semi-dilute experimental channel flows. The first implementation is the MF-MB99-A model, which introduces a fully tensorial frame-invariant form of the same closure. The model's closure parameters were initially calibrated based on some SIM measurements in rheometer flows, but their different values are used in the literature for the simulation of pressure-driven suspension flows. Optimising the MF-MB99-A model aims to find a unique set of parameters that describe experimental SIM behaviour in different pressure-driven flows. The second implementation of the closure is the MF-MB99-B model which partitions the mixture normal stress between the solid and fluid phases. This partitioning aligns with the volume-averaged MF equations and the revisions of SB in the literature. The free parameters of this model are also optimised based on the same experimental flows, and their comparison with those of the MF-MB99-A model illustrates the effect of normal stress partitioning on the viscosities describing the experimental SIM behaviour. 
\section{Governing equations}
The volume-averaged MF conservation equations proposed by \citet{harvie2021tensorial} are implemented to simulate suspension flows in the current study. The MF continuity equations for the fluid and solid phases are respectively as:
\begin{equation}\label{eq:MF-continuity-fluid}
\frac{\partial \phif}{\partial t} + \vnabla \cdot \phif \uf = 0
\end{equation}
\begin{equation}\label{eq:MF-continuity-solid}
\frac{\partial \phis}{\partial t} + \vnabla \cdot \phis \us = 0
\end{equation}
\citet{harvie2021tensorial} has derived the MF momentum equations by first analysing the average fluid flow through a fixed bed of particles with no relative velocity and then examining particles' interaction (collision) due to their relative velocity. The final form of the MF momentum equations for the fluid and solid phases are respectively written as:
\begin{equation}\label{eq:MF-momentum-fluid}
\rhof \left [ \frac{\partial \phif \uf }{\partial t}  + \vnabla \cdot \phif \uf \uf \right ] = -\phif \vnabla\pf  - \vnabla \cdot \left( \vect[dil,m]{\tau}+\tens[int,m]{\tau}-\tens[int,s]{\tau} \right)- \vect[drag,s]{f} - \vect[faxen,s]{f} + \rhof \phif \g
\end{equation}
\begin{equation}\label{eq:MF-momentum-solid}
\rhos \left [ \frac{\partial \phis \us }{\partial t}  + \vnabla \cdot \phis \us \us \right ] = -\phis \vnabla\pf- \vnabla\cdot\tens[int,s]{\tau}-\vnabla\posm + \vect[drag,s]{f} + \vect[faxen,s]{f} + \rhos \phis \g
\end{equation}
\\
where $\phif=1-\phis$ is fluid phase volume fraction, $\phis$ is solid phase volume fraction, $\uf$ is fluid phase velocity, $\us$ is solid phase velocity, $\pf$ is averaged hydro-static pressure in the fluid, $\g$ is the vector of gravitational acceleration. $\vect[drag,s]{f}$ is the drag force, $\vect[faxen,s]{f}$ is the Faxen force, $\vect[dil,m]{\tau}$ is the dilute stress tensor of the mixture, $\tens[int,m]{\tau}$ is the interaction stress tensor of the mixture, $\tens[int,s]{\tau}$ is the interaction stress tensor of the solid phase, and $\posm$ is the solid phase osmotic pressure. The drag, Faxen, and dilute stress forces in the momentum equations result from the average force analysis through the fixed bed of particles. On the other hand, the solid and mixture phase interaction stresses are caused by the collision of the particles. Additionally, in the current study, the osmotic force $(\vnabla\posm)$ is added to the solid phase momentum equation to account for the impact of the Brownian motion of the micron-sized particles on SIM behaviour in the flows.
\subsection{Osmotic force} 
The current study introduces the new osmotic force term into the solid phase momentum equation to account for the impact of Brownian motion on the distribution of $1-3\;\mu\mbox{m}$ sized particles in the flows of average $\PeB$ numbers within $10-500$. \citeauthor{harvie2021tensorial}'s model only considered the hydrodynamic interaction of non-Brownian particles in the volume averaging procedure and deriving the phase stresses. However, the Brownian motion of particles at moderate $\PeB$ numbers can alter the local stress field around the particles, the pair-wise force between the particles, and the suspension microstructure \citep{morris2002microstructure,brady1993brownian}. The random motions in the Brownian suspensions affect particle trajectories' asymmetry, making the suspension flows less anisotropic than non-Brownian suspension flows. In this study, the free parameters of the MF-roughness microstructure model introduced next are adjusted based on the particle size, considering the influence of Brownian motion on the microstructure. In addition, the osmotic force is included in the solid phase momentum equation to account for the direct impact of Brownian motion on the pair-wise interaction forces between particles. This modification assumes that the particles interact in a stationary ensemble at thermal equilibrium through Brownian motion. It also assumes that the Brownian contribution is equal in the solid and mixture phases and only has an isotropic normal force contribution without any shear force contribution.\\
\\
\citeauthor{carnahan1969equation} (CS) osmotic pressure is used for the simulations in the current study. It accurately represents the experimental equilibrium observations for solid volume fractions $(\phis)$ less than $0.5$ \citep{piazza1993equilibrium,buzzaccaro2007sticky}. This is consistent with the suspension flows studied here, as $\phis$ values only exceed this $0.5$ in isolated points of the flows. This osmotic pressure is defined to be equal to the CS osmotic pressure as:
\begin{equation}\label{eq:CS-osmotic-pressure}
\posm=\scal[s]{n}\kB T \frac{\left(1+\phis+\phis^2-\phis^3\right)}{\left(1-\phis\right)^3}
\end{equation}
where $\scal[s]{n}=\phis/\scal[p]{V}$ is the number density of particles, $\scal[p]{V}$ is the volume of a single particle, $\kB$ is Boltzmann’s constant, and $T$ is the temperature. By applying the gradient vector to this pressure, the CS osmotic force can be expressed as:
\begin{equation}\label{eq:CS-osmotic-force}
\vnabla\posm=-\frac{\kB T}{\scal[p]{V}}\scal[CS]{\zeta}\vnabla \phis\quad \text{with} \quad \scal[CS]{\zeta}=1+\frac{2\phis\left(4-\phis\right)}{\left(1-\phis\right)^4}
\end{equation}
Using the CS osmotic force to include the contribution of Brownian motion to interaction forces neglects the shear force contribution of Brownian motion and does not account for the variation of normal force contribution with the local Peclet number, as both seen in DNS results in the literature \citep{yurkovetsky2008particle}. However, the sensitivity analysis in the results section indicates that these two assumptions have negligible impacts on the solid distributions of the studied flows.
\section{MF-Roughness closure}
This section presents the semi-dilute closure of the drag, Faxen, and stress forces by \citet{harvie2021tensorial} used as as a MF closure model in the current study. As the closure relates the interaction stresses to the hydrodynamic roughness value for the particles, its implementation in the MF conservation equations is named the MF-roughness model. Tables \ref{tab:theory-MF-roughness-closure} and \ref{tab:interaction-viscosities-related-to-indexed-integrals} summarise \citeauthor{harvie2021tensorial}'s final formulation of the different force and stresses, which they depend on phase-specific shear rate variables defined in table \ref{tab:general-shear-rate-tensors}. The explicit expressions for drag, Faxen, and dilute stress in this formulation are derived based on the volume average analysis of a fixed bed of particles moving with the same velocity. However, the interaction stresses related to the pair-wise collision of particles are expressed in terms of certain mobility and resistance integrals, for which explicit expressions have not been found. These functional expressions are found in the current study via numerical and curve-fitting procedures. Before describing these procedures, we will first review some of the details of the fixed bed forces (drag, Faxen, and the dilute shear stress) and interaction stresses in \citeauthor{harvie2021tensorial}'s final formulation.\\
\\
\textit{Fixed bed forces}\\
The fixed bed forces arise from the volume averaging analysis of the local fluid stress field around a reference particle moving in a bed of particles with uniform solids velocity and volume fraction, equal to the solid phase's field variables ($\us$ and $\phis$). The local fluid velocity in the bed is also uniform and equal to that of the fluid phase velocity $(\uf)$. In this analysis, a $\lambdafix$ variable is used to account for the increased local shear rate and thus resistance to the flow around the fixed bed of particles with the addition of the solid particles. In the current study, this variable is set to be $\lambdafix=\lambdaRZ=\phif^2/f(\phis)$ so that the settling velocity of the non-Brownian particles from the MF momentum equations is calculated as $\us-\um=\phif\left(\us-\uf\right)=2a^2(\rhos-\rhof)\g f(\phis)/\left(9\muf\right)$ \citep{richardson1954sedimentation,richardson1997sedimentation}. As the particles in the current study are not highly Brownian, \citeauthor{richardson1954sedimentation}'s hindered settling function $f(\phis)=(1-\phis)^n$ with $n=4.5$ close to the values suggested for the non-Brownian particles by \citet{richardson1954sedimentation} and \citet{brzinski2018observation} is used for the simulation of the suspension flows. Thus, this calibration of the model's fixed forces is based on the ratio of the drag force on the solid phase to the Stokes drag force on a single particle. Notably, a $\lambdaint$ is also used in the definition of the interaction stresses. This factor is the ratio of the shear rate local to the interacting reference particle to the solid phase shear rate, accounting for the increased viscosity of an effective fluid medium around the particle due to the addition of the other particles. In the current study, this effective viscosity is also assumed to be equal to the ratio of interphase to the Stokes drags as $\lambdaint=\lambdaRZ=\phif^2/f(\phis)$.\\
\\
\textit{Interaction stresses}\\
The interaction stresses in \citeauthor{harvie2021tensorial}'s model are derived as the volume-averaged collision force and stresslet on a reference particle in a proposed form of the anisotropic suspension microstructure. These stresses for the solid and mixture phases are respectively given as:
\begin{equation}\label{eq:harvie-volume-averaged_tau_int_s}
\tens[int,s]{\tau}(\x)=-\frac{1}{2}\frac{\phis(\x)^2}{\scal[p]{V}^2}
\int_{r=2a}^{r=\infty}rp(\x,\vect{r})\vect[col]{f}(\x,\vect{r})dr
\end{equation}
\begin{equation}\label{eq:harvie-volume-averaged_tau_int_m}
\tens[int,m]{\tau}(\x)=-a\frac{\phis(\x)^2}{\scal[p]{V}^2}\int_{r=2a}^{r=\infty}p(\x,\vect{r})\vect[col]{S}(\x,\vect{r})dr
\end{equation}
where in these volume averaging integrals, the collision force is given as $\vect[col]{f}(\x,\vect{r})=\vect[pair]{f}(\x,\vect{r})-\vect[mono]{f}(\x,\vect{r})$ and the collision stresslet is given as $\vect[col]{S}(\x,\vect{r})=\vect[pair]{S}(\x,\vect{r})-\vect[mono]{S}(\x,\vect{r})$. Here $\vect[pair]{f}(\x,\vect{r})$ and $\vect[pair]{S}(\x,\vect{r})$ represent the force and stresslet on a reference particle centred at $\x$ due to the presence of a secondary particle centred at $\x+\vect{r}$. Also, $\vect[mono]{f}(\x,\vect{r})$ and $\vect[mono]{S}(\x,\vect{r})$ denote the force and stresslet on the isolated reference particle centred at $\x$. \citet{harvie2021tensorial} has derived the relations for these collision variables for spherical particles based on semi-analytical studies by \citet{kim2013microhydrodynamics} and \citet{jeffrey1993pressure} that expressed the forces and stresslets in terms of some mobility and resistance scalars versus the separation distance from the reference particle. This collision model assumes that the interacting particle pairs are, in Stokes flow, equal as well as force- and torque-free.\\
\\
The impact of the suspension microstructure on the volume-averaged interaction stresses is determined by the probability distribution function (PDF) $p(\x,\vect{r})$. This variable denotes the conditional probability of having a reference particle centred at $\x$ interacting with a secondary particle centred at $\x+\vect{r}$. \citet{harvie2021tensorial} has proposed the following mathematical form of PDF to account for the anisotropic suspension microstructure in the linear shear flows:
\begin{equation}\label{eq:harvie-p-definition}
p(\x,\vect{r}) = \left( \scal[2]{f}^*(\hat{r})-\scal[1]{f}^*(\hat{r})\tensgammahatdots(\x):\vect{n}\vect{n}\right) q(\hat{r})
\end{equation}
where $\hat{r}=r/a$ is the centre-to-centre separation distance between the reference and secondary particles normalized by the particle radius $a$. In this PDF model, the expression in the parentheses represents the anisotropic modifier function for the isotropic pair distribution function $q(\hat{r})$ defined by \citet{batchelor1972determination}. Based on the Liouville equation and relating the conservation of smooth particle pairs in Stokes flow and their relative velocity, \citet{batchelor1972determination} have argued that for secondary particles on their trajectories at a large separation distance (thus with a uniform distribution), $q(\hat{r})$ is given by:
\begin{equation}\label{eq:batchelor-q-definition}
q(\hat{r}) = \frac{1}{1-A(\hat{r})} \mbox{exp} \left \{\int_{r^{\prime}=\hat{r}}^{r^{\prime}=\infty} \frac{3\left [B(r^{\prime})-A(r^{\prime})\right ]}{r^{\prime}\left [1-A(r^{\prime})\right ]} \mathrm{d}r^{\prime} \right \}  \\\\ \end{equation}
where $A(\hat{r})$ and $B(\hat{r})$ are defined as the functions of certain resistance scalars used to describe relative velocity of particle pairs in Stokes flows, as seen in the table \ref{tab:interaction-viscosities-related-to-indexed-integrals}. Generally, in sheared flows, the presence of closed streamline regions and surface roughness impacts on particle trajectories can result in not all particles following their paths from infinity. Surface asperities greater than the minimum separation distances of around $\hat{r}\simeq 10^{-4}$, which smooth particles can approach, can hinder the particles from reaching closer, alter their path after the contact point, and result in their transfer between different trajectories \citep{da1996shear,wilson2000viscosity}. Due to these differences, the spatial variation of $p(\x,\vect{r})$ in the flows of rough particles can deviate from the isotropic $q(\hat{r})$ distribution found for the smooth particles. The free expressions $\scal[1]{f}^*(\hat{r})$ and $\scal[2]{f}^*(\hat{r})$ along with the local expression $\tensgammahatdots(\x):\vect{n}\vect{n}$, which captures the relative velocity of the secondary particles relative to the reference particle, is proposed to capture this deviation and represent an anisotropic PDF distribution.\\
\\
\citet{harvie2021tensorial} has derived the different components of the two interaction stresses in table \ref{tab:theory-MF-roughness-closure} by substituting the $p(\x,\vect{r})$ model in Eq. (\ref{eq:harvie-p-definition}) into the volume averaging integrals Eqs. (\ref{eq:harvie-volume-averaged_tau_int_s}, \ref{eq:harvie-volume-averaged_tau_int_m}). Thus, the different components of the two interaction stresses, the interaction viscosities in the table \ref{tab:interaction-viscosities-related-to-indexed-integrals}, are expressed in terms of some volume integrals of the mobility and resistance functions that, as mentioned above, define the collision force and stresslet ($\vect[col]{f}$ and $\vect[col]{S}$). Finally, the interaction viscosities were used, along with a phenomenological form of the deformation tensor found in the early literature \citep{hand1962theory,rivlin1997further}, to define the fully tensorial frame-invariant interaction stresses in the table \ref{tab:theory-MF-roughness-closure}. In the following sections, we use piece-wise constant forms of the expressions $\scal[1]{f}^*(\hat{r})$ and $\scal[2]{f}^*(\hat{r})$ to numerically evaluate the mobility and resistance integrals. Curve fitting is then used to determine the functional forms of these integrals and close the MF-roughness model.\\
\begin{table}
\def~{\hphantom{0}}
\renewcommand{\arraystretch}{1.5}
\begin{center}
\resizebox{\columnwidth}{!}{
\begin{tabular}{|wl{\columnwidth/3}|wl{\columnwidth/3}|wl{\columnwidth/3}|}
\hline
\multicolumn{3}{|c|}{Fixed bed forces and stress}\\
\hline
\multicolumn{3}{|l|}{$\vect[faxen,s]{f} = \phis \muf \left(\frac{3}{4}\lambdafix+1\right)\vnabla \cdot \tensgammadotf - \phis\muf\left(\frac{41}{12}\lambdafix+\frac{4}{3}\right)\vnabla(\vnabla\cdot\uf)+\frac{5}{2}\muf\phis\vnabla\lambdafix\cdot\tensgammadotf$}\\
\multicolumn{3}{|l|}{$\vect[drag,s]{f} = \beta(\uf-\us)$ \qquad with \qquad $\beta=\phis \frac{9\muf}{2a^2}\lambdafix$}\\
\multicolumn{3}{|l|}{$\tens[dil,m]{\tau}=-\muf\tensgammadotm-\muf\munhdil\tensgammadotf$ \qquad with \qquad $\munhdil=\frac{5}{2}\lambdafix\phis$}\\
\hline
\multicolumn{3}{|c|}{Interaction stresses}\\
\hline
\multicolumn{3}{|l|}{$\tauintm=-\muf\lambdaint\phis^2 \left [ \muintm \tensgammadots - \gammadots \left ( \musidonem \tensgammahatdots \cdot \tensgammahatdots + \musidtwom \tens{I} \right )+ \gammadotsphs \left ( \mubulkonem \tens{I} - \mubulktwom \tensgammahatdots \right ) \right ]$}\\
\multicolumn{3}{|l|}{$\tauints=-\muf\lambdaint\phis^2 \left [ \muints \tensgammadots - \gammadots \left ( \musidones \tensgammahatdots \cdot \tensgammahatdots + \musidtwos \tens{I} \right )+ \gammadotsphs \left ( \mubulkones \tens{I} - \mubulktwos \tensgammahatdots \right ) \right ]$}\\
\hline
\multicolumn{3}{|c|}{Some derived viscosity definitions ($i$ = s or m)}\\
\hline
$\musidi=\musidonei+\musidtwoi$&$\mubulki=\mubulkonei+\mubulktwoi$&$\munhinti=\lambdaRZ\phis^2\muinti$\\
\hline
$\munhsidonei=\lambdaRZ\phis^2\musidonei$&$\munhsidtwoi=\lambdaRZ\phis^2\musidtwoi$&$\munhsidi=\munhsidonei+\munhsidtwoi$\\
\hline
$\munhbulkonei=\lambdaRZ\phis^2\mubulkonei$&$\munhbulktwoi=\lambdaRZ\phis^2\mubulktwoi$&$\munhbulki=\munhbulkonei+\munhbulktwoi$\\
\hline
\end{tabular}}
\caption{The definitions for the fixed forces (the Faxen, the drag, and dilute shear stress) and the interaction stresses proposed by \citet{harvie2021tensorial} for the semi-dilute suspension flows used for the MF-roughness model in the current study. Here, we define the fixed bed and interaction factors based on hindered settling function by \citet{richardson1954sedimentation} as $\lambdafix=\lambdaint=\lambdaRZ=\left(1-\phis\right)^{2-n}$ with $n=4.5$.}
\label{tab:theory-MF-roughness-closure}
\resizebox{\columnwidth}{!}{
\begin{tabular}{|l|l|l|l|}
\hline
\multicolumn{2}{|l|}{Solid phase}& \multicolumn{2}{l|}{Mixture} \\
\hline
\multicolumn{2}{|l|}{$\muints=\frac{9}{80} (2\hat{C}_{A2}+3\hat{C}_{B2})$}&\multicolumn{2}{l|}{$\muintm=\frac{15}{2}\hat{D}_{K2}+5\hat{D}_{L2}+\hat{D}_{M2}$}\\
\hline
\multicolumn{2}{|l|}{$\musidones=\frac{9}{280} (4\hat{C}_{A1}+3\hat{C}_{B1})$}&\multicolumn{2}{l|}{$\musidonem=2\hat{D}_{L1}+\frac{4}{7}\hat{D}_{M1}$}\\
\multicolumn{2}{|l|}{$\musidtwos=\frac{9}{140}(\hat{C}_{A1}-\hat{C}_{B1})$}&\multicolumn{2}{l|}{$\musidtwom=2\hat{D}_{S1}-\frac{4}{3}\hat{D}_{L1}-\frac{8}{21}\hat{D}_{M1}$}\\
\hline
\multicolumn{2}{|l|}{$\mubulkones=\frac{9}{16}\hat{C}_{A2}$}&\multicolumn{2}{l|}{$\mubulkonem=\frac{15}{2}\hat{D}_{S2}$}\\
\multicolumn{2}{|l|}{$\mubulktwos=\frac{9}{40}\hat{C}_{A1}$}&\multicolumn{2}{l|}{$\mubulktwom=\frac{3}{2}\hat{D}_{K1}+2\hat{D}_{L1}+\hat{D}_{M1}$}\\
\hline
\multicolumn{4}{|c|}{Indexed integrals}\\
\hline
\multicolumn{4}{|c|}{$\scal[\Theta i]{\hat{C}}=\int_{\hat{r}=2}^{\hat{r}=\infty} f_i^* \hat{r}^4 q \Theta \mathrm{d} \hat{r}. \quad i=1,2. \quad \Theta=A, B$}\\
\multicolumn{4}{|c|}{$\scal[\Theta i]{\hat{D}}=\int_{\hat{r}=2}^{\hat{r}=\infty} f_i^* \hat{r}^2 q \Theta \mathrm{d} \hat{r}. \quad i=1,2. \quad \Theta=K, L, M, S$}\\
\hline
\multicolumn{4}{|c|}{$\Theta$ functions versus the scalars $\xi$}\\
\hline
$A=\frac{2}{r}(x_{11}^g+x_{21}^g)$ &
$B=\frac{4}{r}(y_{11}^g+y_{21}^g)$ &$K=\frac{3}{20\pi a^3}z_1^m-1$&$L=\frac{3}{20\pi a^3}(y_1^m-z_1^m)$\\
\hline
$M=\frac{3}{20\pi a^3}(\frac{3}{2}x_1^m-2y_1^m+\frac{1}{2}z_1^m)$ & $P=\frac{r}{10a}(X_{11}^P-X_{12}^P)$ &
$Q=\frac{2}{5}(X_{11}^Q+X_{12}^Q)$ &
$S=Q-PA$ \\
\hline
\end{tabular}}
\caption{The relation between the interaction viscosities and the volume-averaging integrals of the mobility and resistance variables in \citeauthor{harvie2021tensorial}'s interaction stress closures. The viscosities are defined as a function of the indexed integrals ($\scal[\Theta i]{\hat{C}}$, $\scal[\Theta i]{\hat{D}})$ of $\Theta$ functions as the algebraic combinations of the mobility and resistance scalars $\xi$.}.
  \label{tab:interaction-viscosities-related-to-indexed-integrals}
  \end{center}
\end{table}
\begin{table}
\begin{center}
\renewcommand{\arraystretch}{1.5}
\def~{\hphantom{0}}
\begin{tabular}{|l|l|}
\hline
Description&Shear rate variable\\ 
\hline
Deformation tensor&$\tens[tot,i]{\gammadot}=\vnabla\vect[i]{u} + (\vnabla\vect[i]{u})^\text{T}
$\\
\hline
Trace of the deformation tensor&$\scal[sph,i]{\gammadot}=\frac{1}{3} \text{tr}(\tens[tot,i]{\gammadot}) $\\
\hline
Shear rate tensor&$\tens[i]{\gammadot}=\tens[tot,i]{\gammadot} - \scal[sph,i]{\gammadot} \tens{I} $\\
\hline
Shear rate magnitude&$\scal[i]{\gammadot}=\sqrt{\frac{1}{2}\tens[i]{\gammadot}:\tens[i]{\gammadot}}$\\
\hline
Normalised shear rate tensor &$\tens[i]{\hat{\gammadot}}=\frac{\tens[i]{\gammadot}}{\scal[i]{\gammadot}}$\\
\hline
\end{tabular}
\caption{The definition of different shear rate tensors and their magnitudes for the the fluid phase (i=f), the solids phase (i=s), or the mixture (i=m).}
  \label{tab:general-shear-rate-tensors}
  \end{center}
\end{table}
\\
\subsection{Piece-wise constant form of $\scal[1]{f}^*(\hat{r})$ and $\scal[2]{f}^*(\hat{r})$ expressions}
\begin{figure}
\centering
\begin{subfigure}[b]{0.22\textwidth}
\centering
\includegraphics[width=\textwidth]{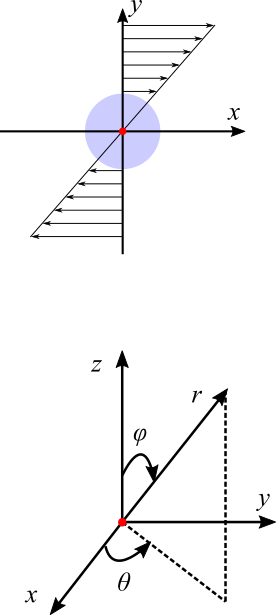}
\caption{}
\label{fig:pdf_coordinate_system}
\end{subfigure}
\begin{subfigure}[b]{0.62\textwidth}
\centering
\includegraphics[width=\textwidth]{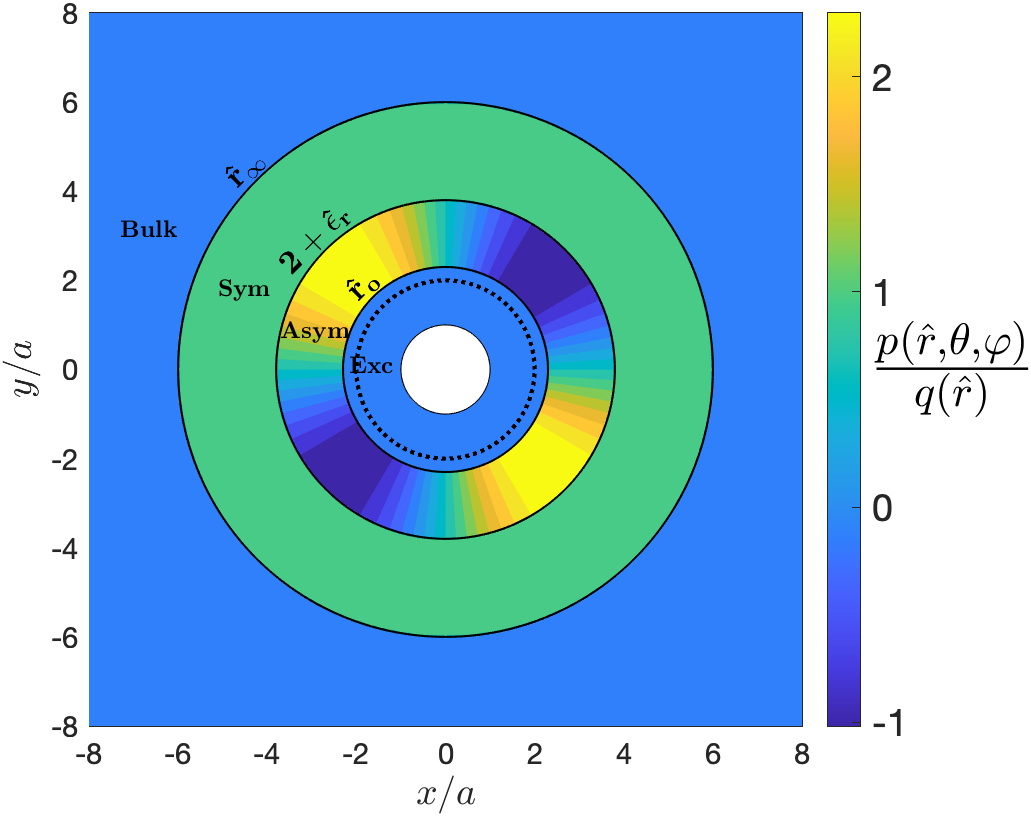}
\caption{}
\label{fig:Mohammad_pdf_2d}
\end{subfigure}
\caption{(a) A schematic of the Cartesian $(x,y,z)$ and spherical $(r,\theta,\varphi)$ coordinate systems local to a reference particle within a linear shear flow. In this unidirectional flow, the velocity, velocity gradient, and vorticity directions align with the x-, y-, and z-axes. The velocity profiles show the local linear shear resulting from the bulk strain rate around the particle. (b) A shear plane view of the PDF model used for the microstructure of the sheared suspensions in the current study. This $p(\hat{r},\theta,\varphi)/q(\hat{r})$ distribution is calculated by substituting the chosen forms of $\scal[1]{f}^*(\hat{r})$ and $\scal[2]{f}^*(\hat{r})$ in Eq. \ref{eq:piece-wise-f1-star-f2-star} into the Eq. \ref{eq:harvie-p-definition}. As discussed earlier, in the local linear shear flow $\tensgammahatdots:\vect{n}\vect{n}=2sin\left(\theta\right)cos\left(\theta\right)sin^2\left(\varphi\right)$ and the relative PDF ratios are related as $\scal[1]{f} = 3\pi\left(\scal[c]{f}-\scal[e]{f}\right)/8$ and $\scal[2]{f} = \left(\scal[c]{f}+\scal[e]{f}\right)/2$. The values of $\scal[c]{f}=1.5$ and $\scal[e]{f}=0$ are chosen based on the optimised simulation results in the subsequent sections. Moreover, some arbitrary screening radii are used to clarify the four regions in the PDF model, where the legends Exc, Asym, Sym, and Bulk denote the exclusion, asymmetric, symmetric, and bulk regions.}
\label{fig:Mohammad_pdf_2d_coordinate_system}
\end{figure}

Finding the functional forms of the interaction viscosities requires defining the expressions $\scal[1]{f}^*(\hat{r})$ and $\scal[2]{f}^*(\hat{r})$ in the PDF model. These expressions are used to calculate the indexed integrals $\scal[\Theta i]{\hat{C}}$ and $\scal[\Theta i]{\hat{D}}$ and thus determine the interaction viscosities in table \ref{tab:interaction-viscosities-related-to-indexed-integrals} for closing the MF-roughness model. In this study, piece-wise constant forms of $\scal[1]{f}^*(\hat{r})$ and $\scal[2]{f}^*(\hat{r})$ expressions are used such that the resulting PDF model aligns with experimental measurements \citep{blanc2011experimental,blanc2013microstructure,gao2010direct} or DNS results \citep{yeo2010simulation,gallier2014rheology} in the literature and effectively captures the influence of surface roughness on creating an anisotropic PDF model. The selected form of the expressions also allows for adjusting the degree of microstructure anisotropy in the simulation of micron-sized particles in the current study based on the particle or the flow properties consistent with the literature's results and observations.\\
\\
Perturbation of particle trajectories due to their surface asperities and near-contact interactions is thought to cause anisotropy of PDF in sheared suspensions. Studies on the impact of surface roughness on particle interactions \citep{da1996shear,gallier2014rheology} indicate that smooth particles interact without direct contact and can come to the close separation distances of around $\hat{r}\approx 10^{-4}$ with symmetrical trajectories. However, surface roughnesses larger than these values result in direct contact between the particles on their approach side, hindering them from getting closer. Surface roughness also causes the trajectories on the receding side to be perturbed to greater separations. Theoretical works by \citet{wilson2000viscosity} and \citet{rampall1997influence} have discussed the impacts of surface roughness inclusion to \citeauthor{batchelor1972determination}'s microstructure model. In shared suspensions, this inclusion along with the closed streamline regions of the flow, can lead to PDF distributions different from from $q(\hat{r})$. They excluded secondary particles from regions with separation distances smaller than the roughness value. These excluded particles aggregate on the contact sheet, creating a high PDF at this separation distance. Similar to trajectories ending at the contacts, the high PDF sheet is also assumed to be perturbed on the receding side. The perturbation of this sheet, along with the closed streamline regions, is thought to make the PDF anisotropic and different from \citeauthor{batchelor1972determination}'s $q(\hat{r})$ formulation.\\
\\
The experimental PDF measurements are consistent with these theoretical models and suggest that the sheared suspensions have an anisotropic microstructure. The measurements of sheared non-Brownian suspensions in the semi-dilute regime by \citet{blanc2011experimental,blanc2013microstructure} and \citet{deboeuf2018imaging} show a narrow layer of high PDF in the shear plane of the flow and near contact. These contact sheets include a secondary particle depletion region that, on the approaching side, is along the velocity direction at approximately $\theta\approx 0$. However, on the receding side, the depletion region rotates towards a positive radial angle, which causes the compression quadrants around the reference particle to have higher average PDF values than the expansion quadrants. \citeauthor{blanc2013microstructure}'s measurements also show that solid fraction variations affect the suspension microstructure. Their results indicate that with an increase in the solid fractions, the depletion region's tilt angle and the contact sheet's maximum PDF, resulting in a more anisotropic PDF near contact. Moreover, a secondary layer of high PDF around $\hat{r}\approx4$ appears at higher concentrations, indicating the importance of triple particle interactions.\\
\\
The PDF measurements of micron-sized Brownian suspensions \citep{gao2010direct,xu2014microstructure} also exhibit a non-uniform contact sheet with a depletion region on the receding side. Consistent with DNS results \citep{morris2002microstructure,nazockdast2012microstructural}, the non-uniformity of the contact sheet in these measurements relatively decreased with the reduction of moderate $Pe$ numbers $(Pe<1000)$. The DNS results show that suspension microstructure and rheology can vary depending on the magnitude of the moderate $Pe$ numbers, representing the significance of the hydrodynamic forces compared to Brownian forces. The random thermal interactions of Brownian particles are thought to reduce the roughness contacts' effect on perturbing the trajectories of secondary particles on the receding side. The contact sheet's non-uniformity thus decreases as the thermal interactions increase in lower Pe numbers, potentially reaching a fully isotropic PDF in no-shear flows $(Pe=0)$ in thermal equilibrium.\\
\\
Piece-wise constant forms of $\scal[1]{f}^*(\hat{r})$ and $\scal[2]{f}^*(\hat{r})$ expressions in Eq (\ref{eq:harvie-p-definition}) can result in an effectively anisotropic microstructure model that is align with the described literature data. We use the following forms of these expressions to partition the PDF model of the suspension microstructure into four distinct regions:
\begin{equation}\label{eq:piece-wise-f1-star-f2-star}
\scal[1]{f}^*(\hat{r}) = \left\{
    \begin{array}{ll}
      0&2<\hat{r}<\hat{r}_o \\[2pt]
      \scal[1]{f}&\hat{r}_o \leq \hat{r} < 2+\scal[r]{\hat{\epsilon}} \\[2pt]
      0&2+\scal[r]{\hat{\epsilon}}\leq \hat{r} < \scal[\infty]{\hat{r}}\left(\phis\right) \\[2pt]
      0&\hat{r} \ge \scal[\infty]{\hat{r}}\left(\phis\right) \\[2pt]
    \end{array} \right.  \qquad \scal[2]{f}^*(r) = \left\{
    \begin{array}{ll}
      0&2<\hat{r}<\hat{r}_o \\[2pt]
      \scal[2]{f}&\hat{r}_o \leq \hat{r} < 2+\scal[r]{\hat{\epsilon}} \\[2pt]
      1&2+\scal[r]{\hat{\epsilon}}\leq \hat{r} < \scal[\infty]{\hat{r}}\left(\phis\right) \\[2pt]
      0&\hat{r} \ge \scal[\infty]{\hat{r}}\left(\phis\right) \\[2pt]
    \end{array} \right.\\
\end{equation}
where $\scal[1]{f}$ and $\scal[2]{f}$ are constant parameters, and $\scal[o]{\hat{r}}$, $2+\scal[r]{\hat{\epsilon}}$, and $\scal[\infty]{\hat{r}}$ are the three screening radii that divide the four regions in the PDF model, as shown in Fig. \ref{fig:Mohammad_pdf_2d_coordinate_system}. The exclusion region is the closest region to the reference particle surface, with the outer radius of $\scal[o]{\hat{r}}$ indicating the minimum separation of the particles and thus having zero PDF values. The width of this region from the pair particle contact surface is assumed to be $\scal[o]{\hat{\hat{r}}}=\scal[o]{\hat{r}}-2=4\times 10^{-5}$ close to the minimum separation distances calculated by \citet{arp1977kinetics} and \citet{da1996shear}. The asymmetric region is the next zone in the PDF model indicating the impact of particle surface asperities on the asymmetry of the trajectories and on the near-contact PDF distribution \citep{blanc2011experimental,gallier2014rheology}. The outer radius of this region defined based on a hydrodynamic roughness value as $\hat{r}=2+\scal[r]{\hat{\epsilon}}$ where $\scal[r]{\hat{\epsilon}}=\scal[r]{\epsilon}/a$. This region captures the effect of direct contact points from surface asperities on hindering particles from getting closer, which causes high pair correlations on the approaching side of particles. It also considers the depletion zone on the receding side of interacting particle pairs due to the surface asperities perturbing the secondary particles' trajectories to higher separation distances \citep{blanc2011experimental,gallier2014rheology}. This trajectory perturbation at the particle scale leads to the SIM behaviour in the macroscopic flow scale. The PDF values in this region due to the difference in the magnitudes of $\scal[1]{f}$ and $\scal[2]{f}$ and the use of $\tensgammahatdots(\x):\vect{n}\vect{n}$ varies around the reference particle. This spherical variation at a specific separation distance $\hat{r}$ from the reference particle centre results in higher average PDF values in the compression quadrants than in the expansion quadrants. To calculate this variation, we refer to the local spherical and Cartesian coordinate systems and flow directions shown in the Fig. \ref{fig:pdf_coordinate_system} and use the projection of the unit normal vector along the particle radius given by $\vect{n}=cos\left(\theta\right)sin\left(\varphi\right)\vect[x]{\delta}+sin\left(\theta\right)sin\left(\varphi\right)\vect[y]{\delta}+cos\left(\varphi\right)\vect[z]{\delta}$. Here, $\vect[x]{\delta}$, $\vect[y]{\delta}$, and $\vect[z]{\delta}$ are the unit normal vectors along the x-, y-, and z-axis of the Cartesian coordinate system. In the unidirectional shear flow around the reference particle, the normalised shear rate tensor is also calculated as $\tensgammahatdot=\begin{bsmallmatrix}
0 & 1 & 0 \\
1 & 0 & 0 \\
0 & 0 &  0 \\
\end{bsmallmatrix}$, where the matrix's first, second, and third directions denote the x-, y-, and z-axis. Thus, it can be shown that in this flow $\tensgammahatdots:\vect{n}\vect{n}=2sin\left(\theta\right)cos\left(\theta\right)sin^2\left(\varphi\right)$, which results in the average $p(r,\theta,\varphi)/q(r)$ in the expansion $(0\le\theta<\pi/2)$ and compression $(\pi/2\le\theta<\pi)$ zones of the asymmetric region given by $\scal[e]{f}= \scal[2]{f}-4\scal[1]{f}/(3\pi)$ and $\scal[c]{f}= \scal[2]{f}+4\scal[1]{f}/(3\pi)$.\\
\\
The third and the fourth zones in the PDF model are the symmetric trajectory region and the bulk region separated by the screening radius $\scal[\infty]{\hat{r}}$. In the symmetric region, the PDF distribution is assumed to be isotropic and equal to $q(\hat{r})$, representing the probability of finding the secondary particles on their original trajectories from infinity. In an effective volume averaging technique, rather than considering the pair interactions to an infinite separation distance, we neglect pair interactions beyond $\scal[\infty]{\hat{r}}$ as the bulk region to capture the effect of multi-body interactions. Notably, the chosen form of isotropic PDF in the symmetric region might differ from some experimental or DNS observations in the literature. Some reported PDF distributions for the suspensions of $\phis<0.2$ show regions with low PDF values (possibly due to closed streamlines) that continue to large separation distances \citep{blanc2011experimental,blanc2013microstructure}. Additionally, high PDF tails emerge along the receding side of interacting particles in these solid fractions. In sheared suspensions of $\phis>0.35$, the observations demonstrate the appearance of a new layer of relatively high pair correlations around $\hat{r}\approx 4$ due to triple particle interactions \citep{morris2002microstructure,blanc2013microstructure}. Although the observations suggest an anisotropic PDF behaviour for the symmetric region, we assume that the effective anisotropic parameters in the asymmetric region capture these.\\
\\
The chosen $f_1^*$ and $f_2^*$ expressions relate the PDF model of sheared suspensions to the average p/q values $\scal[1]{f} = 3\pi\left(\scal[c]{f}-\scal[e]{f}\right)/8$ and $\scal[2]{f} = \left(\scal[c]{f}+\scal[e]{f}\right)/2$, the particles minimum separation screening radius $\scal[o]{\hat{\hat{r}}}=4\times 10^{-5}$, the hydrodynamic roughness value $\scal[r]{\hat{\epsilon}}=\scal[r]{\epsilon}/a$ denoting the region of asymmetric particle trajectories with anisotropic PDF distribution in the microstructure, and $\scal[\infty]{\hat{r}}$ denoting the effective isotropic region of the microstructure with the symmetric particle trajectories. In the current study, we use constant $\scal[c]{f}$, $\scal[e]{f}$, and $\scal[r]{\epsilon}$ values for each type of particle, and discuss the effect of their variation with the flow properties on the interaction viscosities or simulation results by performing sensitivity analysis. Also, as it is discussed in the following, we approximate $\scal[\infty]{\hat{r}}$ as a function of $\phis$ to get a specified number of secondary particles interacting with the reference particles $\scal[{2-\scal[\infty]{\hat{r}}}]{N}$. The impact of varying the constant $\scal[{2-\scal[\infty]{\hat{r}}}]{N}$ value on the interaction viscosities is again studied in a sensitivity analysis. By substituting $f_1^*$ and $f_2^*$ expressions into the definitions of $\scal[\Theta i]{\hat{C}}$ and $\scal[\Theta i]{\hat{D}}$ in the table \ref{tab:interaction-viscosities-related-to-indexed-integrals}, these indexed integrals can be expressed in the form of $\scal[\Theta]{\hat{C}} (\scal[j]{\hat{r}})$ and $\scal[\Theta]{\hat{D}} (\scal[j]{\hat{r}})$ integrals, as shown in table \ref{tab:piece-wise-indexed-integrals}. The latter integrals are evaluated from the surface of the reference particle $\hat{r}=2$ to an upper-bound separation distance of $\scal[j]{\hat{r}}$, which can be one of the three screening radii in the PDF model ($\scal[j]{\hat{r}}=\scal[o]{\hat{r}}$, $2+\scal[r]{\hat{\epsilon}}$, and $\scal[\infty]{\hat{r}}$). The numerical and curve-fitting approaches discussed next allow us to find the functional form of these integrals versus a generic upper-bound separation distance, thus determining the functional forms of the interaction viscosities. 
\begin{table}
\begin{center}
\renewcommand{\arraystretch}{1.75}
\def~{\hphantom{0}}
\begin{tabular}{|l|}
\hline
$\scal[\Theta i]{\hat{C}}$ and $\scal[\Theta i]{\hat{D}}$ indexed integrals\\
\hline
$\scal[\Theta 1]{\hat{C}}=f_1\left (\scal[\Theta]{\hat{C}}(2+\scal[r]{\hat{\epsilon}})-\scal[\Theta]{\hat{C}}(\scal[o]{\hat{r}})\right )$\\
\hline
$\scal[\Theta 2]{\hat{C}}=f_2\left (\scal[\Theta]{\hat{C}}(2+\scal[r]{\hat{\epsilon}})-\scal[\Theta]{\hat{C}}(\scal[o]{\hat{r}})\right )+\left (\scal[\Theta]{\hat{C}}(\scal[\infty]{\hat{r}})-\scal[\Theta]{\hat{C}}(2+\scal[r]{\hat{\epsilon}})\right )$\\
\hline
$\scal[\Theta 1]{\hat{D}}=f_1\left (\scal[\Theta]{\hat{D}}(2+\scal[r]{\hat{\epsilon}})-\scal[\Theta]{\hat{D}}(\scal[o]{\hat{r}})\right )$\\
\hline
$\scal[\Theta 2]{\hat{D}}=f_2\left (\scal[\Theta]{\hat{D}}(2+\scal[r]{\hat{\epsilon}})-\scal[\Theta]{\hat{D}}(\scal[o]{\hat{r}})\right )+\left (\scal[\Theta]{\hat{D}}(\scal[\infty]{\hat{r}})-\scal[\Theta]{\hat{D}}(2+\scal[r]{\hat{\epsilon}})\right )$\\
\hline
$\scal[\Theta]{\hat{C}}$ and $\scal[\Theta]{\hat{D}}$ integrals (for $\scal[j]{\hat{r}}=\scal[o]{\hat{r}}$, $2+\scal[r]{\hat{\epsilon}}$, and $\scal[\infty]{\hat{r}}$)\\
\hline
$\scal[\Theta]{\hat{C}}(\scal[j]{\hat{r}})=\int_{\hat{r}=2}^{\hat{r}=\scal[j]{\hat{r}}} \hat{r}^4 q(\hat{r}) \Theta(\hat{r}) \mathrm{d} \hat{r}, \quad \Theta=A, B$\\
\hline
$\scal[\Theta]{\hat{D}}(\scal[j]{\hat{r}})=\int_{\hat{r}=2}^{\hat{r}=\scal[j]{\hat{r}}} \hat{r}^2 q(\hat{r}) \Theta(\hat{r}) \mathrm{d} \hat{r} , \quad \Theta=K, L, M, S$\\
\hline
\end{tabular}
\caption{The conversion of the indexed integrals $\scal[\Theta i]{\hat{C}}$ and $\scal[\Theta i]{\hat{D}}$, which define the interaction viscosities (table \ref{tab:interaction-viscosities-related-to-indexed-integrals}), to the integrals $\scal[\Theta]{\hat{C}}$ and $\scal[\Theta]{\hat{D}}$ based on the chosen form of the PDF model in current study (Eq. (\ref{eq:piece-wise-f1-star-f2-star})).}
  \label{tab:piece-wise-indexed-integrals}
  \end{center}
\end{table}
\subsection{Functional form of $\scal[\Theta]{\hat{C}}$ and $\scal[\Theta]{\hat{D}}$ integrals}
\begin{figure}
\centering
\includegraphics[width=0.6\textwidth]{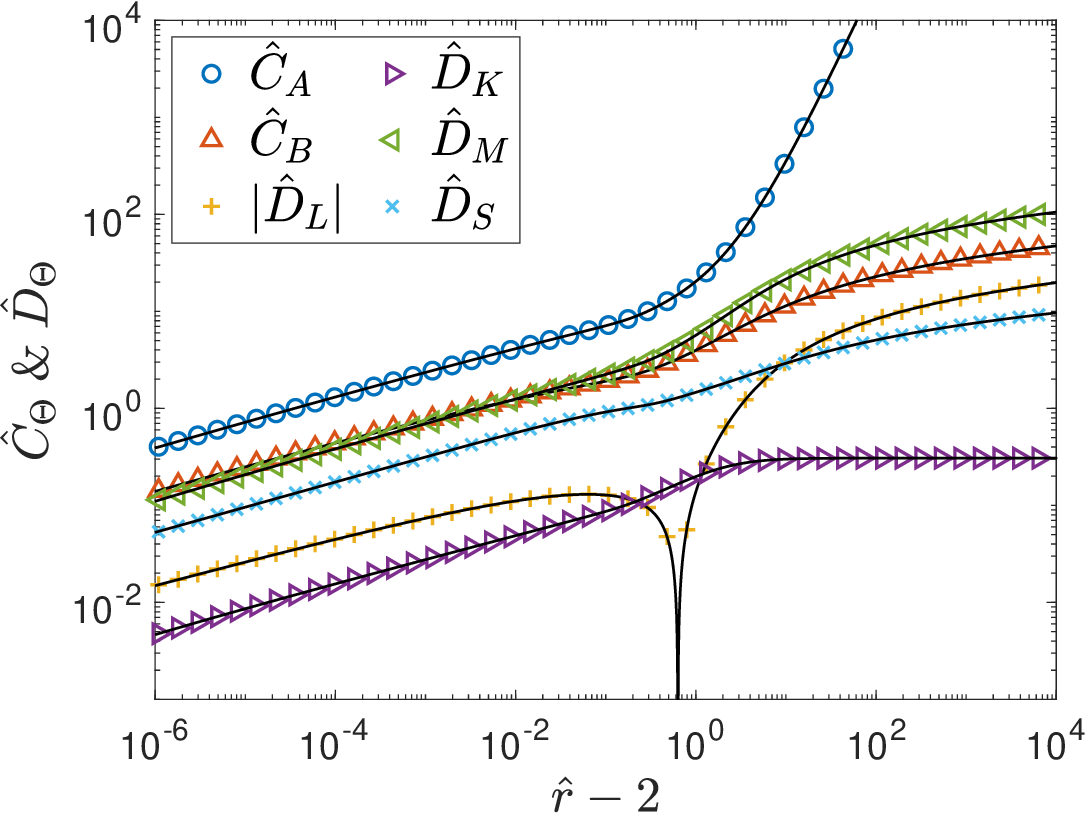}
\caption{The variation of $\scal[\Theta]{\hat{C}}(\hat{r})$ and $\scal[\Theta]{\hat{D}}(\hat{r})$ integrals with the separation distance. The symbols are the numerical data, and the solid lines are the fitting functions whose relative differences are less than $2\%$ (as shown in section S1 in the supplementary document).}
\label{fig:mobility-resistance-integrals}
\end{figure}

Calculating the functional forms of the volume averaging integrals $\scal[\Theta]{\hat{C}}$ and $\scal[\Theta]{\hat{D}}$, requires considering $\Theta$ and $q$ as functions of the mobility or resistance scalars $\xi(\hat{r})$. These scalars describe the different moments of the disturbed fluid on the reference particle due to its Stokes flow interaction with the secondary particle. When dealing with widely separated particle pairs in far-field interactions, the methods of reflection and multipole expansions are used to approximate $\xi(\hat{r})$ as an asymptotic mathematical series \citep{jeffrey1984calculation}. However, for close-contact particles in near-field interactions, where the mathematical series convergence rate is low, $\xi(\hat{r})$ scalars are approximated by analysing the fluid film in the gap between particles using lubrication theory \citep{kim2013microhydrodynamics}. This theory results in a more complex mathematical expression for $\xi(\hat{r})$, which consequently complicates the direct calculation of the expressions for the integrals $\scal[\Theta]{\hat{C}}$ and $\scal[\Theta]{\hat{D}}$. Therefore, we determine the numerical values of $\scal[\Theta]{\hat{C}}$ and $\scal[\Theta]{\hat{D}}$ integrals and then derive their mathematical expressions using curve-fitting. The procedure for calculating a series of numerical values for $\scal[\Theta]{\hat{C}}$ and $\scal[\Theta]{\hat{D}}$ integrated from the surface of the reference particle at $\hat{r}=2$ to a series of $\hat{r}$ values and subsequently performing curve-fitting on these numerical series, is as follows:\\
\begin{enumerate}
  \item Defining the near-field and far-field analytic expressions of the scalars $\xi(\hat{r})$. The expressions for the mobility scalars ($x_{11}^a$, $x_{21}^a$, $y_{11}^a$, $y_{21}^a$, $x_{11}^g$, $x_{21}^g$, $y_{11}^g$, $y_{21}^g$, $x_1^m$, $y_1^m$, and $z_1^m$) collected from \citet{kim2013microhydrodynamics} and the expressions for the resistance scalars ($X_{11}^P$, $X_{12}^P$, $X_{11}^Q$, and $X_{12}^Q$) extracted from \citet{jeffrey1993pressure}.\\
  \item Defining scalar expressions working for an arbitrary separation distance. The collected near- and far-field expressions are merged as:
  \begin{equation}
  \xi\left(\hat{r}\right)=(1-\scal[\xi]{\omega}\left(\hat{r}\right))\scal[small]{\xi}\left(\hat{r}\right)+\scal[\xi]{\omega}\left(\hat{r}\right)\scal[large]{\xi}\left(\hat{r}\right)
  \end{equation}
  where the small and large subscripts denote the far- and near-field separations and the merge function is given by $\scal[\xi]{\omega}=\scal[small]{e}/\left(\scal[small]{e}+\scal[high]{e}\right)$. The two error functions are defined as $\scal[small]{e}(\hat{r})=\scal[small]{a}\left(\hat{r}-2\right)^{\scal[small]{n}}$ and $\scal[large]{e}(\hat{r})=\scal[large]{a}\hat{r}^{\scal[large]{n}}$, where $\scal[small]{a}=0.003$, $\scal[small]{n}=3/2$, $\scal[large]{a}=1$, and $\scal[large]{a}=-13$. These error functions are chosen to smoothly transition the merged expressions between the near- and far-field scalars (as plotted in the section S1 in the supplementary document).\\
  \item Defining the merged analytic forms of $\Theta(\hat{r})$ functions ($A$, $B$, $K$, $L$, $M$, $P$, $Q$, and $S$) by substituting the calculated merged $\xi(\hat{r})$ expressions into the formulations in the table \ref{tab:interaction-viscosities-related-to-indexed-integrals}.\\
  \item Calculating the numerical values of $q(\hat{r})$ by using the merged expressions of functions $A(\hat{r})$ and $B(\hat{r})$ in Eq. \ref{eq:batchelor-q-definition}, and solving the integral for each value of $\hat{r}$ numerically. A series of numerical $\hat{r}$ values increasing from 2, with an initial step size of $1\times 10^{-10}$ and a growth factor of $1.001$, to the maximum value set at $10000$ is used. Section S1 in the supplementary document presents a sensitivity analysis on the step size growth factor revealing that a five times lower value of this parameter changes the values of $\scal[\Theta]{\hat{C}}(\hat{r})$ and $\scal[\Theta]{\hat{D}}(\hat{r})$ integrals in the practical $\hat{r}$ values by less than $3\%$.\\
  \item Calculating the numerical values of $\scal[\Theta]{\hat{C}}(\hat{r})$ and $\scal[\Theta]{\hat{D}}(\hat{r})$ using the merged $\Theta(\hat{r})$ expressions and $q(\hat{r})$ numerical values in the integrals defined in table \ref{tab:piece-wise-indexed-integrals}, and solving them numerically for each value of $\hat{r}$. The lower and upper bounds of these integrals are respectively $\hat{r}=2$ and a $\hat{r}$ value chosen from the numerical series described in the previous step. A sensitivity analysis in the first section of the supplementary data demonstrates that using a higher resolution of the numerical $\hat{r}$ values or changing the shape of the merge function $\scal[\xi]{\omega} \left(\hat{r}\right)$ has very limited impact on the calculated integral values.\\
  \item Finally, the resulting numerical values of the integrals versus the separation distance were fitted as $\scal[\Theta]{\hat{C}}(\hat{r})\ \&\  \scal[\Theta]{\hat{D}}(\hat{r})=(1-\scal[\Theta]{\omega})\scal[\Theta,small]{F}+\scal[\Theta]{\omega}\scal[\Theta,large]{F}$ using the curve fitting toolbox in MATLAB \citet{matlab2020curvefitting}. Fig. \ref{fig:mobility-resistance-integrals} compares numerical and fitting function profiles of different mobility and resistance integrals $\scal[\Theta]{\hat{C}}(\hat{r})$ and $\scal[\Theta]{\hat{D}}(\hat{r})$, which based on the supplementary document (section S1), all sets of the numerical data were fitted with relative errors less than $2\%$. As detailed in the table \ref{tab:Chat-Dhat-fitting-functions}, two different expressions $\scal[\Theta,small]{F}$ and $\scal[\Theta,large]{F}$ are used for the small and large separation distances merged with the expression $\scal[\Theta]{\omega}$ to get the functional forms for an arbitrary separation distance. Power law functions as $\scal[\Theta,small]{F}=a\left(\hat{r}-2\right)^b+c\left(\hat{r}-2\right)^d$ are used for the low separations. We used these power functions as all the $\Theta$ functions, as well as $q$, were found to follow a powered function of $\hat{r}-2$ for the separation distances lower than $\hat{r}<2.01$. The variable $q$ scales as $\hat{r}^{-2}(\hat{r}-2)^{-0.75}$, which is consistent with nearly field assymtotes used by \citet{brady1997microstructure} as $(\hat{r}-2)^{-0.78}$ and by \citet{batchelor1972determination} as $ -(\log(\hat{r}-2))^{-0.29}(\hat{r}-2)^{-0.78}$. Using two different power-law functions in low separations helped us to perfectly fit the numerical values. All the $F_{\Theta,large}$ expressions, except for their constant term $e$, are derived based on the far-field limits of the integrals using a Taylor expansion. For this, we have used the expansion of far-field $\Theta$ functions along with $q(\hat{r})=1+25/(2\hat{r}^6)$ as estimated by \citet{batchelor1972determination}, then analytically calculated the integrals\footnote{Using Taylor series expansions of the far-field $A$ and $B$ functions, we could find higher order approximation of far-field $q$ to be as $1+25/(2\hat{r}^6)-165/(8\hat{r}^8)+125/\hat{r}^9-75//\hat{r}^{10}+...$, which fits the numerical $q$ values for the separation distances $\hat{r}>2.2$ with a relative error less than $2\%$.}. The form of the fitting functions for the small and large separation distances are merge together with expressions of the form $\scal[\theta]{\omega} =  (\hat{r} -2)^n/\left(\beta+(\hat{r} -2)^n\right)$ to get the fitting functions for an arbitrary separation distance, which then the listed parameter values $a$, $b$, $c$, $d$, $e$, $\beta$, and $n$ are found by fitting the merged expression to the numerical values.
\end{enumerate}
\begin{table}
\begin{center}
\renewcommand{\arraystretch}{2}
\def~{\hphantom{0}}
\resizebox{\columnwidth}{!}{\begin{NiceTabular}{|c|cccc|lc|cc|}
\cline{1-9}
&\Block{1-4}{
Small $\hat{r}$ region\\
$\scal[\Theta,small]{F}=a\left(\hat{r}-2\right)^{b}+c\left(\hat{r}-2\right)^{d}$}&&&&\Block{1-2}{
Large $\hat{r}$ region}&&\Block{1-2}{Merge $(\scal[\Theta]{\omega})$}&\\
$\Theta$ & $a$ & $b$ & $c$ & $d$ & $\scal[\Theta,large]{F}$& $e$ & $\beta$ & $n$ \\\hline 
$A$ & -25 & 0.3361 & 36.51 & 0.2937 &$5\hat{r}^2/2-25/\hat{r}-8\log(\hat{r})+e$& 15.08 & 0.0129 & 3.084 \\ 
$B$ & -4.0 & 0.4096	& 6.375 & 0.2698 &$(16/3)\log(\hat{r})+e$&-1.849 & 0.0261& 2.303 \\
$K$ & -0.1511 & 0.2428 & 0.00062 & 0 &$1/\hat{r}^2-5/(8\hat{r}^5)+e$ & -0.307 & 0.0075 & 3.093 \\
$L$ & 0.4794 & 0.25 & -0.4959 & 0.5441 &$-(5/2)\log(\hat{r})-5/\hat{r}^2-25/(12\hat{r}^3)+e$ & 3.229 & 0.100 & 4.099 \\
$M$ & -5.822 & 0.1998 & 9.742 & 0.2181 &$(25/2)\log(\hat{r})+35/(2\hat{r}^2)+e$ & -9.661 & 0.1964 & 2.549 \\
$S$ & 1.901 & 0.2594 & -1.216 & 0.95 &$\log(\hat{r})+e$ & 0.4375 & 0.0927 & 2.559 \\
\hline
\end{NiceTabular}}
\caption{Functional forms of the mobility and resistance integrals obtained using curve fitting. The integral expressions for an arbitrary separation distance are given as $\scal[\Theta]{\hat{C}}(\hat{r})\ \&\  \scal[\Theta]{\hat{D}}(\hat{r})=(1-\scal[\Theta]{\omega})\scal[\Theta,small]{F}+\scal[\Theta]{\omega}\scal[\Theta,large]{F}$. The fitting functions in the small $(\scal[\Theta,small]{F})$ and large $(\scal[\Theta,large]{F})$ separations are shown in the table. The merging functions are defined as $\scal[\Theta]{\omega} =  (\hat{r} -2)^n/\left(\beta+(\hat{r} -2)^n\right)$.
}
\label{tab:Chat-Dhat-fitting-functions}
\renewcommand{\arraystretch}{1.5}
\def~{\hphantom{0}}
\begin{tabular}{|c|c|}
\hline
$\scal[\infty]{\hat{r}}$ & $2+\left (1-\omega \right)\scal[\infty,small]{\hat{r}} + \omega \scal[\infty,large]{\hat{r}}$\\
\hline
$\scal[\infty,small]{\hat{r}}$&$a\left (\frac{\scal[{2-\scal[\infty]{\hat{r}}}]{N}}{\phis}\right)^b$\\
\hline
$\scal[\infty,large]{\hat{r}}$&$\left(c+\frac{\scal[{2-\scal[\infty]{\hat{r}}}]{N}}{\phis} \right)^d-2$\\
\hline
$\omega$&$\left(\frac{\scal[{2-\scal[\infty]{\hat{r}}}]{N}}{\phis}\right)^{n}\left[\beta+\left(\frac{\scal[{2-\scal[\infty]{\hat{r}}}]{N}}{\phis}\right)^n\right]^{-1}$\\
\hline
\end{tabular}
\caption{The relation between the screening radius $\scal[\infty]{\hat{r}}$ and the total number of particles located between the surface of the reference particle and this radius, $\scal[{2-\scal[\infty]{\hat{r}}}]{N}$ at a specific $\phis$ value obtained using a curve fitting procedure. The parameter values are found as $a=1\times 10^{-4}$, $b=4$, $c=2.1366$, $d=1/3$, $\beta=1.3955\times 10^7$, and $n=8.329$.}
\label{tab:r_hat_infty_fitting_function}
\end{center}
\end{table}
\subsection{Functional form of $\scal[\infty]{\hat{r}}$}
\begin{figure}
\centering
\includegraphics[width=0.6\textwidth]{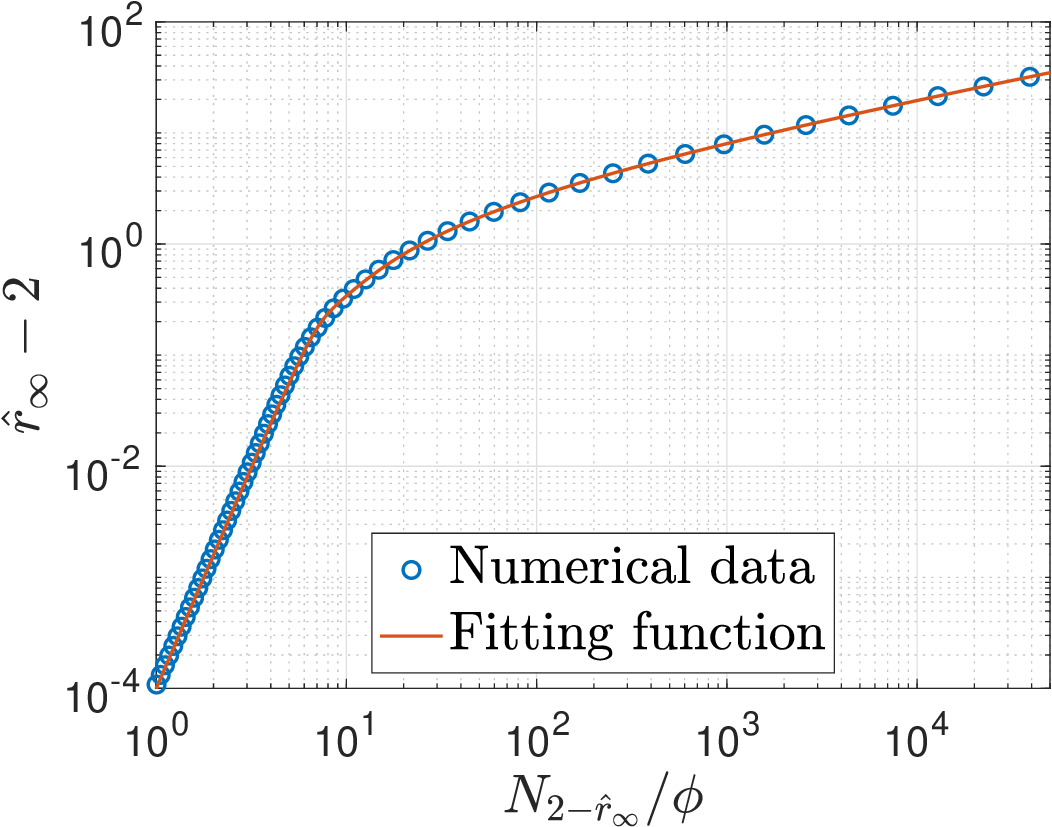}
\caption{The variation of the calculated $\scal[\infty]{\hat{r}}$ values as the outer radius of the symmetric region in the PDF model versus $\scal[{2-\scal[\infty]{\hat{r}}}]{N}/\phi$. The relative differences between numerical and fitting function values are less than $1\%$ (as shown in section S1 in the supplementary document).}
\label{fig:r_hat_infty}
\end{figure}

The screening length $\scal[\infty]{\hat{r}}$ can be defined in a way that only the interaction of the reference particle with a specified number of secondary particles is considered in the interaction viscosities. This effective radius denotes the fictitious outer surface of the symmetric region in the PDF model beyond which, in an effective averaging technique, the pair interactions are neglected rather than using a re-normalisation method to account for the complexities arising from the multi-particle or long-range interactions in the semi-dilute suspension flows. By neglecting the difference between the average $q(\hat{r})$ and $p(\hat{r},\theta,\varphi)$ in the asymmetric region of the PDF model, we can estimate the number of secondary particles $\scal[{2-\scal[\infty]{\hat{r}}}]{N}$ within the surface of the reference particle to outer surface of the symmetric region as:
\begin{equation}\label{eq:N_r_integral}
\scal[{2-\scal[\infty]{\hat{r}}}]{N}= 3\phis\int_{\hat{r}=2}^{\hat{r}=\scal[\infty]{\hat{r}}} \hat{r}^2 q\left(\hat{r}\right)\mathrm{d}\hat{r}
\end{equation}
Using the numerical series of $q(\hat{r})$ discussed above, the numerical values of $\scal[{2-\scal[\infty]{\hat{r}}}]{N}/\phis$ for each substituted value of $\scal[\infty]{\hat{r}}$ are calculated. Having calculated the numerical series of $\scal[{2-\scal[\infty]{\hat{r}}}]{N}/\phis$ values versus $\scal[\infty]{\hat{r}}$ values, we used curve-fitting tools and fitted two functions to these numerical data at the small and large separation distances, merged with an intermediate function. Fig. \ref{fig:r_hat_infty} shows the variation of $\scal[\infty]{\hat{r}}$ values with $\scal[{2-\scal[\infty]{\hat{r}}}]{N}/\phis$ and the table \ref{tab:r_hat_infty_fitting_function} lists the fitting functions. The forms of these functions in the low and high separation distances are again chosen based on the far and near field forms of $q$ discussed above implemented into the integral in the Eq. (\ref{eq:N_r_integral}). To close the MF-roughness model, we use a constant $N_{2-\scal[\infty]{\hat{r}}}=1$ and study its variation impact on interaction viscosities by conducting a sensitivity analysis. It is also noteworthy that in the current study, we implement $\scal[\infty]{\hat{r}}+\scal[\infty,min]{\hat{r}}$ with $\scal[\infty,min]{\hat{r}}=0.2$ as the outer radius of the symmetric region of the PDF model instead of $\scal[\infty]{\hat{r}}$. This prevents the overlap between the symmetric and asymmetric regions of the PDF model, thus avoiding numerical problems that can occur at high concentrations when $\scal[\infty]{\hat{r}}$ approaches $2+\scal[r]{\hat{\epsilon}}$. The added minimum separation results in a nearly constant value of $\hat{r}=2.2$ as the outer radius of the symmetric region in the solid fractions higher than $\phis>0.3$. It can be justified to incorporate a minimum effective screening in high concentrations as multi-particle effects can induce higher screening of interactions compared to the pairwise $\scal[\infty]{\hat{r}}$ values. A sensitivity analysis is presented in the subsequent sections to discuss how variations in $\scal[\infty,min]{\hat{r}}$ affect the model viscosities.\\
\subsection{MF-roughness viscosity plots}
\begin{figure}
\centering
\begin{subfigure}[b]{1\textwidth}
\centering
\includegraphics[width=1\textwidth]{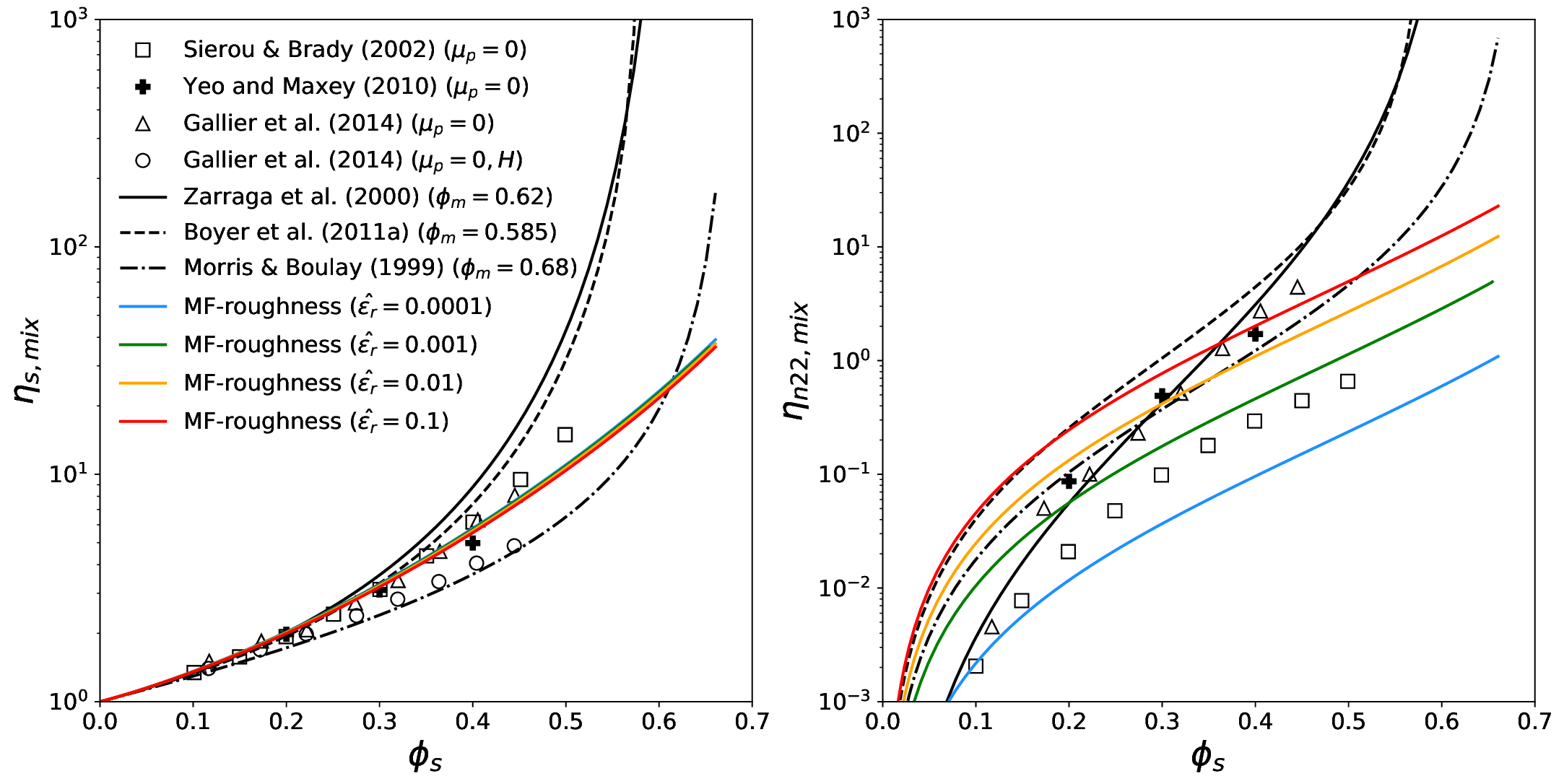}
\caption{The relative shear and the normal viscosities of the mixture phase versus $\phis$}
\label{fig:theory_viscosities_etas_etan}
\end{subfigure}
\begin{subfigure}[b]{\textwidth}
\centering
\includegraphics[width=1\textwidth]{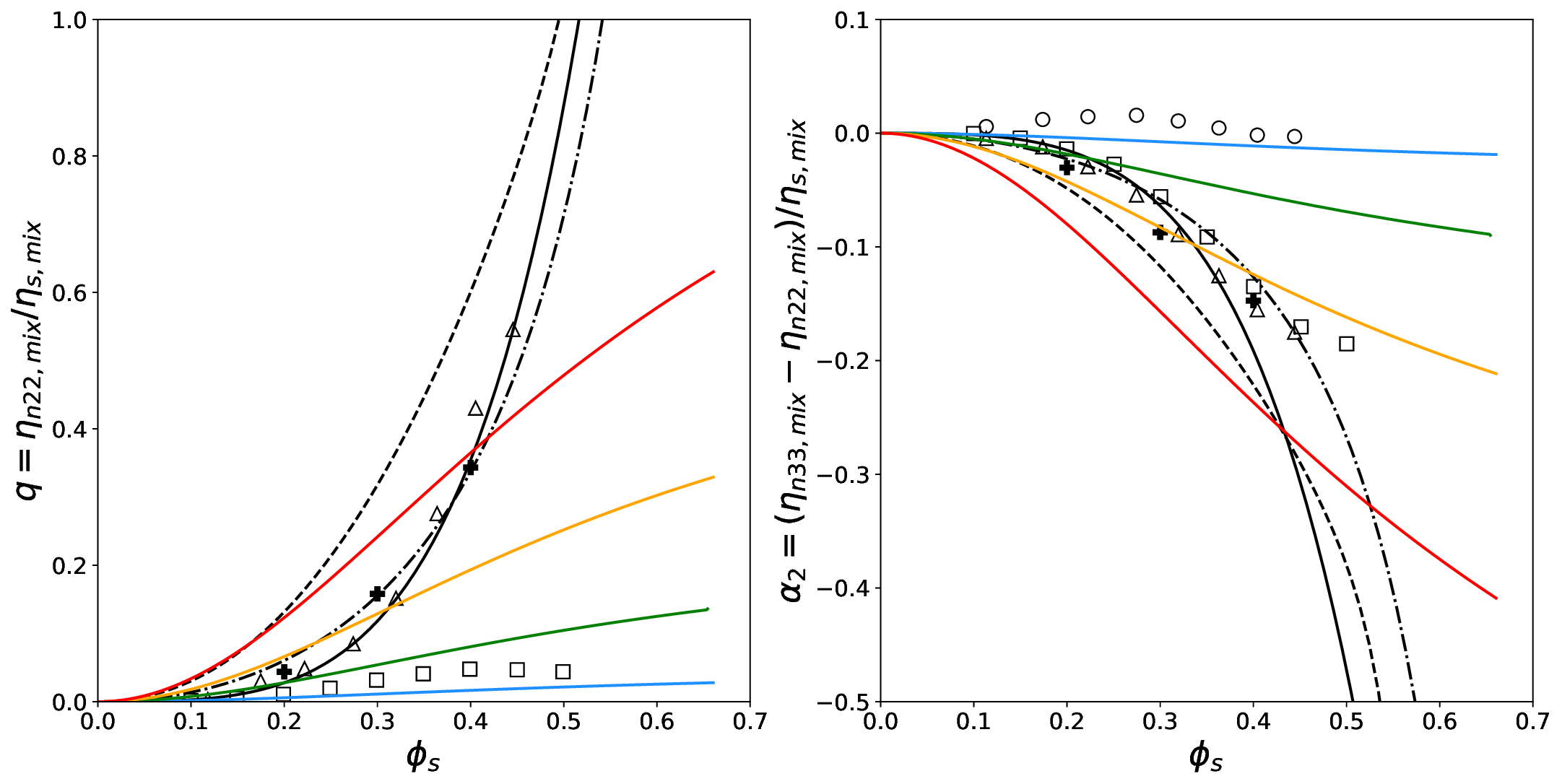}
\caption{The viscosity ratios $q=\scal[n22,mix]{\eta}/\scal[s,mix]{\eta}$ and $\scal[2]{\alpha}=(\scal[n33,mix]{\eta}-\scal[n22,mix]{\eta})/\scal[s,mix]{\eta}$ versus $\phis$}
\label{fig:theory_viscosities_q_alpha2}
\end{subfigure}
\caption{The MF-roughness mixture phase viscosities (a) and viscosity ratios (b) compared with the literature's DNS results, as well as experimental and phenomenological correlations. The legend parameter $\scal[p]{\mu}$ represents the friction coefficient used in the DNS techniques, while the label $H$ represents the hydrodynamic contribution to the DNS results reported by \citet{gallier2014rheology}. The MF-roughness viscosities are calculated based on the parameter values $\scal[o]{\hat{\hat{r}}}=4\times 10^{-5}$, $\scal[{2-\scal[\infty]{\hat{r}}}]{N}=1$, $\scal[c]{f}=1.5$, $\scal[e]{f}=0$ and the labeled $\scal[r]{\hat{\epsilon}}$ values.}
\label{fig:theory_viscosities_etas_etan_q_alpha2}
\end{figure}

\begin{figure}
\centering
\includegraphics[width=0.6\textwidth]{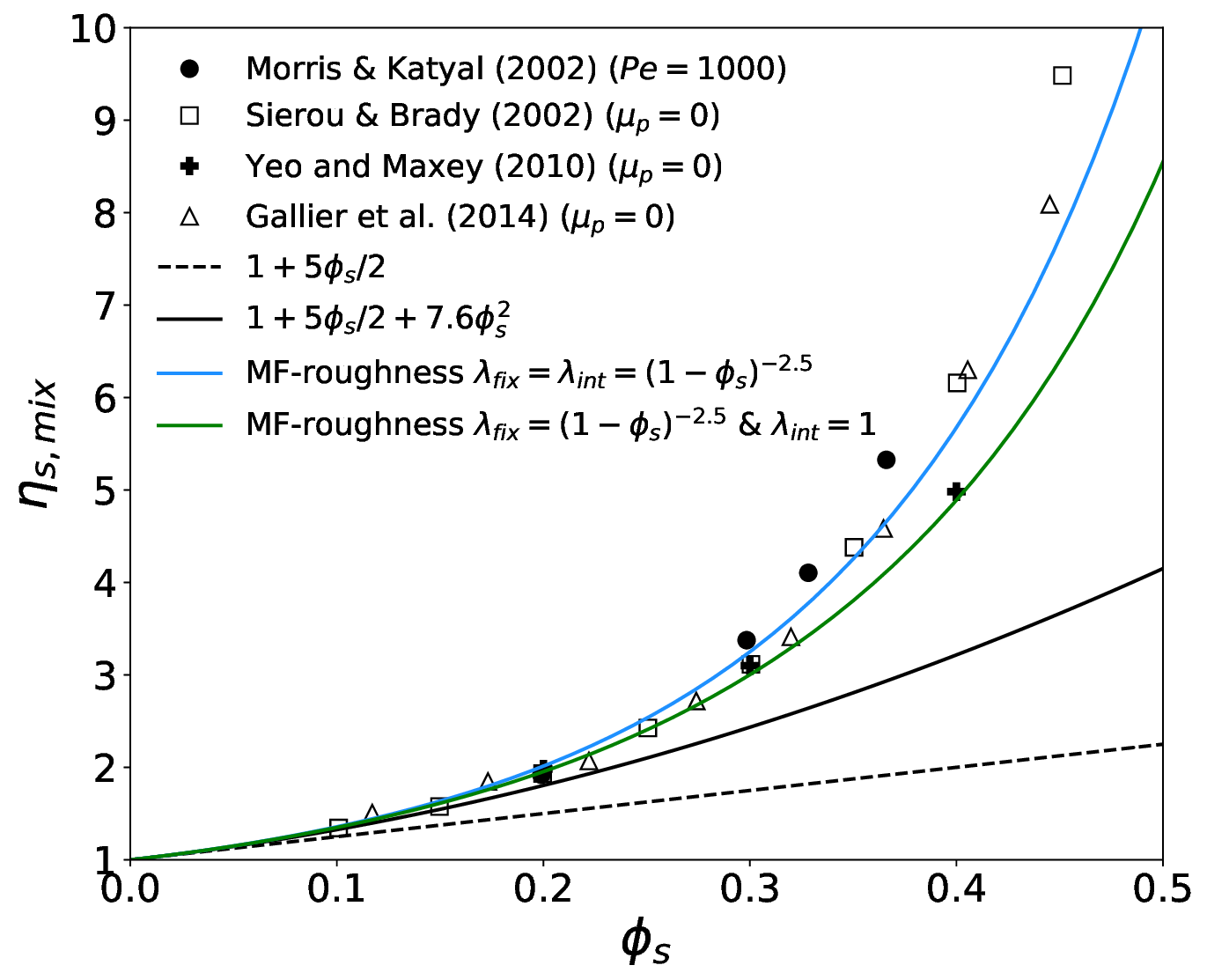}
\caption{The mixture phase shear viscosity of the MF-roughness model in a linear shear flow compared to the theoretical correlations and the DNS results. The MF-roughness viscosities are calculated based on the parameter values $\scal[o]{\hat{\hat{r}}}=4\times 10^{-5}$, $\scal[{2-\scal[\infty]{\hat{r}}}]{N}=1$, $\scal[c]{f}=1.5$, $\scal[e]{f}=0$, and $\scal[r]{\hat{\epsilon}}=0.005$.}
\label{fig:theory_viscosities_etas_effective_viscosity_tests}
\end{figure}

\begin{figure}
\centering
\includegraphics[width=0.6\textwidth]{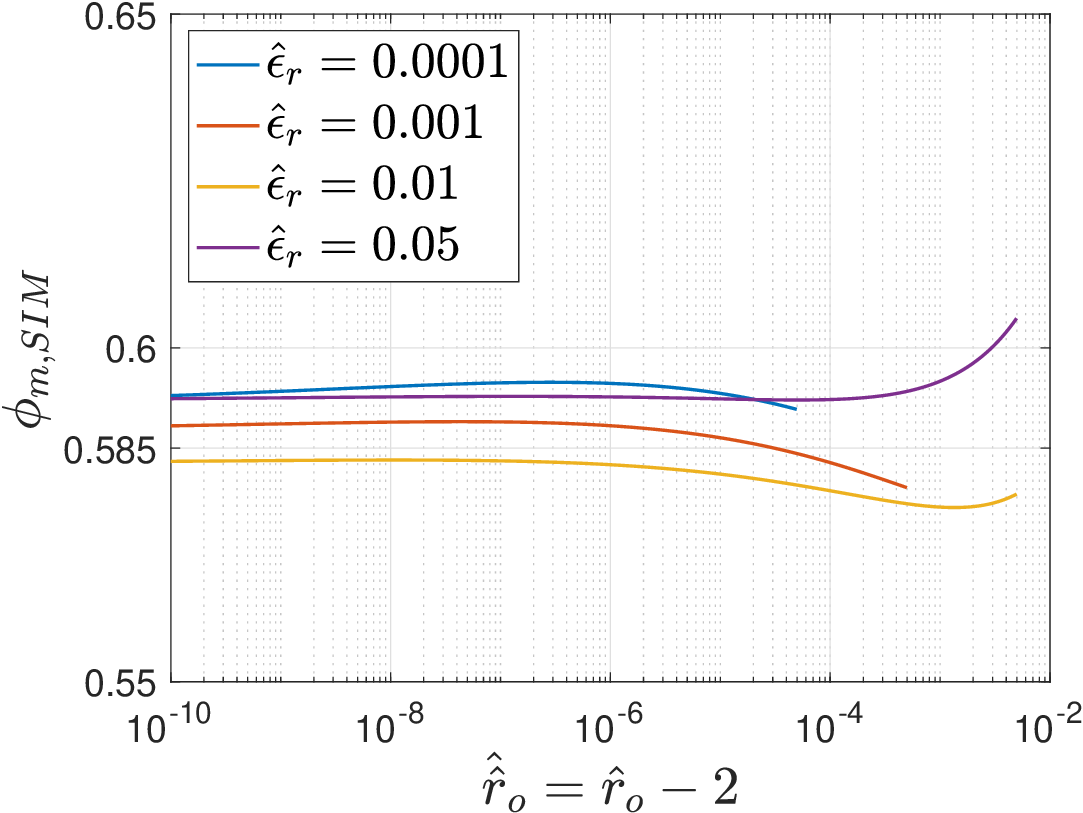}
\caption{The solid-to-mixture phase second normal viscosity $\scal[m,SIM]{\phi}=\scal[n22,sol]{\eta}/\scal[n22,mix]{\eta}$ in the MF-roughness model for a wide range of $\scal[o]{\hat{\hat{r}}}$ and $\scal[r]{\hat{\epsilon}}$ values. This viscosity ratio is independent of the other closure parameters $\scal[c]{f}$, $\scal[e]{f}$, and $\scal[{2-\scal[\infty]{\hat{r}}}]{N}$.}
\label{fig:roughness_model_phi_m_SIM}
\end{figure}

In this section, the MF-roughness relative viscosities are compared to DNS results from the literature \cite{morris2002microstructure, sierou2002rheology, yeo2010dynamics, gallier2014rheology}, as well as phenomenological correlations by \citeauthor{morris1999curvilinear} and experimental correlations suggested by \citet{zarraga2000characterization} and \citet{boyer2011unifying}. The total solid and mixture phase stresses and the relative shear and normal viscosities are defined in table \ref{tab:different-stress-shear-normal-viscosity-variables}, and the relative viscosities can be related to the interaction viscosities in unidirectional shear flows, as shown in table \ref{tab:harvie-stress-closures-in-unidirectional-flows}. The functional forms of the MF-roughness interaction viscosities described in the previous sections were related to the parameters $\scal[1]{f} = (3\pi/8)\left(\scal[c]{f}-\scal[e]{f}\right)$, $\scal[2]{f} = (1/2)\left(\scal[c]{f}+\scal[e]{f}\right)$, $\scal[o]{\hat{\hat{r}}}=4\times 10^{-5}$, $\scal[r]{\hat{\epsilon}}$, and $\scal[{2-\scal[\infty]{\hat{r}}}]{N}=1$. Here, the relative viscosities are calculated using $\scal[c]{f}=1.5$ and $\scal[e]{f}=0$, based on the optimization results of SIM observations in the following sections for different $\scal[r]{\hat{\epsilon}}$ values, potentially describing particles ranging from nearly smooth to very rough by varying the width of the asymmetric region in the PDF model. Fig. \ref{fig:theory_viscosities_etas_etan_q_alpha2} shows the relative shear and normal viscosities $\scal[s,mix]{\eta}$ and $\scal[n22,mix]{\eta}$ of the mixture along with the normal to shear viscosity ratios $q=\scal[n22,mix]{\eta}/\scal[s,mix]{\eta}$ and $\scal[2]{\alpha}=(\scal[n33,mix]{\eta}-\scal[n22,mix]{\eta})/\scal[s,mix]{\eta}$.\\
\begin{table}
\def~{\hphantom{0}}
\renewcommand{\arraystretch}{1.5}
\begin{center}
\begin{tabular}{|l|l|l|}
\hline
Description & Solid phase & Mixture\\
\hline
Total stress & $\tens[s]{\tau}=\tauints$ & $\tens[m]{\tau}=\taudilm+\tauintm$ \\
\hline
Shear viscosity&
$\scal[s,sol]{\eta}=\left|\scal[s12]{\tau}\right|/\left(\muf\gammadotm\right)$ & $\scal[s,mix]{\eta}=\left|\scal[m12]{\tau}\right|/\left(\muf\gammadotm\right)$ \\
\hline
Normal viscosity&$\scal[nii,sol]{\eta}=\scal[sii]{\tau}/\left(\muf\gammadotm\right)$ & $\scal[nii,mix]{\eta}=\scal[mii]{\tau}/\left(\muf\gammadotm\right)$\\
\hline
\multicolumn{3}{|c|}{Viscosity ratios}\\
\hline
$q=\scal[n22,mix]{\eta}/\scal[s,mix]{\eta}$&$\scal[2]{\alpha}=\left(\scal[n33,mix]{\eta}-\scal[n22,mix]{\eta}\right)/\scal[s,mix]{\eta}$&$\scal[m,SIM]{\phi}=\scal[n22,sol]{\eta}/\scal[n22,mix]{\eta}$\\
\hline
\end{tabular}
\caption{The relation between the different relative viscosities and the components of the total solid and mixture phase stress tensors. The scalars $\scal[sij]{\tau}$ and $\scal[mij]{\tau}$ respectively represent the components of the solid and mixture stress tensors (where i,j = 1, 2, or 3), with the indices 1, 2, and 3 indicating the velocity, velocity gradient, and vorticity directions in a linear shear flow, respectively.}
  \label{tab:different-stress-shear-normal-viscosity-variables}
\end{center}
\def~{\hphantom{0}}
\renewcommand{\arraystretch}{1.75}
\begin{center}
\begin{tabular}{|c|}
\hline
\citeauthor{harvie2021tensorial}'s frame invariant stresses in a unidirectional flow\\
\hline
$\tens[m]{\tau}=\muf\gammadot\begin{bmatrix}
\munhsidonem+\munhsidtwom & -1-\munhdil-\munhintm & 0 \\
-1-\munhdil-\munhintm & \munhsidonem+\munhsidtwom & 0 \\
0 & 0 & \munhsidtwom \\ \end{bmatrix}$ \\
\hline
$\tens[s]{\tau}=\muf\gammadot\begin{bmatrix}
\munhsidones+\munhsidtwos & -\munhints & 0 \\
-\munhints & \munhsidones+\munhsidtwos & 0 \\
0 & 0 & \munhsidtwos \\ \end{bmatrix}$ \\
\hline
\end{tabular}
\caption{The matrices of the solid and mixture phase stress tensors resulting from \citeauthor{harvie2021tensorial}'s closure in a unidirectional flow. The stresses' first, second, and third directions are the velocity, velocity gradient, and vorticity directions in the unidirectional flow. The normalised shear rate tensor and its inner dot product in this flow are respectively calculated as $\tensgammahatdot=\begin{bsmallmatrix}
0 & 1 & 0 \\
1 & 0 & 0 \\
0 & 0 &  0 \\
\end{bsmallmatrix}$ and $\tensgammahatdot\cdot\tensgammahatdot=\begin{bsmallmatrix}
1 & 0 & 0 \\
0 & 1 & 0 \\
0 & 0 &  0 \\
\end{bsmallmatrix}$. For simplicity, the different shear rate variables used in the closure are shown without a subscript and assumed to be equal in the different phases.}
\label{tab:harvie-stress-closures-in-unidirectional-flows}
\end{center}
\end{table}
\\
The model's shear viscosity, $\scal[s,mix]{\eta}$, shows a good correspondence with the DNS results and the experimental correlations in the semi-dilute regime. However, it does not exhibit a diverging behaviour at high solid fractions and shows a low sensitivity to changes in $\scal[r]{\hat{\epsilon}}$. As compared in Fig. \ref{fig:theory_viscosities_etas_effective_viscosity_tests}, the model viscosities for $\phis>2$ align better with the DNS results than the smaller viscosities from the theoretical correlation by \citet{batchelor1972determination}. They analysed the hydrodynamic interaction of smooth particles in semi-dilute extensional shear flows and found that the mixture shear viscosity is given by $\scal[s,mix]{\eta}=1+5\phis/2+k\phis^2+O(\phis^3)$ with $k=7.6\pm 0.8$. This factor for the smooth particles in the extensional flows using $q(\hat{r})$ as the PDF model of the suspension microstructure is defined as
\begin{equation}
k-\frac{5}{2}\simeq \left(\frac{15}{2}\scal[K]{\hat{D}}+5\scal[L]{\hat{D}}+\scal[M]{\hat{D}}\right)\Big|_{\hat{r}\to 2}^{\hat{r}\to \infty}
\end{equation}\\
where the constant value of $5/2$ on the left hand side is added in a renormalization technique to account for the increased viscosity of the effective fluid medium around the reference particle due to the presence of the other particles. The subsequent theoretical studies \citep{zinchenko1984effect,kim1985resistance,wilson2000viscosity} used the same equation but with more recent near-field $\xi$ scalars, which resulted smaller values of this factor within the range of $k=6.9-7.1$. Thus, the calculated integral expression (in the right hand side of the above equation) in the later studies falls within $4.4-4.6$, relatively close to value of $\approx 4.18$ resulting from the fitting functions in the current study. This value can be obtained by evaluating the far-field limit of the integrals (the constant factors $e$ in the table \ref{tab:Chat-Dhat-fitting-functions}) or by examining the expression plot for the entire range of the separation distance in the supplementary document (section S1). By using this value and assuming an isotropic PDF model equal to $q(\hat{r})$, the shear viscosity of the MF-roughness model for a smooth particle in the extensional flow can be determined as
\begin{equation}
\scal[s,mix]{\eta}=1+\left(\frac{5}{2}\phis+4.18\phis^2\right)\scal[RZ]{\lambda}
\end{equation}
Using the Taylor expansion of $\scal[RZ]{\lambda}=(1-\phis)^{-2.5}$, this shear viscosity can be written as $\scal[s,mix]{\eta}=1+5\phis/2+10.43\phis^2+O(\phis^3)$. The relatively larger coefficient of $\phis^2$ in the MF-roughness model compared to mentioned literature is due to different methods used to factor in the increased shear rate near the reference particle interacting with a pair in the presence of other particles. In the effective fluid medium method by \citet{batchelor1972determination}, the ratio of the local shear rate to the mixture average shear rate is only applied to interacting particle pairs without relative velocity, corresponding to the shear rate used in the dilute stress term in the fixed bed analysis. This ratio assumes the pair interaction occurs within an effective fluid medium with Einstein's viscosity and is equal to $(1+\phis)$. However, in the current study, the hindered drag coefficient $\scal[RZ]{\lambda}=(1-\phis)^{-2.5}$ is used as the ratio of the local to the phase shear rates affecting both the dilute and interaction stresses. As seen in Fig. \ref{fig:theory_viscosities_etas_effective_viscosity_tests}, by using this the hindered drag function in the dilute stress term, whether it is applied to the interaction stress term, the resulting shear viscosity for $\phis$ within $0.2-0.4$ shows good correspondence with the DNS results which are larger than the correlation by \citet{batchelor1972determination}.\\
\\
The model's normal viscosities exhibit a considerable change with the hydrodynamic roughness value. At a constant shear viscosity, the model's viscosity ratios $q$ and $\scal[2]{\alpha}$ show a similar variation to the roughness as the normal viscosities. The dependency of the normal viscosities on roughness can be explained by the relation between the interaction viscosities and the integral expressions defined in the table \ref{tab:interaction-viscosities-related-to-indexed-integrals}. The analysis of the near-field integral expressions $2\scal[L]{\hat{D}}+(4/7)\scal[M]{\hat{D}}$ and $2\scal[S]{\hat{D}}-(4/3)\scal[L]{\hat{D}}-(8/21)\scal[M]{\hat{D}}$ in the table \ref{tab:Chat-Dhat-fitting-functions} that respectively define $\musidonem$ and $\musidtwom$, show that they approximately scale as $(\hat{r}-2)^{0.26}$ in the separation distances $\hat{r}-2<0.05$ (section S1 in the supplementary document). This near-field scaling implies the normal viscosities scale with the roughness as $\scal[r]{\hat{\epsilon}}^{0.26}$, which aligns with the theoretical model by \citet{brady1997microstructure}. They also found that the self-diffusion of the solid particles, which is related to the normal viscosity, scale as $\hat{\epsilon}^{0.22}$. The roughness in their model refers to the surface-to-surface separation distance between particles, where a hard-sphere force is implemented when particles are closer to this distance. It is noteworthy that the second normal viscosity of the solid phase in the MF-roughness model has a similar variation with the solid fraction and roughness value as that of the mixture phase. This is because the ratio of the solid to the mixture phase for this viscosity is nearly constant at $\scal[m,SIM]{\phi}=\scal[n22,sol]{\eta}/\scal[n22,mix]{\eta}\simeq 0.585$ independent of the closure parameters, as shown in Fig. \ref{fig:roughness_model_phi_m_SIM}.\\
\\
Despite the relatively good agreement of the MF-roughness viscosities with the DNS results by \citet{gallier2014rheology}, one must note some important differences between the two methods. In the reported DNS results, shear viscosity decreases (by less than $10\%$) as the roughness value increases, which is not observed in the MF-roughness model. Another difference is that the DNS normal viscosities show a relatively low sensitivity to the roughness value, unlike the high sensitivity of the MF-roughness model. The reported DNS results for a suspension of $\phis=0.4$ show variations of around $35\%$ with the change in the roughness within $10^{-4}-10^{-2}$. Finally, the contribution of contact to normal stress in the DNS results is significant, while these interactions are not included in the MF-roughness model. The reported DNS results for the suspensions with $\phis>0.3$ show that the contact interactions could contribute more than $80\%$ of the total normal stress. Interestingly, the combined hydrodynamic and contact contribution of shear or normal viscosity in the DNS results are more consistent with the MF-roughness model or the experimental correlations than considering only the hydrodynamic contribution.\\
\\
Despite the significant impact of the frictional contacts on \citeauthor{gallier2014rheology}'s normal viscosities, they are close to \citeauthor{yeo2010dynamics}'s results and exhibit similarities with MF-roughness results. This can suggest that direct inclusion of the contact force by \citet{gallier2014rheology} effectively accounts for the near-field interaction of particles. It is worth noting that in \citeauthor{yeo2010dynamics}'s DNS method, direct contact interactions are not included in the simulations. Instead, a repulsive interaction force is used for the near-field interactions of particles. This force is non-zero for particles closer than $\hat{r}=2.01$, leading to a minimum separation distance of around $\hat{r}=2.003$. Similar to the sensitivity of the viscosities to the roughness in \citeauthor{gallier2014rheology}'s results, changing the parameters of the repulsive force's effective range has resulted in variations in the resulting shear and normal viscosities. Despite the similarities in the semi-dilute regime, the frictional contact forces in the concentrated regime significantly affected the diverging shear and normal viscosity behaviour in \citeauthor{gallier2014rheology}'s DNS method. Unlike the non-diverging MF-roughness viscosities, these frictional DNS results are found to be consistent with the experimental correlations.\\
\\
To sum up, the MF-roughness semi-dilute viscosity results show a reasonable correspondence with the DNS results and the experimental correlations in the literature. The differences in this flow regime may arise from how the near-field hydrodynamic interactions of the particles or how the hydrodynamic roughness values are implemented in the different methods. At high concentrations, the MF-roughness viscosities demonstrate a non-diverging behaviour, unlike the experimental correlations or the DNS frictional results, indicating the need to investigate the frictional contacts and multi-particle interaction impacts on the PDF model and the effective viscosity function used in the MF-roughness model.
\section{MF-MB99 closures}\label{MF-MB99-closures}
The stress closure proposed by \citet{morris1999curvilinear} is commonly used in studying suspension flows with SB modelling. As formulated in the table \ref{tab:MB99-stresses-frame-variant}, the closure lacks a fully tensorial form and can only be implemented in unidirectional flows where the geometric coordinate system aligns with the flow directions in the domain. Unlike the averaging techniques and MF modelling, this SB closure assumes equal interaction stresses in the solid and mixture phases. Moreover, the closure-free parameters are defined to predict SIM measurements in some curvilinear and rheometer flows, resulting in shear and normal viscosities differing from direct experimental measurements. Thus, in the present study, we propose two modified forms of this phenomenological closure based on the mathematical form of the MF-roughness model, in which their free parameters are ultimately optimized based on simulations of pressure-driven suspension flow. These two models, named MF-MB99-A and MF-MB99-B models, employ the average continuity and momentum equations (Eqs. (\ref{eq:MF-continuity-fluid}-\ref{eq:MF-momentum-solid})) alongside with the CS osmotic force from the Eq. (\ref{eq:CS-osmotic-force}) and a zero Faxen force. The implemented drag force for the models is similar to that of the MF-roughness model and is expressed as $\vect[drag,s]{f}=\beta(\uf-\us)$, with $\beta=9\muf\phis\phif^{2-n}/\left(2a^2\right)$ and $n=4.5$. As noted above, this drag force definition ensures that the MF solid settling velocity is consistent with the empirical correlation by \citet{richardson1954sedimentation}. Finally, the following two subsections define the phase-specific stresses for these MF models.\\
\begin{table}
\def~{\hphantom{0}}
\renewcommand{\arraystretch}{1.5}
\begin{center}
\begin{NiceTabular}{|lc|}
\hline
\multicolumn{2}{|c|}{\citeauthor{morris1999curvilinear}'s stress closure}\\
\hline
$\taum = -\muf\etas\tensgammadottm+ \muf\etan\left|\tensgammadottm\right|\tens{Q}$ &\Block{4-1}{$\tens{Q} = 
\begin{blockarray}{[ccc]}
\scal[1]{\lambda} & 0 & 0 \\
0 & \scal[2]{\lambda} & 0 \\
0 & 0 &  \scal[3]{\lambda} \\
\end{blockarray}$}\\
$\taus = -\muf\left(\etas-1\right)\tensgammadottm+\muf\etan\left|\tensgammadottm\right|\tens{Q}$&\\
$\etas = 1+2.5\phis(1-\phis/\phim)^{-1}+\ks\phis^2(\phim-\phis)^{-2}$&\\
$\etan = \kn\phis^2(\phim-\phis)^{-2}$\\
\hline
\multicolumn{2}{|c|}{The elements of the stress matrices}\\
\hline
$\tens[m]{\tau}=\muf\gammadot\begin{bmatrix}
\scal[1]{\lambda}\etan & -\etas & 0 \\
-\etas & \scal[2]{\lambda}\etan & 0 \\
0 & 0 & \scal[3]{\lambda}\etan \\ \end{bmatrix}$&$\tens[s]{\tau}=\muf\gammadot\begin{bmatrix}
\scal[1]{\lambda}\etan & -\etas+1 & 0 \\
-\etas+1 & \scal[2]{\lambda}\etan & 0 \\
0 & 0 & \scal[3]{\lambda}\etan \\ \end{bmatrix}$\\
\hline
\end{NiceTabular}
\caption{The formulation and matrix form of the solid and mixture phase stresses resulting from \citeauthor{morris1999curvilinear}'s closure. The first, the second, and the third directions of the anisotropic flow-aligned tensor $\tens{Q}$ denote the velocity, the velocity gradient, and the vorticity direction of a unidirectional flow. In this closure, $\left|\tens[tot,i]{\gammadot}\right|$ is the magnitude of the mixture phase deformation tensor $\tens[tot,m]{\gammadot}=\vnabla\um + (\vnabla\um)^\text{T}$. To estimate the matrix form of the stresses in the second row, we assumed equal deformation tensors in different phases, which are deviatoric and symmetric in a unidirectional flow with a magnitude equal to $\gammadot$. \citet{morris1999curvilinear} have suggested the parameter values of this closure to be as $\ks=0.1$, $\kn=0.75$, $\phim=0.68$, $\scal[1]{\lambda}=1$, $\scal[2]{\lambda}=0.8$, and $\scal[3]{\lambda}=0.5$.}
\label{tab:MB99-stresses-frame-variant}
\end{center}
\end{table}
\subsection{MF-MB99-A model}\label{MF-MB99-A-model}
The first modification of the \citeauthor{morris1999curvilinear}'s closure is named MF-MB99-A model. It resolves the lack of a frame-invariant fully tensorial form of closure and partitioning of the dilute stress term. To define the fully tensorial frame invariant form of the stresses, we use \citeauthor{harvie2021tensorial}'s stress closure forms, which can also be found in the closures proposed \citet{buyevich1996particle} for the phase stresses or the migration flux closure proposed by \citet{lhuillier2009migration}. These closures use similar phenomenological forms of deformation tensors found in the early literature, such as \citet{hand1962theory} and \citet{rivlin1997further}. Thus, the MF-MB99-A stresses are assumed to be given as follows:
\begin{equation}\label{eq:MF-MB99-AB-tau-m}
\taum=-\muf\left(1+\munhdil\right)\tensgammadotm-\muf \left[\munhintm \tensgammadots - \gammadots \left ( \munhsidonem \tensgammahatdots \cdot \tensgammahatdots + \munhsidtwom \tens{I} \right)\right]
\end{equation}
\begin{equation}\label{eq:MF-MB99-AB-tau-s}    
\taus=-\muf \left[ \munhints \tensgammadots - \gammadots \left ( \munhsidones \tensgammahatdots \cdot \tensgammahatdots + \munhsidtwos \tens{I} \right )\right ]
\end{equation}
The expressions for the dilute and interaction viscosities are now defined so that the proposed form is consistent with the original \citeauthor{morris1999curvilinear}'s closures in a unidirectional linear shear flow. To achieve this, we need to compare the elements of the proposed stress matrices previously shown in the table \ref{tab:harvie-stress-closures-in-unidirectional-flows} to those of the original form of \citeauthor{morris1999curvilinear}'s closures in the table \ref{tab:MB99-stresses-frame-variant}. After comparing the two, we have chosen the expressions presented in the table \ref{tab:MF-MB99-A-model-viscosities} as the most consistent with the \citeauthor{morris1999curvilinear}'s closures, although they are not an exact match. These expressions are obtained by matching only the components of the mixture stress. Then, the solid phase interaction stress components are assumed to be equal to those of the mixture phase.\\
\\
The dilute stress term in the proposed expressions for the MF-MB99-A stresses is not partitioned between the solid and mixture phases. Unlike in the SB models and aligned with the volume averaging and \citeauthor{harvie2021tensorial}'s closure, this shear stress term only appears in the mixture phase stress. Thus in defining the MF-MB99-A stresses, we first assumed that $\munhdil=\left(5\phis/2\right)\left(1-\phis/\phim\right)^{-1}$ and then by matching the shear component of the mixture stress tensors $\scal[m12]{\tau}$, the expression $\munhintm=\ks\phis^2(\phim-\phis)^{-2}$ is obtained. By comparing the third diagonal components $\scal[m33]{\tau}$, the expression $\munhsidtwom=\lambda_3\kn\phis^2(\phim-\phis)^{-2}$ is obtained. Afterwards, By comparing the second diagonal components $\scal[m22]{\tau}$, the expression $\munhsidonem=(\lambda_2-\lambda_3)\kn\phis^2(\phim-\phis)^{-2}$ is obtained. However, the first stress diagonal components $\scal[m11]{\tau}$ can not be matched. This is due to the limitation of the implemented fully tensorial form that does not distinguish the normal stress in the velocity direction from that of the velocity gradient direction, unlike \citeauthor{morris1999curvilinear}'s closures. This causes a constraint on the proposed fully tensorial unlike \citeauthor{morris1999curvilinear}'s closures here that requires $\scal[1]{\lambda}=\scal[2]{\lambda}$. Along with this constraint, we observe that the solid phase shear components do not match, attributed to the different partitioning of the dilute stress term mentioned earlier. Thus, the table \ref{tab:MF-MB99-A-model-viscosities} along with the Eqs. (\ref{eq:MF-MB99-AB-tau-m}), and (\ref{eq:MF-MB99-AB-tau-s}) define the MF-MB99-A closures that are a fully tensorial frame-invariant form of \citeauthor{morris1999curvilinear}'s closures and are consistent in the unidirectional shear flows except for a constraint on $\scal[1]{\lambda}$ and an introduced improvement in the phase partitioning of the dilute shear stress.
\begin{table}
\begin{center}
\renewcommand{\arraystretch}{1.5}
\def~{\hphantom{0}}
\begin{tabular}{|l|}
\hline
MF-MB99-A model viscosities\\
\hline
$\munhdil=\left(5\phis/2\right)\left(1-\phis/\phim\right)^{-1}$\\
\hline
$\munhintm=\munhints=\ks\phis^2(\phim-\phis)^{-2}$\\
\hline
$\munhsidtwom=\munhsidtwos=\lambda_3\kn\phis^2(\phim-\phis)^{-2}$\\
\hline
$\munhsidonem=\munhsidones=(\lambda_2-\lambda_3)\kn\phis^2(\phim-\phis)^{-2}$\\
\hline
\end{tabular}
\caption{The substituting shear and normal viscosities of the MF-MB99-A model. These viscosities, along with the solid and mixture phase stress formulations in Eqs. (\ref{eq:MF-MB99-AB-tau-m}) and (\ref{eq:MF-MB99-AB-tau-s}) fully define the model stresses. This model is almost identical to a SB model substituted with the stress closure suggested by \citet{morris1999curvilinear}.}
\label{tab:MF-MB99-A-model-viscosities}
\end{center}
\begin{center}
\renewcommand{\arraystretch}{1.5}
\def~{\hphantom{0}}
\begin{tabular}{|l|l|}
\hline
\multicolumn{2}{|c|}{MF-MB99-B model viscosities}\\
\hline
\multicolumn{2}{|l|}{$\munhdil=\left(5\phis/2\right)\left(1-\phis/\phim\right)^{-1}$}\\
\hline
$\munhintm=\ksmix\phis^2(\phim-\phis)^{-2}$&$\munhints=\kssol\phis^2(\phim-\phis)^{-2}$\\
\hline
$\munhsidtwom=\lambda_3\knmix\phis^2(\phim-\phis)^{-2}$&$\munhsidtwos=\lambda_3\knsol\phis^2(\phim-\phis)^{-2}$\\
\hline
$\munhsidonem=\left(\lambda_2-\lambda_3\right)\knmix\phis^2(\phim-\phis)^{-2}$&$\munhsidones=\left(\lambda_2-\lambda_3\right)\knsol\phis^2(\phim-\phis)^{-2}$\\
\hline
\end{tabular}
\caption{The substituting shear and normal viscosities of the MF-MB99-B model. These viscosities, along with the solid and mixture phase stress formulations in Eqs. (\ref{eq:MF-MB99-AB-tau-m}) and (\ref{eq:MF-MB99-AB-tau-s}) fully define the model stresses.}
\label{tab:MF-MB99-B-model-viscosities}
\end{center}
\end{table}
\subsection{MF-MB99-B model}\label{MF-MB99-B-model}
The MF-MB99-B model is proposed as an extension of the MF-MB99-A model that partitions the mixture interaction stress between the solid and fluid phases. This extension aligns with the MF-roughness model and the SB revisions found in \cite{nott2011suspension} and \cite{guazzelli2018rheology}. The MF-MB99-B model differs from the \citeauthor{morris1999curvilinear}'s stress closure proposed for a SB model and utilizes the same interaction stresses for the solid and mixture phases. In the MF-MB99-B model, it is assumed that the interaction stresses of the solid and mixture phases have the same forms but with different coefficient factors. Therefore, as summarised in the table \ref{tab:MF-MB99-B-model-viscosities}, two different coefficients are used for the shear ($\kssol$ and $\ksmix$) and normal ($\knsol$ and $\knmix$) viscosities.\\
\\
The study of SIM in general flow geometries can involve examining the three diagonal components of the stresses. Distinguishing the solid phase normal stresses in the three directions from those of the mixture phase necessitates using phase-specific anisotropic coefficients (for at least two of the $\scal[1]{\lambda}$, $\scal[2]{\lambda}$, and $\scal[3]{\lambda}$ parameters along with the phase-specific $\scal[n]{K}$ values). However, only the normal stress in the velocity gradient direction affects SIM in the channel flows analyzed here. Using $\knsol$ and $\knmix$ is sufficient to differentiate this normal stress component between the phases.
\section{MF SIM in the channel flows}\label{MF-SIM-channel-flows}
The volumetric flux of particles relative to the mixture can be used to assess the rate of SIM development in different flow conditions. In the current study, this flux is referred to as the migration flux, denoted by $\js=\phis\left(\us-\um\right)$. The relationship between the migration flux and the development of solid distribution in pressure-driven flows can be demonstrated by substituting this flux definition into the solid phase continuity equation at steady-state conditions as $\um\cdot\vnabla\phis+\vnabla\cdot\js=0$. Notably, in an order of magnitude analysis, \cite{miller2006normal} has argued that in a unidirectional flow where the velocity component in the vertical direction vanishes, the solid phase continuity equation can be approximated as $\scal[m,x]{u}\p\phis/\p x+\p \scal[sm,z]{j}/\p z=0$. Here, x and z are the axes along the axial length and height of a 2D channel, respectively. Consequently, for the channel flows studied here, the focus is on the cross-stream component of the migration flux vector, as a larger magnitude of this component results in a larger axial development of the solid distribution in the flow. Furthermore, due to the zero-flux condition of the solid and fluid phases into the wall or centerline, a zero cross-stream migration flux represents the fully developed condition of the solid distributions in the flow, $\p\phis/\p x=0$. Thus, this section aims to relate the cross-stream migration flux to the phase-specific viscosity of each MF model under different flow conditions, enabling the comparison of the development and fully developed SIM results of the MF models for the experimental cases of the channel flows.\\
\\
The migration flux relation obtained here applies to steady-state and Stokes flow, where transient and inertial terms in the momentum equations are neglected. Additionally, the Faxen force and dilation stress terms, which are only included in the MF-roughness model, are not considered in this formulation due to the low sensitivity of the steady-state solid distribution results of the studied suspension flows. By substituting the drag force definition $\vect[drag,s]{f}=\beta\left(\uf-\us\right)$ into the solid phase momentum equation in Eq. (\ref{eq:MF-momentum-solid}) and combining it with the mixture momentum equation (sum of Eqs. (\ref{eq:MF-momentum-fluid}) and (\ref{eq:MF-momentum-solid})) to eliminate the fluid pressure force term, the migration flux can be expressed as:
\begin{equation}\label{eq:theory-migration-flux-definition-1}
\js=\frac{\phis}{\beta}\left(\phis\vnabla\cdot\tens[m]{\tau}-\vnabla\cdot\tens[s]{\tau}+\phif\vnabla\posm\right)
\end{equation}
where the drag force coefficient is given as $\beta=\left(9\muf\phis\phif^{2-n}\right)/\left(2a^2\right)$ with $n=4.5$ used for all the three MF models. By evaluating the z-axis components of resulting vectors from the three terms in the parentheses, the cross-stream migration flux can be expressed as:
\begin{equation}
\label{eq:theory-migration-flux-definition-2}
\scal[sm,z]{j}=\frac{\phis\muf}{\beta}\left[\left(\phis-\scal[m,SIM]{\phi}\right)\frac{\p}{\p z}\left(\scal[n22,mix]{\eta}\gammadot\right)+\posm^{'}\frac{\phif}{\muf}\frac{\p \phis}{\p z} \right]    
\end{equation}
where $\gammadot$ is the magnitude of the shear rate tensor assumed to be equal in all the phases. In deriving this equation, the forces due to the gradient of shear stress terms ($\scal[m12]{\tau}$ and $\scal[s12]{\tau}$) along the z-axis are neglected compared to the normal stress forces. Also, the viscosity ratio $\scal[m,SIM]{\phi}=\scal[n22,sol]{\eta}/\scal[n22,mix]{\eta}$ is assumed to be a constant value. In the MF-MB99-A and MF-MB99-B models, it is equal to $1$ and $\scal[n,sol]{K}/\scal[n,mix]{K}$ respectively while for the MF-roughness model, it is calculated to be around $0.585$ as shown in Fig. \ref{fig:roughness_model_phi_m_SIM}. By using the partial derivative of CS osmotic pressure as $\posm^{'}=\p \posm/ \p \phis=\left(\kB T \scal[CS]{\zeta}\right)/\scal[p]{V}$ and defining the local Peclect number as $Pe=\left(9\scal[f]{\mu}\scal[p]{V}\gammadot\right)/\left(2\kB T\right)$, the second form of flux can be written as:
\begin{equation}\label{eq:theory-migration-flux-definition-3}
\scal[sm,z]{j}=\frac{\phis\muf\gammadot}{z\beta}\left[\left(\phis-\scal[m,SIM]{\phi}\right)\scal[n22,mix]{\eta}\left(\frac{z}{\gammadot}\frac{\p \gammadot}{\p z}+z\frac{\p \phis}{\p z}\frac{\scal[n22,mix]{\eta}^{'}}{\scal[n22,mix]{\eta}}\right)+\frac{9\phif\scal[CS]{\zeta}}{2 Pe}z\frac{\p \phis}{\p z}\right]
\end{equation}
with $\scal[n22,mix]{\eta}^{'}={\p \scal[n22,mix]{\eta}}/{\p \phis}$. The expression for the variation of the shear rate magnitude along the z-axis can be determined from the mixture momentum equation. After neglecting the inertial and transient terms, the momentum equation in the axial direction along the x-axis can be written as:
\begin{equation}
-\frac{\p }{\p z}\left(\muf\scal[s,mix]{\eta}\gammadot\right)=\frac{\p \scal[f]{p}}{\p x}
\end{equation}
By assuming a uniform axial pressure gradient across the channel height and a zero shear rate at the centerline, the pressure gradient is given by integrating the equation as $\p \pf/\p x=-\muf\scal[s,mix]{\eta}\gammadot/z$. Thus the relation for the dimensionless shear rate gradient is written as: 
\begin{equation}\label{eq:theory-shear-rate-formulation}
\frac{z}{\gammadot}\frac{\p \gammadot}{\p z}=1-\frac{{\scal[s,mix]{\eta}}^{'}}{\scal[s,mix]{\eta}}z\frac{\p \phis}{\p z}
\end{equation}
Using this relation, the cross-stream migration flux can be expressed as:
\begin{equation}\label{eq:theory-migration-flux-definition-4}
\scal[sm,z]{j}=\frac{\phis\muf\gammadot}{z\beta}\left[\left(\phis-\scal[m,SIM]{\phi}\right)\scal[n22,mix]{\eta}\left(1+z\frac{\p \phis}{\p z}\left(\frac{\scal[n22,mix]{\eta}^{'}}{\scal[n22,mix]{\eta}}-\frac{\scal[s,mix]{\eta}^{'}}{\scal[s,mix]{\eta}}\right)\right)+\frac{9\phif\scal[CS]{\zeta}}{2 Pe}z\frac{\p \phis}{\p z}\right]
\end{equation}
where $\scal[s,mix]{\eta}^{'}={\p \scal[s,mix]{\eta}}/{\p \phis}$. This flux relation can also be expressed in terms of the ratio of mixture normal to shear viscosity $q=\scal[n22,mix]{\eta}/\scal[s,mix]{\eta}$ as:
\begin{equation}\label{eq:theory-migration-flux-definition-5}
\scal[sm,z]{j}=\frac{\phis\muf\gammadot}{z\beta}\left[\left(\phis-\scal[m,SIM]{\phi}\right)\scal[s,mix]{\eta}\frac{\p \left(zq\right)}{\p z}+\frac{9\phif\scal[CS]{\zeta}}{2 Pe}z\frac{\p \phis}{\p z}\right]
\end{equation}
As discussed earlier, the zero-fluxes of solid and fluid to the wall or centerline necessitate the cross-stream migration to be zero in the fully developed condition. Therefore, setting $\scal[sm,z]{j}=0$, yields the fully developed solid distribution as:
\begin{equation}\label{eq:theory-non-dimensional-phi-grad-1} 
-z\frac{\p \phis}{\p z}=\frac{\scal[s,mix]{\eta}\left(\phis-\scal[m,SIM]{\phi}\right)q}{\scal[s,mix]{\eta}\left(\phis-\scal[m,SIM]{\phi}\right)q^{'}+\frac{9\phif\scal[CS]{\zeta}}{2Pe}}
\end{equation}
where for non-Brownian suspension flows (and neglecting the singularity at $\phis=\scal[m,SIM]{\phi}$), the expression simplifies as:
\begin{equation}\label{eq:theory-non-dimensional-phi-grad-2}    
-z\frac{\p \phis}{\p z}=\frac{q}{q^{'}}
\end{equation}
The formulations of the cross-stream migration flux and the fully developed solid distributions show that the mixture's normal stress impacts the migration via the $\phis$ dependency of the viscosity ratio $q=\scal[n22,mix]{\eta}/\scal[s,mix]{\eta}$, which its competition with the solid phase normal stress appears in the constant viscosity ratio $\scal[m,SIM]{\phi}$. Furthermore, the importance of osmotic forces versus shear forces in the migration is reflected in the ratio $9\phif\scal[CS]{\zeta}/(2Pe)$. A larger magnitude of this ratio leads to a decrease in the migration flux towards the centerline, resulting in a smaller gradient in the solid distribution. These formulations will be used in section \ref{simulation-results} to discuss the simulation results of the different closure models and their correspondence with the experimental flow data in different $\PeB$ numbers.
\section{Experimental case details}
We conduct the simulation and the optimisation of the three models based on eleven monodisperse experimental channel results reported by \citet{semwogerere2007development} and \citet{semwogerere2008shear}. These experiments investigated the development of the SIM in the suspension of neutrally buoyant $1-3\; \mu\mbox{m}$ diameter PMMA particles in a wide channel. The particles suspended in a cyclohexylbromide/decalin solution, with matched fluid density and refractive index, and flowed in a channel with a total height of $2H=50\; \mu$m (along z-axis), a large width of $500\; \mu\mbox{m}$ (along y-axis), and a total axial length of $10$ cm (along x-axis). Using a confocal microscope, the number of dyed particles in 2D $55\times 55\;\mu\mbox{m}^2$ slices (parallel to the yz plane) of the suspension were counted to determine the local solid volume fraction at a specific point in the channel. Multiple slices at intervals of $0.2\;\mu\mbox{m}$ were then used to create a graph showing the variation of $\phis$ across the channel height at a particular distance from the entry. Also, the displacement of the particles in the successive slices taken at a rate of $94$ frames/s was measured to estimate the particle velocity variation across the channel height.\\
\\
A syringe pump controlled the flows to the channel at a constant inflow rate conditions, with a $5\%$ standard deviation. The impact of the particle size or the flowrate on the SIM of particles in the flow is studied based on the flow $\PeB$ number given as $\PeB=6 \pi\muf\scal[ave]{\gammadot}a^3/(\kB T)$. The average shear rate magnitude for a parabolic velocity profile is calculated as $\scal[ave]{\gammadot}=\scal[m,max]{u}/H$ where $\scal[m,max]{u}$ is the centre-line velocity of the suspension given as $\scal[m,max]{u}=3\Uave/2=3Q/\left(2A\right)$, where $\Uave$ is the average inlet velocity, $Q$ is the inflow-rate, and $A=25\times 10^{-9}\; m^2$ is the cross-sectional area of the channel. The reported experimental data includes the solids volume fraction and velocity distributions across the channel height and the graphs for the variation of evolution parameter $\Ep$ with the axial distance from the channel entry. The change in this parameter value with the axial distance can show the development of solids redistribution towards its fully-developed value. An $\Ep$ value is specified for a $\phis$ distribution along the channel height as follows:
\begin{equation}\label{eq:Semwogerere_Ep_definition}
\Ep (x)=\frac{1}{H}\int_0^{H} \left|\frac{\phis(z,x)}{\scal[s]{\bar{\phi}}(x)}-1 \right| dz
\end{equation}
where $\scal[s]{\bar{\phi}}(x)$ is the average solid fraction determined from the same solids distribution profile as:
\begin{equation}
\scal[s]{\bar{\phi}}(x)=\frac{1}{H}\int_0^{H} \phis(z,x) dz
\end{equation}
The different parameters for the experimental cases are summarised in table \ref{tab:Semwogerere-exp-cases}. The experimental references do not include the fluid viscosity and flow rates for cases C8 and C9. The fluid viscosities calculated using the $\PeB$ definition at $T=295\;\mbox{K}$ can vary within the range of $1.853-2.786\;\mbox{mPa.s}$. This significant variation of the fluid viscosity between different cases is unexpected, considering that the same fluid solution is used in all these experiments. In addition, the integration of experimental particle velocity profiles in case C7 yields a flow rate of around $0.471\;\mu \mbox{L/min}$, which differs from the reported flow rate of $0.51\;\mu \mbox{L/min}$. Different factors could have contributed to these discrepancies, including particle slip velocity or difficulty in identifying the channel walls in microscope images. Without further experimental information, we assume that errors in the volumetric flow rates may be causing the differences between the back-calculated and reported fluid viscosities in the flow cases referenced in \cite{semwogerere2007development}. Also, in the experimental flows from \cite{semwogerere2008shear} with identical reported flow rates, we attribute the fluid viscosities discrepancies to inaccuracies in the reported $\PeB$ numbers. Therefore, we used a consistent fluid viscosity of $2.2494\;\mbox{mPa.s}$ obtained from the Weeks Soft Matter Laboratory webpage \cite{WeekslabJune2009} to simulate all eleven cases. The flow rates of the cases referenced in \cite{semwogerere2007development} and the $\PeB$ numbers of the cases referenced in \cite{semwogerere2008shear} are then adjusted to align with the updated viscosity value. This viscosity also aligns with the value of $2.2471 \pm 0.0202\;\mbox{mPa.s}$ calculated for similar suspensions studied by \citet{frank2003particle} in the same group/lab (assuming $T=295K$).
\begin{table}
\def~{\hphantom{0}}
\begin{center}
\begin{NiceTabular} {|c|c|c|c|c|c|}
\CodeBefore
\cellcolor{gray}{2-1}
\cellcolor{brown}{3-1}
\cellcolor{blue!80}{4-1}
\cellcolor{cyan!50}{5-1}
\cellcolor{red}{6-1}
\cellcolor{orange}{7-1}
\cellcolor{yellow}{8-1}
\cellcolor{violet}{9-1}
\cellcolor{magenta}{10-1}
\cellcolor[cmyk]{1,0,0.85,0}{11-1}
\cellcolor{green}{12-1}
\Body
\hline
Case&Fig. in [Ref.]&$a\;(\mu \text{m})$&$\PeB$&$Q\; (\mu \text{L/min})$&$\phio$\\
\hline
C1&P.C. [8]&$1.5$&$480, 421.603^\dagger$&$0.3$&$0.07$\\
\hline
C2&3c [8]&$0.69$&$47, 41.037^\dagger$&$0.3$&$0.10$\\
\hline
C3&7b,4a [8]&$1.5$&$480, 421.603^\dagger$&$0.3$&$0.10$\\
\hline
C4&4a [8]&$1.5$&$480, 421.603^\dagger$&$0.3$&$0.15$\\
\hline
C5&10 [7]&$0.70$&$129$&$1.02, 0.943^\dagger$&$0.26$\\
\hline
C6&7c,7d [8]&$0.69$&$47, 41.037^\dagger$&$0.3$&$0.25$\\
\hline
C7&4 [7]&$0.70$&$60$&$0.51, 0.439^\dagger$&$0.33$\\
\hline
C8&8 [7]&$0.70$&$70$&$0.512^\dagger$&$0.35$\\
\hline
C9&8 [7]&$0.70$&$15$&$0.110^\dagger$&$0.35$\\
\hline
C10&5 [7]&$1.15$&$80$&$0.102, 0.126^\dagger$&$0.28$\\
\hline
C11&5 [7]&$1.15$&$80$&$0.102, 0.126^\dagger$&$0.23$\\
\hline
\end{NiceTabular}
\caption{The suspension properties and the experimental flow conditions of the different monodisperse channel flow cases are being examined in this study. The first column lists an index assigned to each experimental case. The second column indicates the figure number in each reference where the information about the experimental case can be found. P.C. denotes personal communication, [8] refers to the reference by \citet{semwogerere2008shear}, and [7] refers to the reference by \citet{semwogerere2007development}. The $\PeB$ and $Q$ values, indicated with a superscript $^\dagger$, are the values used in the current study to calculate the fluid viscosity $\muf=2.249\; \mbox{mPa.s}$ across all the cases (based on the $\PeB$ definition), which are implemented in the simulations.}
  \label{tab:Semwogerere-exp-cases}
  \end{center}
\end{table}
\section{Numerical simulations}
\begin{figure}
\centering
\includegraphics[width=0.6\textwidth]{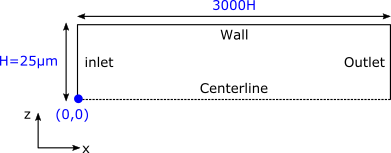}
\caption{The axisymmetric 2D computational domain used for the MF simulations of the channel flows.}
\label{fig:Semwogerere_2D_geometry}
\end{figure}

The finite volume solver \textit{Arb} \citep{harvie2010implicit} is used to solve the MF conservation equations numerically. These equations include the continuity and momentum conservations of the fluid phase in Eqs. (\ref{eq:MF-continuity-fluid}) and (\ref{eq:MF-momentum-fluid}), as well as those of the solid phase in Eqs. (\ref{eq:MF-continuity-solid}) and (\ref{eq:MF-momentum-solid}). After neglecting the influence of gravity on the neutrally buoyant particles, a symmetrical 2D computational domain in the Cartesian coordinate system is employed for the simulations, as shown in Fig. \ref{fig:Semwogerere_2D_geometry}. Using a 2D computational domain is based on the dimensions of the experimental flow chamber, which has a width ten times greater than its height. The distance between the wall and the centre line in the computational domain is $H=25\;\mu$m, equal to half the height of the experimental channel flow. The domain length is chosen to be $3000H$ long enough compared to the experimental channel length to cover all experimental data for comparison purposes.\\
\\
The simulations used a non-uniform structured mesh of the computational domain. This mesh includes $60$ cells with a growth rate of $1$ in the vertical (z-axis) direction and $100$ cells with a growth rate of $1.02$ in the axial (x-axis) direction. The cell size growth along the x-axis accounts for a large migration flux near the channel entry compared to the developed flow condition away from the entry, as observed in the experimental $\Ep$ graphs. Also, as the magnitudes of $\p \phis/\p z$ could be significant at any point across the channel height, a uniform cell size is used in this direction. A mesh sensitivity analysis was conducted to ensure the results were independent of the cell sizes (section S2 in the supplementary document).\\
\\
The boundary conditions applied to the computational domain for the simulations are described as follows:
\begin{enumerate}
    \item At the inlet, a uniform solids volume fraction equal to $\phio$ was applied. At this boundary, the phase-specific velocities are assumed to be equal, unidirectional, and fully developed at the inlet. Also, a uniform inlet pressure $\pf=\scal[inlet]{p}$ is applied to ensure that the average velocities are consistent with the flowrate.\\
    \item At the outlet, one-point zero fluid pressure is used, and the pressure and $\phis$ distributions, along with the unidirectional phase-specific velocities, are assumed to be fully developed.\\
    \item No fluid and solid fluxes are applied to the wall. Zero gradients of $\phis$ and stress forces (the first and the second terms on the right-hand side of each momentum equation) are applied to the normal direction at the wall to calculate the $\phis$ and $\pf$ on this boundary.\\
    \item Symmetric boundary conditions are used for the centreline. The conditions sets the normal component of the velocities along with the normal components of the gradients in the velocities, $\pf$, and $\phis$ to zero.
\end{enumerate}
\section{Global optimisation}
\begin{figure}
\centering
\includegraphics[width=0.6\textwidth]{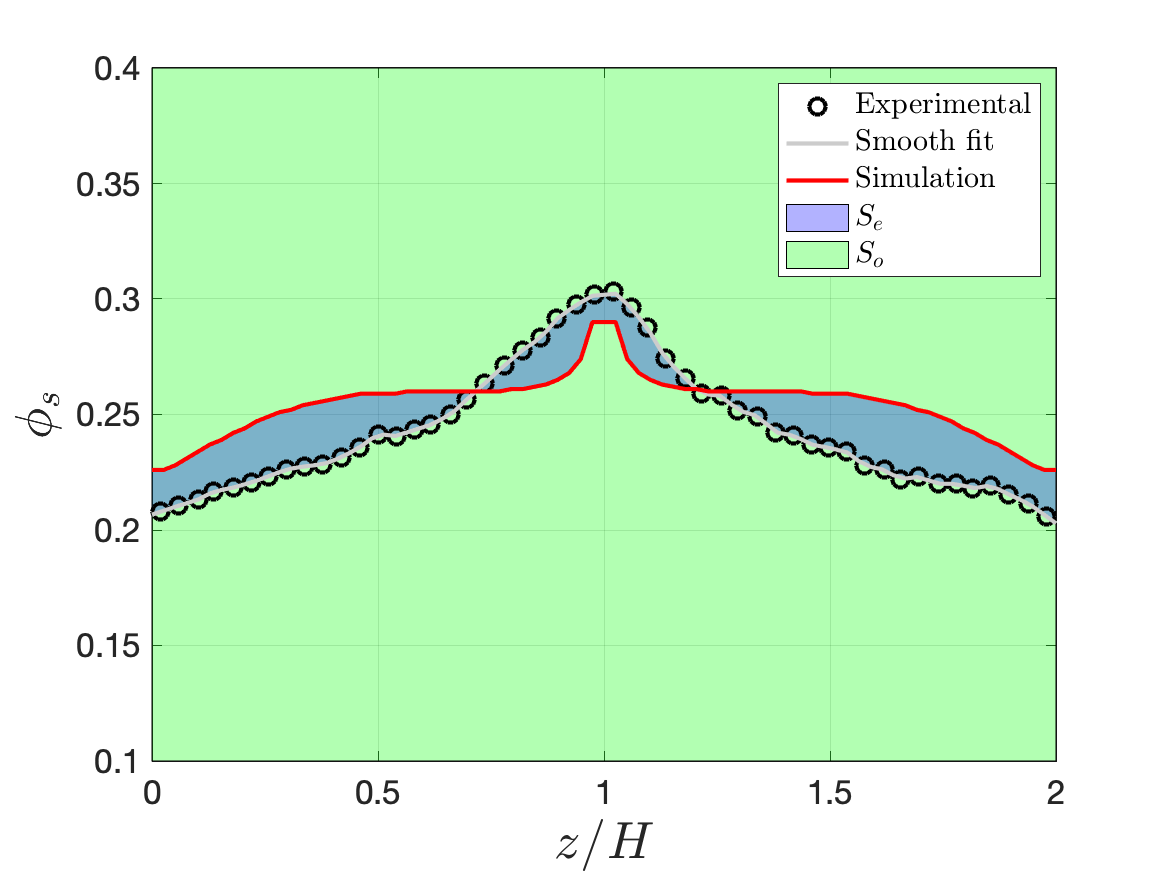}
\caption{A schematic showing the calculation of a single residual value $(\scal[i]{R})$ by comparing experimental and simulation data in a subplot based on the Eq. (\ref{eq:single_R_value_calculation}). The $\scal[i]{R}$ value is defined as $\scal[e]{S}/(\scal[e]{S}+\scal[o]{S})\times 100$ where $\scal[e]{S}$ is the area of the enclosed region between the two solid lines, and $\scal[o]{S}$ is the subplot area outside of the enclosed region.}
\label{fig:single-residual-calculation}
\end{figure}

A global gradient descent optimisation method (included in \textit{Arb} package \citep{harvie2010implicit}) is used to determine the optimal closure parameters of the MF models. The objective is to minimise the total squared residual value defined as $TSRV=\scal[i]{\sum} \scal[i]{R}^2$, measuring the total deviation of the simulation results from all the experimental solids volume fraction and velocity distributions. In this summation of squared residuals, $\scal[i]{R}$ represents the residual value obtained by comparing the simulation results with each experimental distribution. As schematically shown in Fig. \ref{fig:single-residual-calculation}, it is calculated as:
\begin{equation}\label{eq:single_R_value_calculation}
\scal[i]{R}=\frac{\int\left|\scal[s,sim]{\phi}-\scal[s,exp]{\phi}\right|\text{d}\left(\frac{z}{H}\right)}{\Delta \phis\Delta\left(\frac{z}{H}\right)}\times 100
\end{equation}
where $\scal[s,sim]{\phi}$ and $\scal[s,exp]{\phi}$ represent the $\phis$ values on the simulation and smoothed experimental graphs, respectively. Also, the parameters $\Delta \phis$ and $\Delta (z/H)$ denote the ranges of $\phis$ and $z/H$ axes considered in a simulation-experimental graph comparison. As seen in Fig. \ref{fig:Semwogerere_all_models_optimized_phi}, we chose $\Delta \phis = 0.3$ and $\Delta \left(z/H\right)=2$ for the comparison of all the solids volume fraction distributions. The residual $\scal[i]{R}$ values related to the solids velocity distributions are calculated using an equation analogous to Eq. (\ref{eq:single_R_value_calculation}) $(\scal[s,x]{u})$, but with $\Delta \scal[s,x]{u}=600\;\mu \text{m/s}$. Notably, the $\Ep$ graphs were not considered in optimisations since the parameter values were highly sensitive to local variations in $\phis$ within the range of experimental data accuracy, as is further discussed in section \ref{the-evolution-parameter-distributions}. Nevertheless, we will compare the optimised simulation results with the experimental $\Ep$ graphs.\\
\\
A random algorithm followed by a quadratic gradient descent method was used for the optimisation. Initially, the random algorithm, which samples the parameters randomly within specified ranges, is used to conduct the simulations of each MF model with 200-400 sets of randomly chosen closure parameters. Subsequently, some random parameter sets resulting in low $TSRV$s are implemented as the initial values of the quadratic gradient descent method to find the local minimum of $TSRV$. In this method, the step sizes of the parameters are determined based on the change in $TSRV$ magnitude and the optimised results achieved when the non-dimensional step size for each parameter was less than $1\times 10^{-5}$. The variation of the three model closure parameters in the optimisation procedure can be described as follows.
\subsection{MF-roughness optimisation}
The MF-roughness model interaction viscosities and thus the closure of the model can be defined based on the tables \ref{tab:interaction-viscosities-related-to-indexed-integrals}, \ref{tab:piece-wise-indexed-integrals}, \ref{tab:Chat-Dhat-fitting-functions} and \ref{tab:r_hat_infty_fitting_function} that relate the viscosities to the parameters $\scal[o]{\hat{\hat{r}}}=4\times 10^{-5}$, $\scal[{2-\scal[\infty]{\hat{r}}}]{N}=1$, $\scal[1]{f}$, $\scal[2]{f}$, and $\scal[r]{\hat{\epsilon}}$. Therefore, the last three parameters, which define the width and the magnitude of the pair correlations in the asymmetric region in the PDF model, must be specified in the optimisation procedure. However, the normal viscosities do not depend on the magnitude of $\scal[2]{f}=(1/2)\left(\scal[c]{f}+\scal[e]{f}\right)$, and sensitivity analysis shows that varying this parameter has only an insignificant impact on the shear viscosities of the model. Thus, focusing only on the magnitude of $\scal[1]{f} = (3\pi/8)\left(\scal[c]{f}-\scal[e]{f}\right)$, we set $\scal[e]{f}=0$ and varied $\scal[c]{f}$ within the range of $1-2$ for the optimisation. The non-dimensional hydrodynamic roughness value $\scal[r]{\hat{\epsilon}}=\scal[r]{\epsilon}/{a}$ determines the outer radius of the asymmetric region in the PDF model. This effective radius is expected to be associated with the asperities or any non-uniformities on the surface of particles that lead to particle trajectory perturbation and create the anisotropic high pair correlation ring in the microstructure. In the optimisation, it is assumed that same-sized particles involved in different flow cases have a similar dimensional $\scal[r]{\epsilon}$ (nm) value. Thus, a particle-size dependent $\scal[r]{\epsilon}$ (nm) value is used, which is varied within the range of $1-35\; \text{nm}$. This range covers the particles of very low dimensional roughness value to the highly roughened particles with non-dimensional roughness values of $\scal[r]{\hat{\epsilon}}\simeq 0.05$.
\subsection{MF-MB99-A,B optimisation}
The closure of stresses in the MF-MB99-A and MF-MB99-B models depends on the parameters defining the interaction viscosities. These paramters in the MF-MB99-A model are as $\phim$, $\scal[s]{K}$, $\scal[n]{K}$, $\scal[2]{\lambda}$, and $\scal[3]{\lambda}$ values, whereas in the MF-MB99-B model are as $\phim$, $\scal[n,sol]{K}$, $\scal[n,mix]{K}$, $\scal[s,sol]{K}$, $\scal[s,mix]{K}$, $\scal[2]{\lambda}$, and $\scal[3]{\lambda}$. The optimisation procedure varied all these closure parameters except for the anisotropic coefficients, which are fixed at $\scal[2]{\lambda}=0.8$ and $\scal[3]{\lambda}=0.5$ (as suggested by \citet{morris1999curvilinear}). Using the fixed values is because $\scal[3]{\lambda}$ does not affect the second normal stress defining the cross-stream SIM in the channel flows (as discussed earlier in the section \ref{MF-SIM-channel-flows}). Moreover, this component of the phase-specific normal stress depends on $\scal[2]{\lambda}$ multiplied by the normal viscosity coefficients ($\scal[n]{K}$ in MF-MB99-A and $\scal[n,sol]{K}$ and $\scal[n,mix]{K}$ in MF-MB99-B). Hence, tuning only one of the two types of parameters (i.e., the normal viscosity coefficient or the anisotropic coefficient) is sufficient for optimising the SIM behaviour.\\
\\
The mixture shear viscosity coefficients, $\scal[s]{K}$ in the MF-MB99-A and $\scal[s,mix]{K}$ in the MF-MB99-B model, are varied as the optimisation parameters. However, in the MF-MB99-B model, the solid phase shear viscosity coefficient is assumed to be the same as that of the mixture phase $\scal[s,sol]{K}=\scal[s,mix]{K}$. This assumption is based on the sensitivity analysis, which shows that varying $\scal[s,sol]{K}$ within a wide range has minimal impact on the simulation results. Additionally, in both models, the squared of a residual value related to the deviation of the mixture shear viscosity from the Maron-Pierce correlation $\scal[s,mix]{\eta}=(1-\phis/\scal[m]{\phi})^{-2}$ is added to $TSRV$. Notably, the maximum solid fraction $\phim$ value in this correlation is assumed to equal each MF model. The shear viscosity from this correlation is consistent with the direct experimental measurements and larger than the original \citeauthor{morris1999curvilinear}'s correlation. Considering the shear viscosity graph in the optimisation is because the flow simulation results show a low sensitivity to this viscosity; thus, this addition ensures finding the closure parameters that result in the shear viscosities in the range of the Maron-Pierce correlation. To calculate the residual value related to this deviation, a similar procedure to that of Eq. \ref{eq:single_R_value_calculation} and Fig. \ref{fig:single-residual-calculation} is employed. The $\scal[i]{R}$ value is calculated as the surface area enclosed between the Maron-Pierce correlation and a MF model graph of log$(\scal[s,mix]{\eta})$ versus $\phis$ within $0-0.6$ normalised by a reference area of $0.6\times\text{log}(60)$.
\section{Simulation results}\label{simulation-results}
\begin{figure}
\centering
\begin{subfigure}[a]{1\textwidth}
\centering
\includegraphics[width=\textwidth]{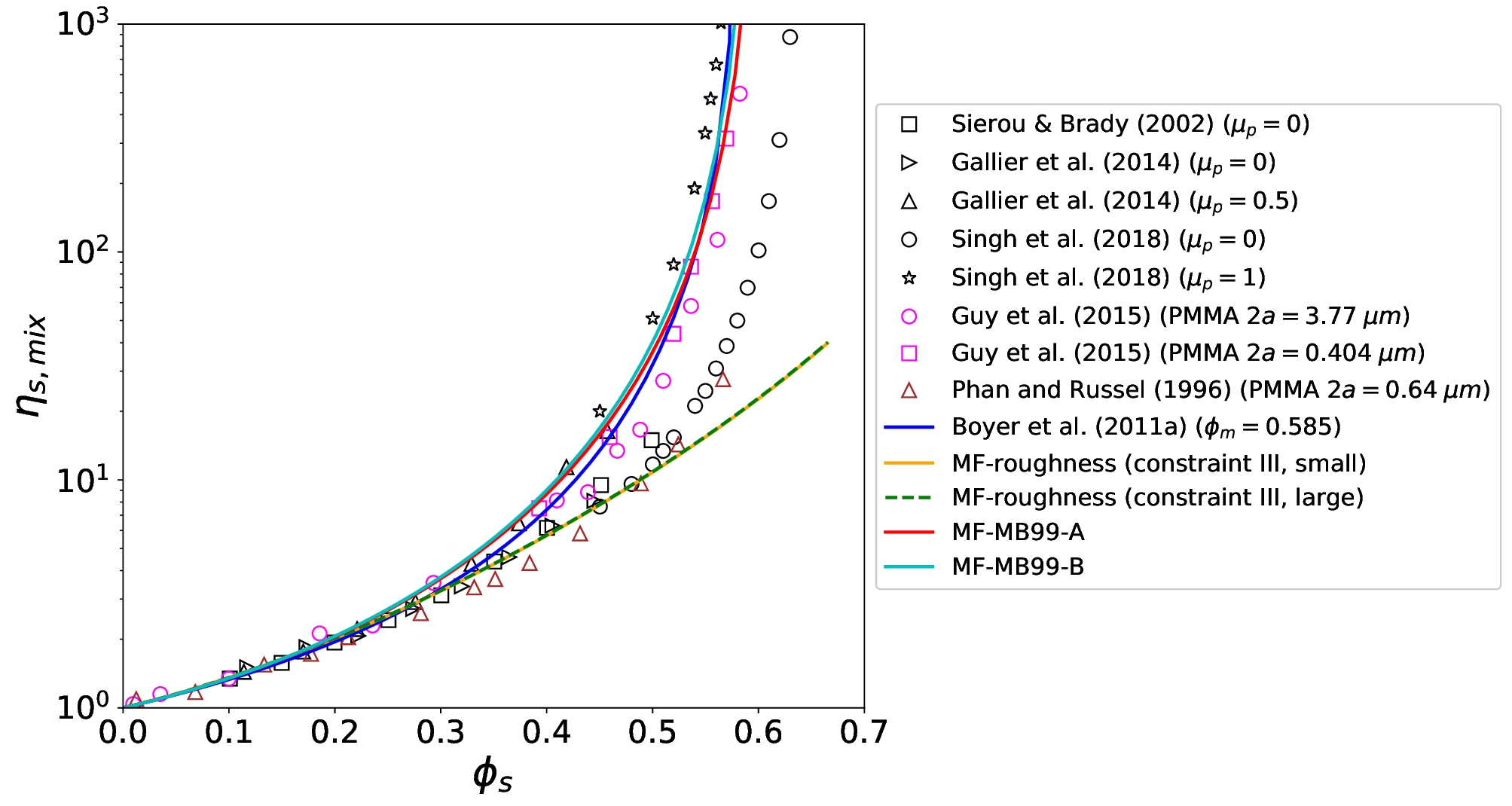}
\caption{}
\label{fig:Semwogerere_all_models_optimized_viscosities_etas}
\end{subfigure}
\begin{subfigure}[b]{1\textwidth}
\centering
\includegraphics[width=\textwidth]{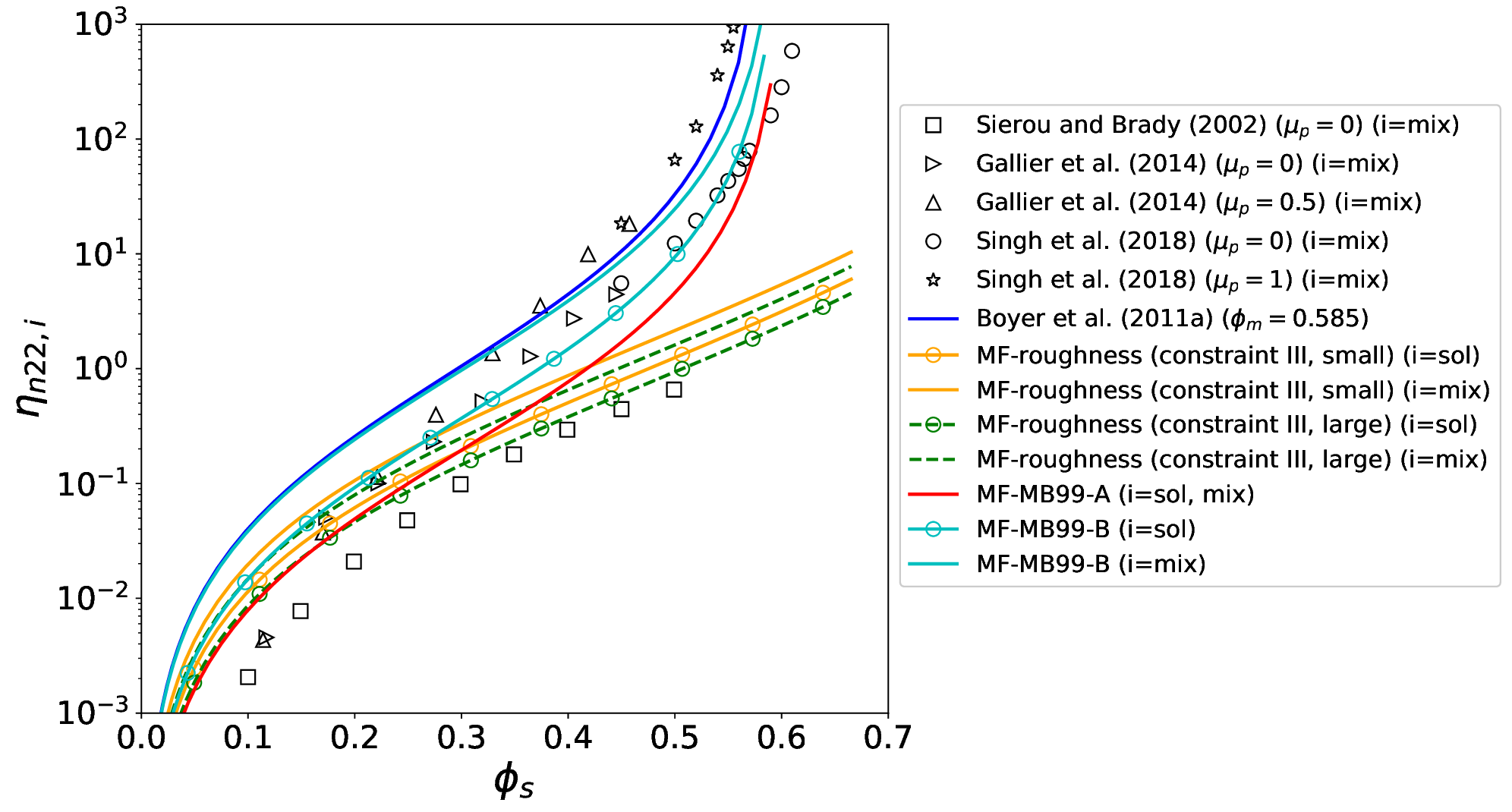}
\caption{}
\label{fig:Semwogerere_all_models_optimized_viscosities_etan}
\end{subfigure}
\caption{The mixture shear viscosity (a) and phase-specific normal viscosities in the shear direction (b) for the different optimised MF models (table \ref{tab:Semwogerere_optimized_closure_paramters}) compared to the reported experimental measurements \citep{guy2015towards,phan1996phase} and DNS results \citep{sierou2002rheology,gallier2014rheology,singh2018constitutive}.}
\label{fig:Semwogerere_all_models_optimized_viscosities_etas_etan}
\end{figure}

The optimisation procedure, involving the random parameter variation followed by a quadratic descent approach, was used to minimise the $TSRV$ in each model. Table \ref{tab:Semwogerere_optimized_closure_paramters} lists the optimised viscosity closure parameters of the three models. Also, the Figs. \ref{fig:Semwogerere_all_models_optimized_viscosities_etas_etan} and \ref{fig:Semwogerere_roughness_optimized_viscosities_etan_types_I_II_III} compares the related shear and normal viscosity graphs, while the Fig. \ref{fig:Semwogerere_all_models_optimized_phi} compares the simulation results of the optimised models for the different channel flow cases. The viscosity plots show that the three models have similar mixture shear viscosities up to a semi-dilute concentration of around $\phis=0.35$, consistent with existing literature. However, there are notable differences in their optimised phase-specific normal viscosities. Despite this, the optimised models show small variations in the $TSRV$s and in their resulting solid and velocity distributions, except for some differences in the $\Ep$ graphs observed for the most concentrated cases.\\
\\
The closure parameters in the MF-roughness model are linked to the suspension's microstructure and may vary based on the particle size. In this model, one set of $(\scal[r]{\epsilon},\scal[c]{f})$ values for the small $(a=0.69\;\mu\mbox{m})$ and another set for the large $(a=1.5\;\mu\mbox{m})$ particles needed to be determined by minimising the $TSRV$. Three choices can be used to specify these parameters. We can either use different $(\scal[r]{\epsilon},\scal[c]{f})$ values for each of the two-particle sizes or use identical values for both particles. Another approach is to use different parameters for the two particles but assuming that the non-dimensional roughness value $\left(\scal[r]{\hat{\epsilon}}=\scal[r]{\epsilon}/a\right)$ is the same for them. As a result, the table presents three constraint types of the model's closure parameters, each conforming to one of the approaches. The largest $TSRV$ is for constraint III, where identical microstructure parameters are used for both particle sizes. However, the difference in the $TSRV$s value is not considerable around $7\%$, so it can be concluded that using particle-dependent microstructure parameters in the constraint types I and II does not improve the correspondence of the simulation results with the experimental data.\\
\begin{table}
\def~{\hphantom{0}}
\renewcommand{\arraystretch}{1}
\begin{center}
\begin{NiceTabular}{|cc|ccc|ccc|}
\toprule
\Block{1-2}{MF-roughness}&&\Block{1-3}{Fixed}&&&\Block{1-3}{Optimised}&&\\
\midrule
Constraint&$TSRV$&$\scal[o]{\hat{\hat{r}}}$&$N_{2-\scal[\infty]{\hat{r}}}$&$\scal[e]{f}$&particle size&$\scal[c]{f}$&$\scal[r]{\epsilon} (\mbox{nm})$\\
\midrule
\Block{2-1}{I}&\Block{2-1}{$75.966$}&\Block{2-1}{$4\times10^{-5}$}&\Block{2-1}{$1$}&\Block{2-1}{$0$}&small&$1.941$&$2.046$\\
&&&&&large&$1.228$&$11.625$\\
\midrule
\Block{2-1}{II}&\Block{2-1}{$76.242$}&\Block{2-1}{$4\times10^{-5}$}&\Block{2-1}{$1$}&\Block{2-1}{$0$}&small&$1.439$&$4.796$\\
&&&&&large&$1.166$&$10.223$\\
\midrule
III&$80.651$&$4\times10^{-5}$&$1$&$0$&small, large& $1.565$&$3.09$\\
\bottomrule
\end{NiceTabular}
\begin{NiceTabular}{|c|cc|c|ccc|}
\toprule
MF-MB99-B&\Block{1-2}{Fixed}&&\Block{1-4}{Optimised}&&\\
\midrule
$TSRV$&$\scal[2]{\lambda}$&$\scal[3]{\lambda}$&$\phim$&i&$\scal[n,\text{i}]{K}$&$\scal[s,i]{K}$\\
\midrule
\Block{2-1}{$75.551$}&\Block{2-1}{$0.8$}&\Block{2-1}{$0.5$}&\Block{2-1}{$0.599$}&mix&$1.203$&$1.239$\\
&&&&sol&$0.459$&$=\scal[s,mix]{K}$\\
\bottomrule
\end{NiceTabular}
\begin{NiceTabular}{|c|cc|ccc|}
\toprule
MF-MB99-A&\Block{1-2}{Fixed}&&\Block{1-3}{Optimised}&&\\
\midrule
$TSRV$&$\scal[2]{\lambda}$&$\scal[3]{\lambda}$&$\phim$&$\scal[n]{K}$&$\scal[s]{K}$\\
\midrule
$79.585$&$0.8$&$0.5$&$0.605$&$0.252$&$1.239$\\
\bottomrule
\end{NiceTabular}
\caption{The optimised closure parameters of the three MF models. The radius of the small and large particles in the different flow cases are $a=0.69\;\mu\mbox{m}$ and $a=1.5\;\mu\mbox{m}$, respectively. The reported $TSRV=\sum_{\mbox{i}} \scal[i]{R}^2$ here exclude $\scal[i]{R}^2$ values calculated for the deviation of $\scal[s,mix]{\eta}$ graphs in the MF-MB99-A and MF-MB99-B models from the Maron-Piece correlation.}
\label{tab:Semwogerere_optimized_closure_paramters}
\end{center}
\end{table}
\\
The non-dimensional roughness values $(\scal[r]{\hat{\epsilon}})$ in the three constraint types of the MF-roughness model are within the range of $0.002-0.008$. In type I constraint, the larger particle has a higher $\scal[r]{\hat{\epsilon}}$, while in type III, the smaller particle has a higher $\scal[r]{\hat{\epsilon}}$ value. Additionally, in type II, the $\scal[r]{\hat{\epsilon}}$ values of the two particles are very similar, calculated to be $0.007$ and $0.0068$ for the small and large particles, respectively. The calculated $\scal[c]{f}$ values of the particles in the different constraint types are within $1-2$, where a lower non-dimensional roughness value corresponds with a higher $\scal[c]{f}$ value. This suggests that the model viscosities are over-conditioned when varying $\scal[r]{\hat{\epsilon}}$ and $\scal[c]{f}$ simultaneously. Increasing one parameter while reducing the other can result in the same normal stress and thus lead to the same simulation results. The Fig. \ref{fig:Semwogerere_roughness_optimized_viscosities_etan_types_I_II_III} illustrates that while the particles in the three constraint types use different microstructure parameters, their resulting normal viscosities are very close. Consequently, the simulation results and their $TSRV$s are very similar. Notably, the model shear viscosity is almost independent of the microstructure parameters, and also the simulation results are not sensitive to this viscosity variation.\\
\begin{figure}
\centering
\includegraphics[width=\textwidth]{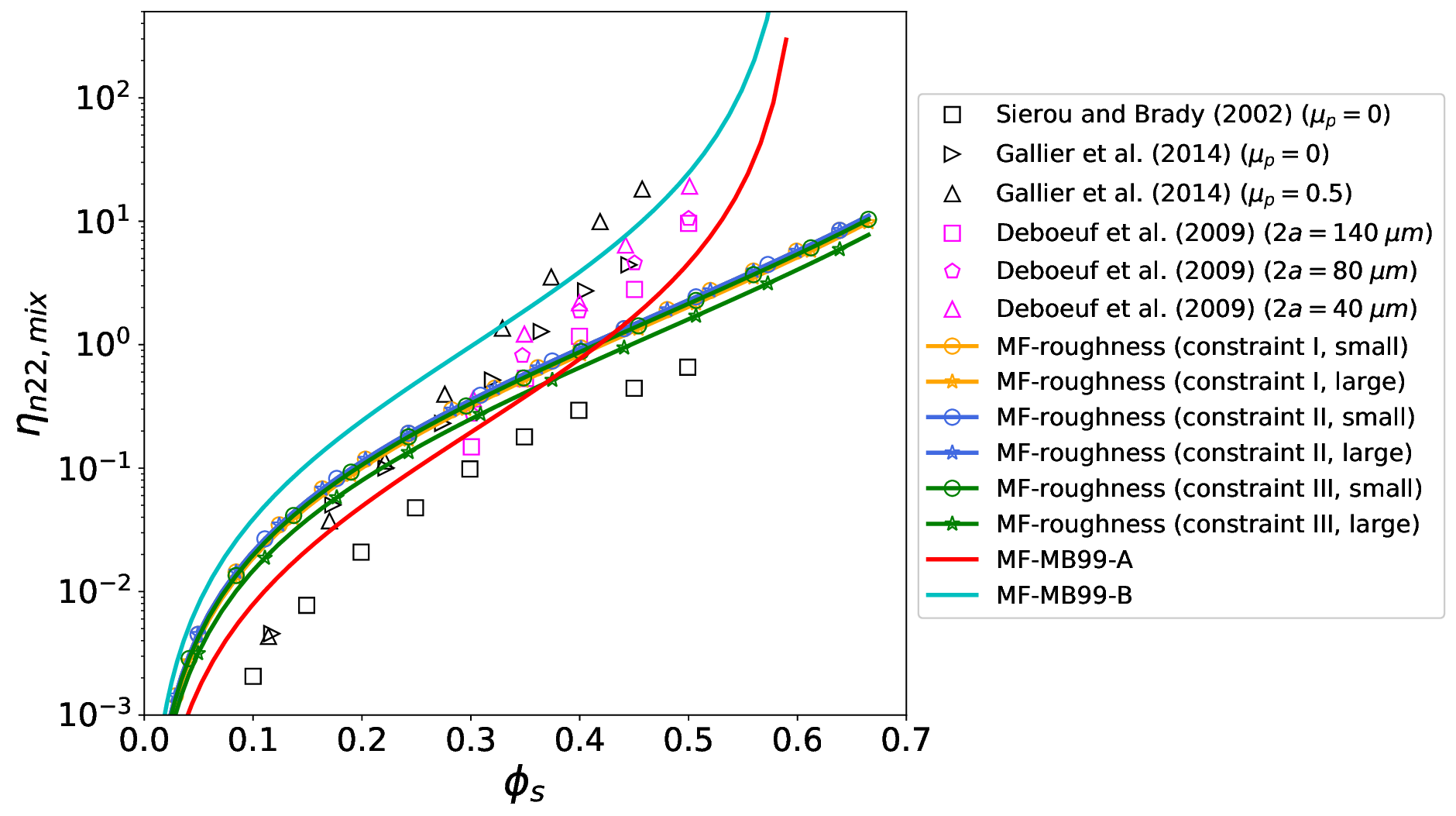}
\caption{The mixture phase normal viscosity in the shear direction for the different optimised MF models (table \ref{tab:Semwogerere_optimized_closure_paramters}) compared to the reported experimental measurements \citep{deboeuf2009particle} and DNS results \citep{sierou2002rheology,gallier2014rheology}.}
\label{fig:Semwogerere_roughness_optimized_viscosities_etan_types_I_II_III}
\end{figure}
\\
The normal viscosity of the three constraint types of the MF-roughness model in Fig. \ref{fig:Semwogerere_roughness_optimized_viscosities_etan_types_I_II_III} show a relatively good comparison with the literature's data in the semi-dilute regime. The mixture's normal viscosity obtained for the different particles with $\scal[r]{\hat{\epsilon}}$ within $0.002-0.008$ at low concentrations is higher than the DNS results by \citet{gallier2014rheology} for particles with $\scal[r]{\hat{\epsilon}}=0.005$ in frictionless flows ($\mu=0$). However, they match in the semi-dilute concentrations for $\phis$ within the range $0.15-0.32$. In constraint type III, using similar $\scal[r]{\epsilon}$ (nm) and $\scal[c]{f}$ values for both particles have led to the smaller particle having a larger $\scal[r]{\hat{\epsilon}}=\scal[r]{\epsilon}/a$ and thus normal viscosities approximately $1.33$ times that of larger particles. The larger normal viscosity for small particles of the same material agrees with the experimental measurements of PS particles by \citet{deboeuf2009particle}. This experimental data shows that the normal viscosity of $40\; \mu\mbox{m}$ diameter particles can be around $1.2-1.8$ of $140\; \mu\mbox{m}$ diameter particles. This agreement suggests using a unique pair of ($\scal[r]{\epsilon},\scal[c]{f}$) as an option for linking the microstructure of similar particles in different sizes to their normal viscosity. In future work, it is important to assess how these parameter with the particle type, size, or other flow variables. This might require a detailed comparison of the MF-roughness variables and viscosity results with experimental particle roughness, suspension microstructure, and viscosity measurements.\\
\\
The MF-MB99-A and MF-MB99-B models have similar optimal values for the maximum solid fraction and shear viscosity coefficient. As the variation of these parameters has a very low impact on the simulation results, their optimal values are primarily determined by the deviation of the models' mixture shear viscosity from the Maron-Pierce correlation, which is considered in minimising the models' $TSRV$s. However, the normal viscosity coefficient of the two models is significantly different. The solid and mixture phase normal viscosity coefficients of the MF-MB99-B model are nearly two and five times larger than those of the MF-MB99-A model. Thus, partitioning the mixture's normal viscosity in the MF-MB99-B model has significantly changed the optimum normal viscosities compared to the MF-MB99-A model. Despite this difference, the minimised $TSRV$s of the two models are very similar, with around a $5\%$ difference. This implies that multiple $(\scal[n,sol]{K},\scal[n,mix]{K})$ pair values in the MF-MB99-B model can produce nearly similar $TSRV$s and thus optimised results. Considering the MF-MB99-A model as a certain case of the MF-MB99-B model, obtained by setting $\scal[n,sol]{K}=\scal[n,mix]{K}=\scal[n]{K}$, by tuning $\scal[n,mix]{K}$ within $0.252-1.203$ a $\scal[n,sol]{K}$ value in range of $0.252-0.459$ can also be found, such that the resulting $TSRV$ is in range of $75-80$. The formulation for the migration flux in the Eq. (\ref{eq:theory-migration-flux-definition-1}) attributes this to the competition between the normal stresses ($\phis\vnabla\cdot\tens[m]{\tau}-\vnabla\cdot\tens[s]{\tau}$). When $\scal[n,mix]{K}$ value increases, the migration flux can remain nearly the same if the value of $\scal[n,sol]{K}$ is also increased. However, due to the $\phis$ prefactor in the term $\phis\vnabla\cdot\tens[m]{\tau}$, the relative increase in $\scal[n,mix]{K}$ value must be greater than that in $\scal[n,sol]{K}$ to maintain the SIM flux balance.
\subsection{Particle and velocity distributions}
The optimised simulation results in Fig \ref{fig:Semwogerere_all_models_optimized_phi} show the solid and velocity distributions of the different models are nearly similar. The residuals $\scal[i]{R}$ values calculated for these distributions across different cases are around $2-3$. Among the different flow cases, the distributions for large particle cases (C1, C3, and C4) in a high $\PeB$ number correspond better with the experimental data. For these flow cases, the solid distributions far from the channel entry, which defines the maximum $\Ep$ value, and the shape of the $\Ep$ graphs, representing the SIM development, correspond well with the experimental data. The graphs of $\Ep$, which are not considered when optimising the closure parameters, show significant variations among the three models or differ from the experimental data in some flow cases of the small or medium sized particles. These differences and the reliability of the parameter in discussing the development of the SIM behaviour are further discussed in the next section. In the case C7, the increased shear viscosity in the centerline, caused by the SIM migration of the particles toward this region, leads to a relatively flattened velocity profile in the simulation results. This agrees with the experimental results, where the maximum velocity decreases from $434.8\;\mu\mbox{m/s}$ to $405.1\;\mu \mbox{m/s}$ with the change in the axial distance from $x=20H$ to $x=1040H$. \\
\\
Using a lower flow rate in case C7 for the simulations resulted in some velocity deviations from the experimental data. For this flow case, the simulations used a flow rate of $0.439\;\mu\mbox{L/min}$, which is relatively smaller than the calculated value of $0.471\; \mu\mbox{L/min}$ from the experimental velocity distributions. This difference has resulted in the simulation velocities being smaller than the experimental data, particularly on one side of the channel cross-section. Also, the experimental data near the walls indicates non-zero velocities greater than $100\;\mu\mbox{m/s}$. If we assume that, similar to the simulation results, the interphase slip velocities in the experiments are negligible, then the non-zero wall velocities and the difference between the reported experimental flowrate for this case ($0.51\;\mu\mbox{L/min}$) and the calculated flowrate from the experimental velocity profiles ($0.471\;\mu\mbox{L/min}$) may indicate some inaccuracies in the velocity measurements near the wall regions. The analysis of the simulation results for all the flow cases shows small interphase slip velocities less than $0.05\;\mu \mbox{m/s}$. Moreover, in a similar experimental study, \citet{frank2003particle} found slip velocities less than measurement accuracy of around $20\;\mu\mbox{m/s}$ for the flows of $2.2\;\mu\mbox{m}$ diameter particles with an average velocity of around $933\;\mu\mbox{m/s}$, $\phio=0.22$, and $\PeB=770$. The experimental measurements in the flows studied here may have been affected by inaccuracies in locating the exact position of the wall. For example, the wall may have been placed closer to the centerline than in reality, leading to a shift in the velocity profile. This could potentially explain the reported non-zero wall velocities.
\subsection{The evolution parameter $\Ep$ distributions}\label{the-evolution-parameter-distributions}
\begin{figure}
\centering
\begin{subfigure}[a]{0.9\textwidth}
\centering
\includegraphics[width=\textwidth]{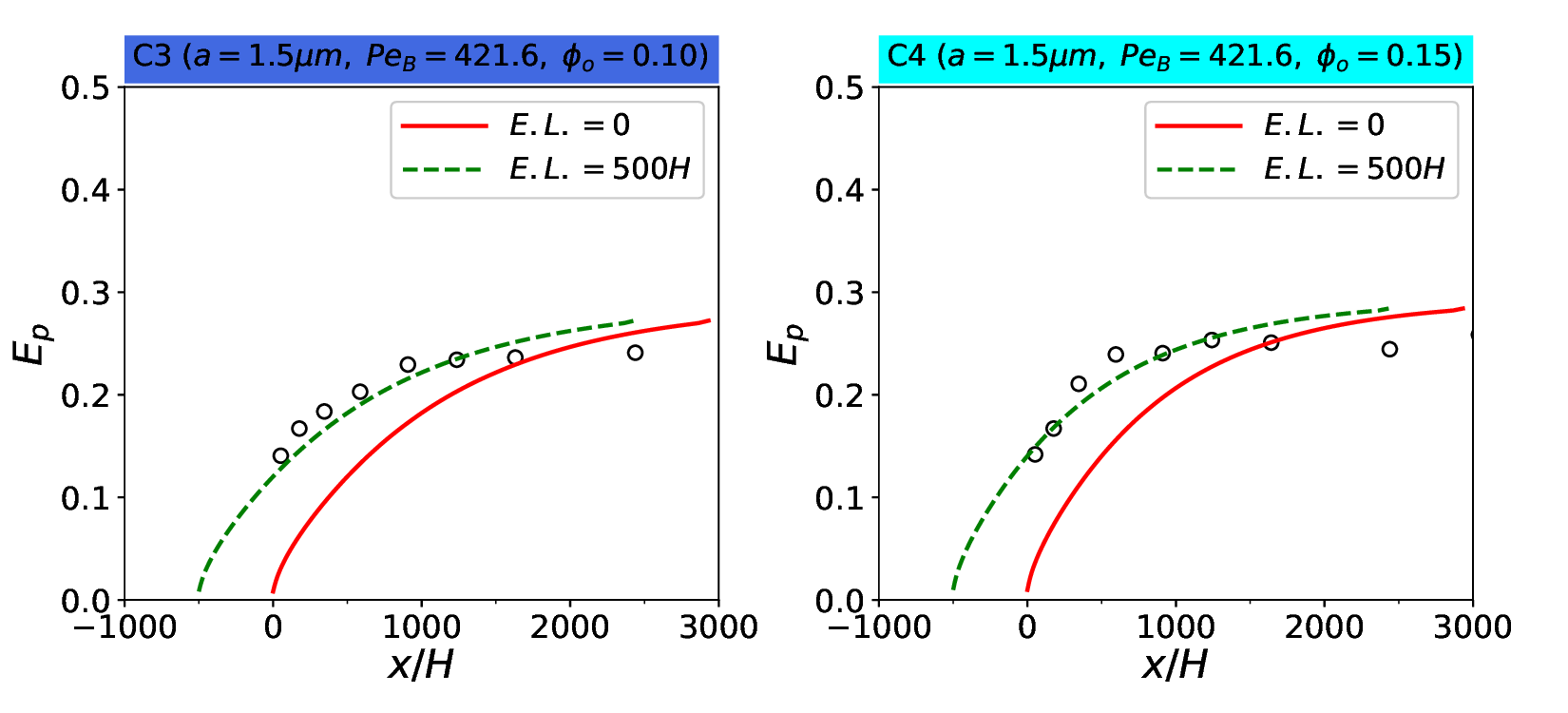}
\caption{}
\label{fig:Semwogerere_roughness_optimized_EL_effect_on_C3_C4}
\end{subfigure}
\begin{subfigure}[b]{0.9\textwidth}
\centering
\includegraphics[width=\textwidth]{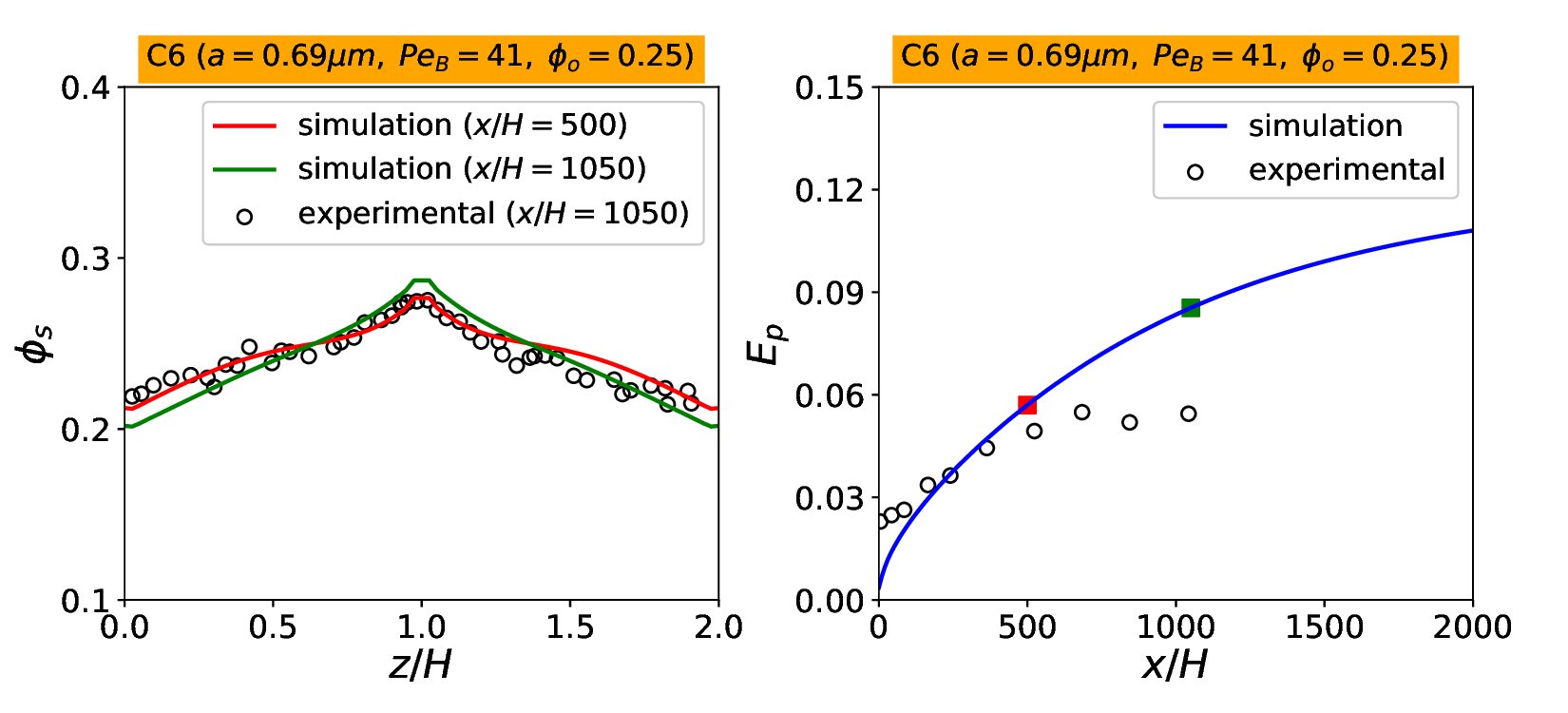}
\caption{}
\label{fig:Semwogerere_roughness_optimized_EP_reliability}
\end{subfigure}
\caption{(a) The impact of using an entrance axial length of $E.L.=500H$ in the optimised MF-roughness model (constraint III) on the correspondence of $\Ep$ results with the experimental data. (b) The impact of slight variation in $\phis$ distributions on $\Ep$ simulation results (optimised MF-roughness model constraint type III). The red and green squares on the right subplot represent the $\Ep$ values calculated for the corresponding $\phis$ distributions shown by the red and green solid lines on the left subplot.}
\label{fig:Semwogerere_roughness_optimized_EL_effect_EP_reliability}
\end{figure}

The increase in the $\Ep$ values with the axial length indicates the progression of SIM along the channel length. The experimental data points of this parameter show an asymptotic behaviour and suggest that the solid distributions reach a fully developed condition after the axial lengths within $200H-1000H$ (depending on $\PeB$). However, the simulation results mostly do not exhibit this behaviour. The results develop slower but continue to distances even larger than $1500H$ with the maximum $\Ep$ values larger than the asymptotic experimental values. For case C9, which has a small $\PeB$ number and where the expansive osmotic force plays a significant role in limiting SIM, the simulation $\Ep$ profiles show asymptotic behaviour.\\
\\
The $\Ep$ graphs show that the MF-MB99-B model exhibits faster development in the semi-dilute flows of $\phio<0.25$ than other models. This difference is primarily attributed to the former model's higher normal viscosity in the solid phase. However, in high-concentration flow cases (C8, C9, and C10), this model's larger normal viscosity in the mixture phase becomes significant and limits the migration flux. This SIM behaviour can be explained by the competition between the solid and mixture phase stress forces, $\phis\vnabla\cdot\tens[m]{\tau}-\vnabla\cdot\tens[s]{\tau}$ in Eq. \ref{eq:theory-migration-flux-definition-1}. For low $\phis$ values, which $-\vnabla\cdot\tens[s]{\tau}$ dominates and the competition is less important, the larger solid phase normal viscosity of the MF-MB99-B model than other models leads to a larger migration flux toward the centerline. However, at high $\phis$ values, when $\phis\vnabla\cdot\tens[m]{\tau}$ is comparable to $-\vnabla\cdot\tens[s]{\tau}$, the significantly larger magnitudes of the mixture normal viscosity in the MF-MB99-B model limits its prediction of the migration flux toward the centerline.\\
\\
In some of the experimental $\Ep$ graphs, the initial parameter values at the channel entry are non-zero. This value can indicate a non-uniform solid distribution at the channel entry. This could be due to the suspension never being uniform before pumping, or it may have partially developed through the piping connections before entering the flow chamber. Fig. \ref{fig:Semwogerere_roughness_optimized_EL_effect_on_C3_C4} addresses the impact of the non-uniform solid distribution at the channel entry on the simulation $\Ep$ results of the cases C3 and C4. In this figure, we have plotted $\Ep$ versus $x/H$ alongside the same graphs but sifted by $500$ on the horizontal axis to account for an extra entrance length of $500H$ before the channel entry. Using this entrance length improves the correspondence of the simulation results with the experimental data in the developing region of the flow.\\
\begin{figure}
\centering
\begin{subfigure}{1\textwidth}
\centering
\includegraphics[width=0.9\textwidth]{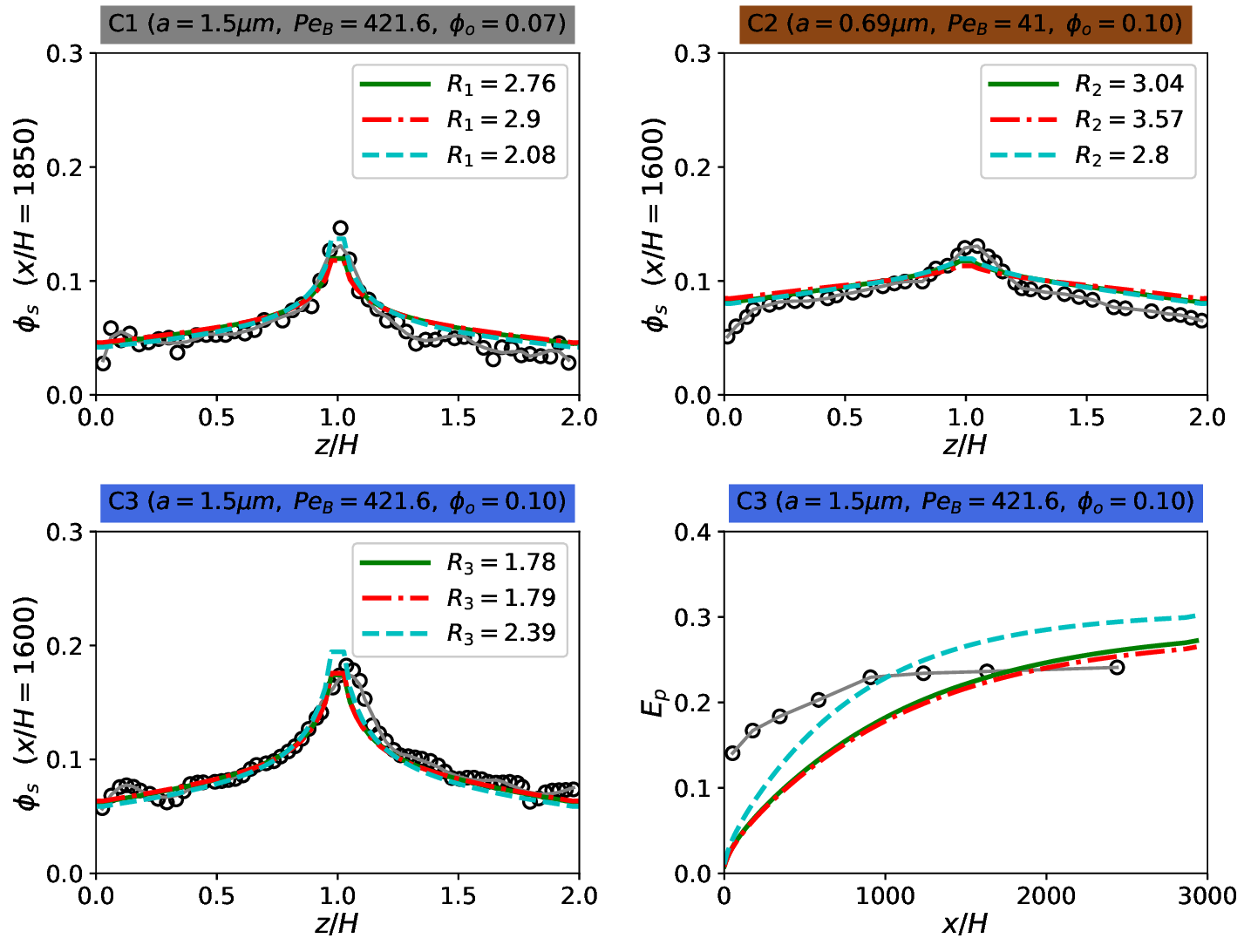}
\end{subfigure}
\begin{subfigure}{1\textwidth}
\centering
\includegraphics[width=0.9\textwidth]{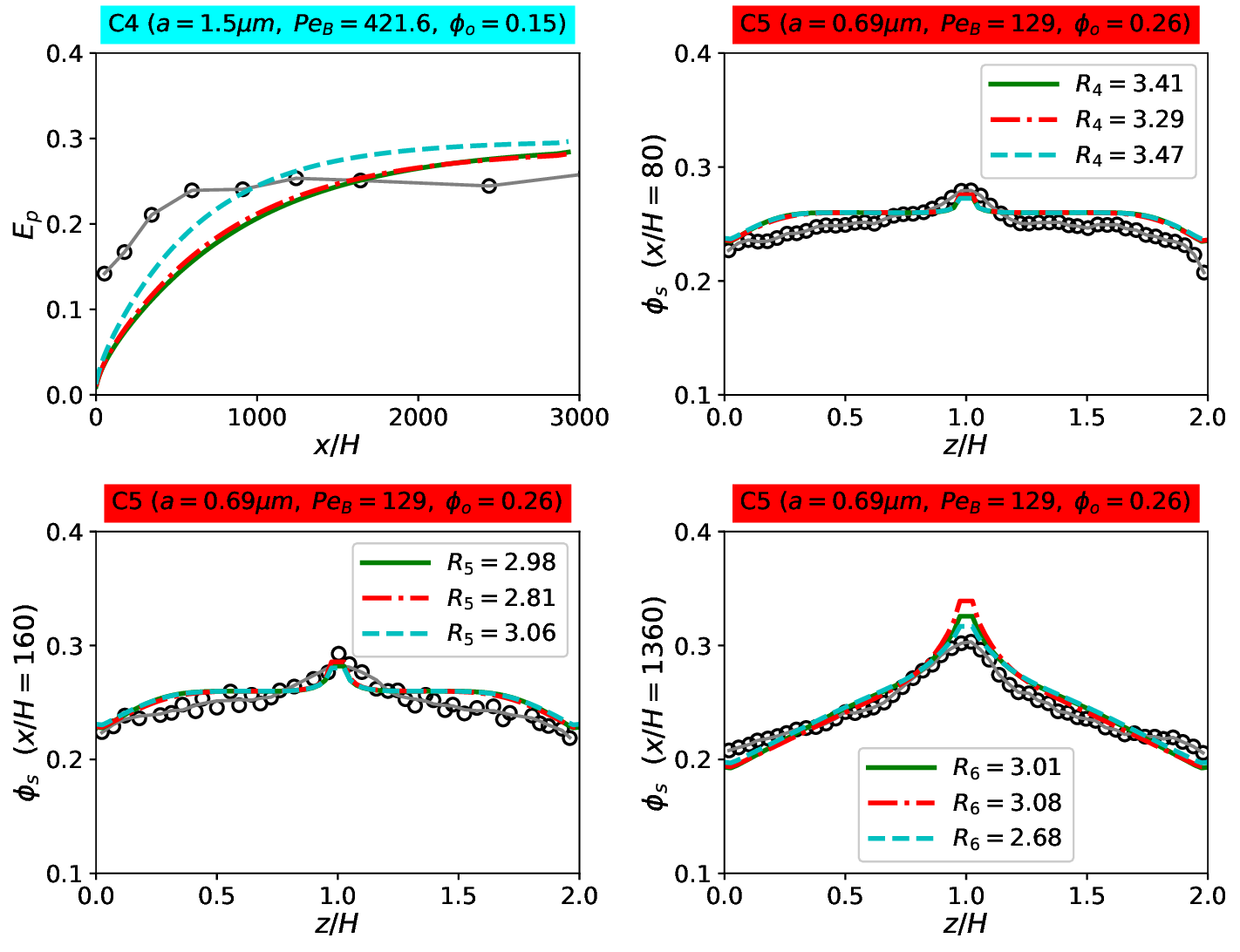}
\end{subfigure}
\caption{Continued}
\vspace{8 cm}
\end{figure}
\begin{figure}
\ContinuedFloat
\centering
\begin{subfigure}{1\textwidth}
\centering
\includegraphics[width=0.9\textwidth]{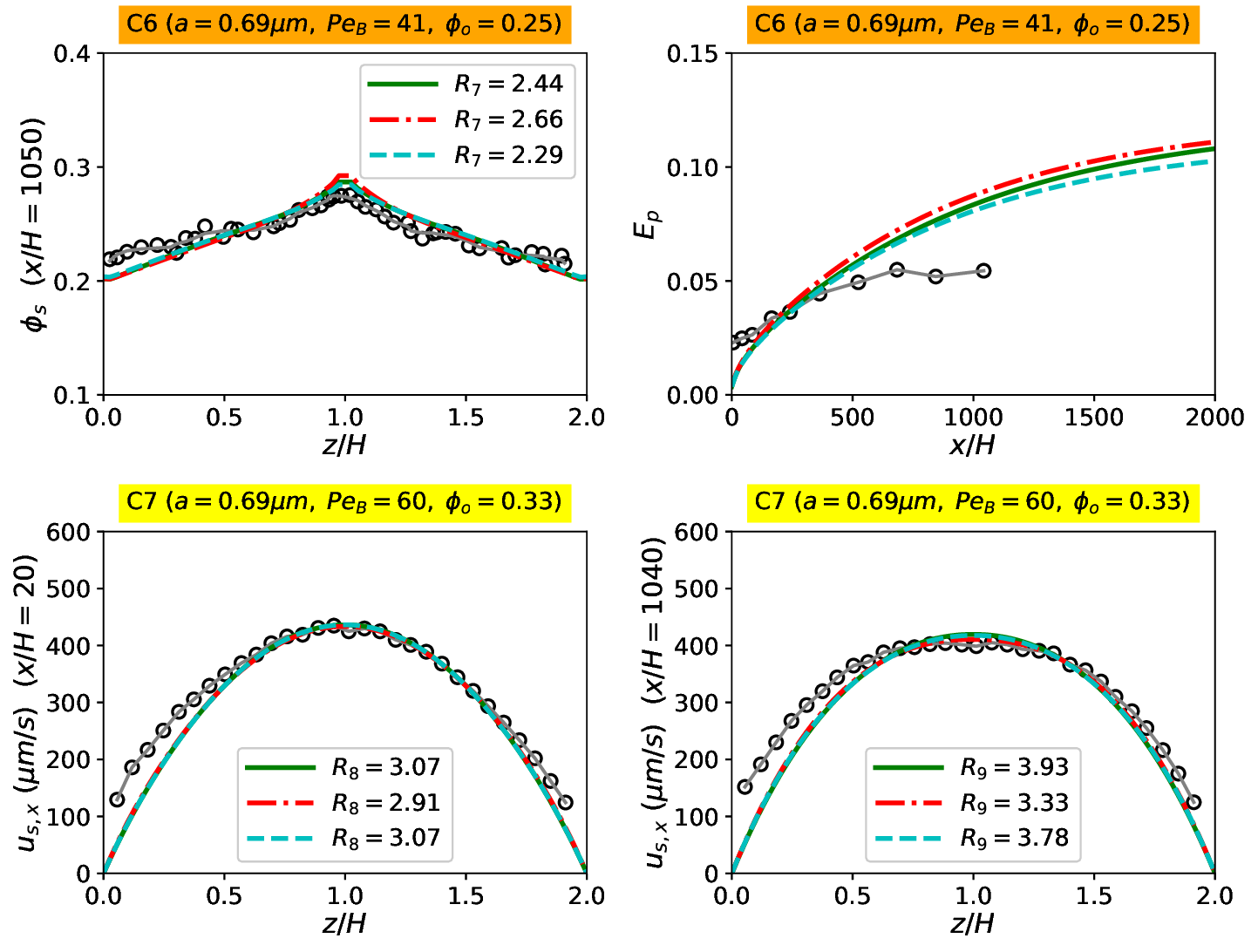}
\end{subfigure}
\begin{subfigure}{1\textwidth}
\centering
\includegraphics[width=0.9\textwidth]{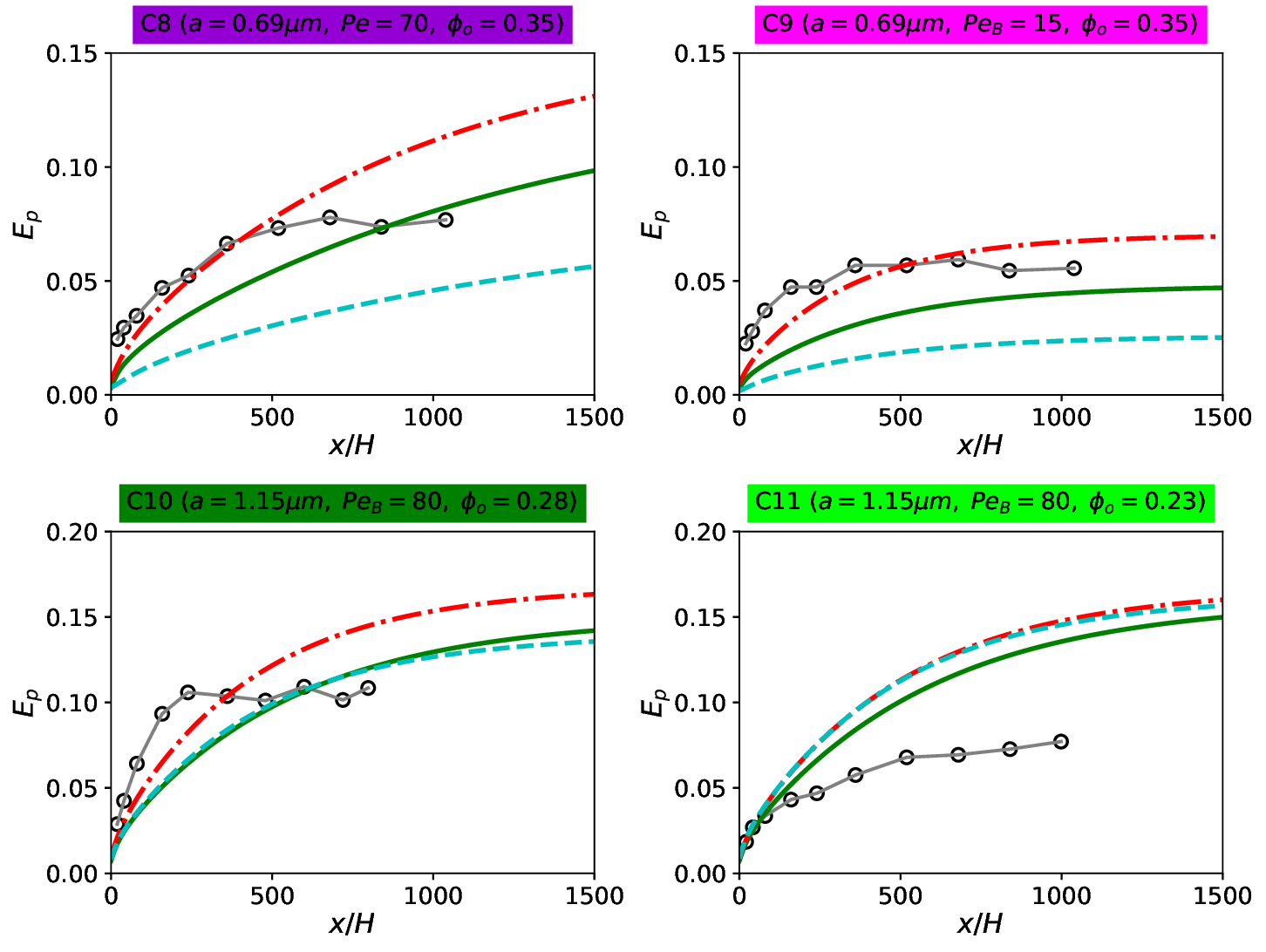}
\end{subfigure}
\caption{The optimised simulation results of the MF-roughness constraint III (green solid line), the MF-MB99-A (red dash-dotted line), and the MF-MB99-B (cyan dashed line) models (table \ref{tab:Semwogerere_optimized_closure_paramters}). The optimised MF-roughness model obtained by minimising $TSRV=\sum_{\mbox{i}=1}^{\mbox{i}=9} \scal[i]{R}^2$, where the residual $\scal[i]{R}$ values are indicated in the legends. Each of the optimised MF-MB99-A and MF-MB99-B models obtained by minimising $TSRV=\sum_{\mbox{i}=1}^{\mbox{i}=10} \scal[i]{R}^2$, where $\scal[10]{R}$ values calculated for the deviation of the model $\scal[s,mix]{\eta}$ graph from the Maron-Piece correlation.}
\label{fig:Semwogerere_all_models_optimized_phi}
\end{figure}
\\
The difference between experimental and simulated $\Ep$ values could result from the high sensitivity of the parameter to the changes in the $\phis$ distribution, particularly the measured $\phis$ profiles. To assess this sensitivity, the calculated $\Ep$ values for two solid distributions results (in the case C6) with slight local differences are shown in the Fig. \ref{fig:Semwogerere_roughness_optimized_EP_reliability}. The figure indicates that a mean $12\%$ difference in the local $\phis$ values can lead to a $60\%$ difference in the corresponding $\Ep$ values. This suggests that using $\Ep$ values for the development study of solid distributions with certain accuracy levels may not be valid. An analysis of Eq. (\ref{eq:Semwogerere_Ep_definition}) shows that for the solid distributions with a fixed absolute error of $\scal[\phi]{\delta}$, the absolute error in the evolution parameters $\scal[E]{\delta}$ could be approximated as $\scal[E]{\delta}\simeq\scal[\phi]{\delta}$. Assuming that the experimental measurements have a constant $\scal[\phi]{\delta}\simeq 0.1$ \citep{semwogerere2007development}, it can be said the larger experimental $\Ep$ data points could have a higher accuracy with a smaller relative error than the lower magnitudes of $\Ep$ values.\\
\subsection{Parameter sensitivity analysis}
\begin{figure}
\centering
\includegraphics[width=0.8\textwidth]{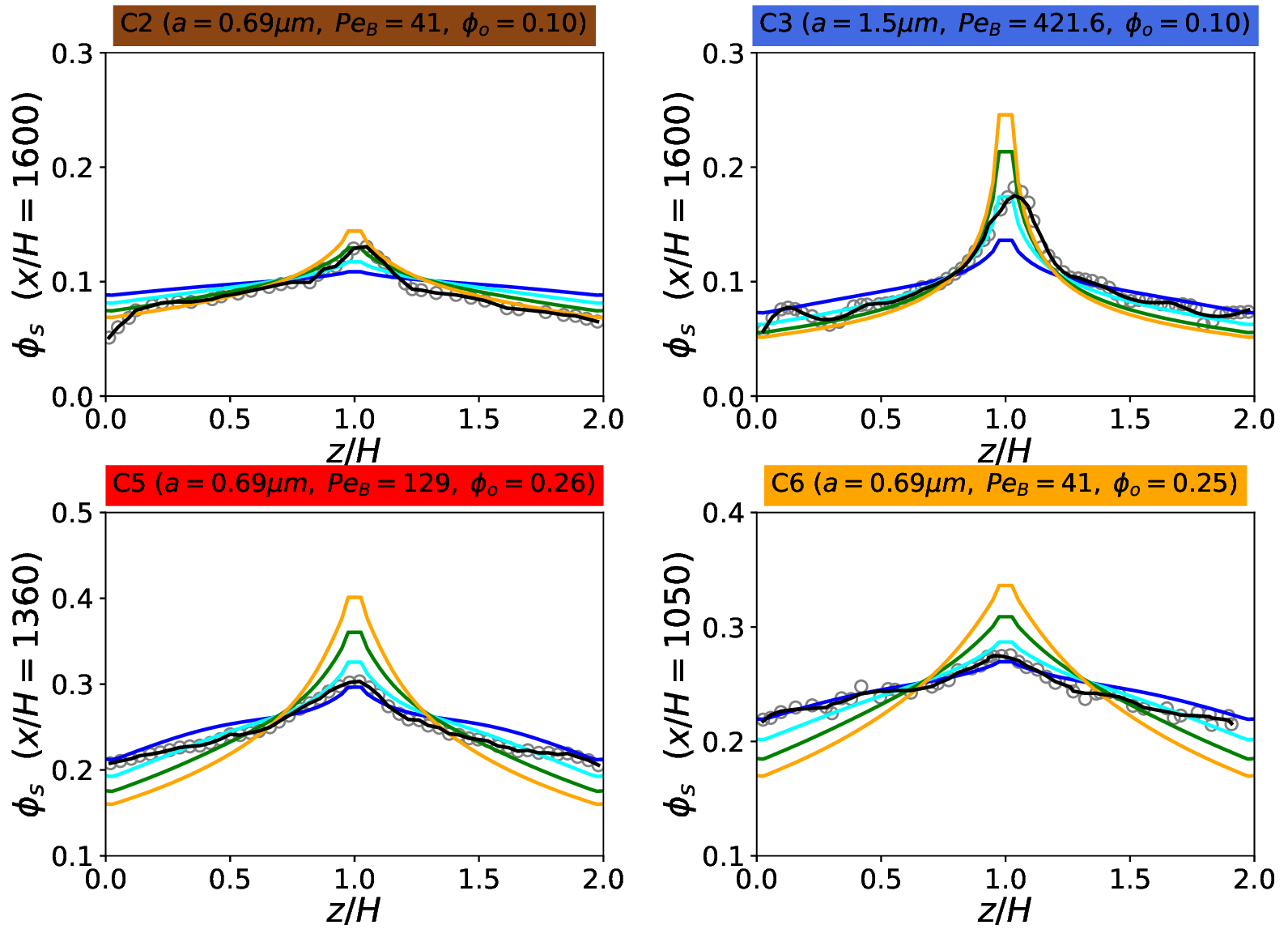}
\caption{The sensitivity of the optimised MF-roughness (constraint type III) to the variation of the dimensional roughness value. The coloured blue, cyan, green, and orange solid lines represent $\scal[r]{\epsilon}$ values $0.5\;\mbox{nm}$, $3.09\;\mbox{nm}$, $20\;\mbox{nm}$, and $100\;\mbox{nm}$, respectively.}
\label{fig:Semwogerere_roughness_sensitivity_on_eps_r_phi_cases2356}
\end{figure}

A series of sensitivity analyses is conducted to examine the impact of varying the MF-roughness and MF-MB99-B model closure parameters on viscosities and simulation results. Additionally, we performed simulations of different flow cases to observe the effects of the osmotic and Faxen forces and bulk stresses on the MF-roughness results of different flow cases. The simulation results and viscosity plots related to these analyses are all presented in section S2 in the supplementary document. The simulations indicate that excluding the Faxen force and bulk stress has minimal impact on the results. Excluding or doubling the osmotic force mainly affects the centerline value of the solids volume fraction, with limited impact on solid distribution results in other regions of the channel. The variation of this force can make considerable changes in the $\Ep$ graphs, attributed to the high sensitivity of the parameter magnitude to the change in the centerline value of the solid fraction. The analysis of MF-roughness shows that varying $N_{2-\scal[\infty]{\hat{r}}}$ within the range of $0.5-10$, as well as varying $\scal[o]{\hat{\hat{r}}}$ within the range of $3-5\times 10^{-5}$, has minimal impact on the model's shear or normal viscosities. Additionally, the model analysis shows that an increase in $\left(\scal[c]{f}+\scal[e]{f}\right)$ at a fixed $\left(\scal[c]{f}-\scal[e]{f}\right)$ value does not affect the normal viscosities but may lead to a relatively small increase in the shear viscosities.\\
\\
The sensitivity analysis shows that increasing the $\scal[c]{f}$ and $\scal[r]{\epsilon}$ values has little impact on shear viscosities but can significantly increase the phase-specific normal viscosities. This normal viscosity variation leads to a greater SIM flux and particles' fully developed central enrichment. Moreover, the downstream velocity profile of the case C7 shows a higher degree of flattening with an increase in the two parameters due to the increased shear viscosity at the centerline resulting from the enhanced SIM of particles. The variation of the cross-stream migration flux $(\scal[sm,z]{j})$ and the fully developed solid distributions (discussed based on $-z\p \phis/\p z$) with $\scal[c]{f}$ and $\scal[r]{\epsilon}$ can be explained by the analytical analysis of the SIM in the Eqs. (\ref{eq:theory-migration-flux-definition-4}) and (\ref{eq:theory-non-dimensional-phi-grad-1}). The two closure parameters are the prefactors to the mixture's normal viscosity $\scal[n22,mix]{\eta}$. Thus, their variations change the magnitude of this viscosity without affecting the ratio $\scal[n22,mix]{\eta}^{'}/\scal[n22,mix]{\eta}$. Consequently, as the parameters have an insignificant impact on the mixture shear viscosity $\scal[s,mix]{\eta}$, their increase results in larger magnitudes of $\scal[sm,z]{j}$ and $-z\p \phis/\p z$. It is noteworthy that by changing the two closure parameters, $\scal[m,SIM]{\phi}$ is found to be nearly fixed at $0.586$, and the shear rate distribution remains linear and unchanged, except for regions near the centerline as seen for the case C7.\\
\\
A similar SIM behaviour is observed in the MF-MB99-B model with the variation of $\scal[n,sol]{K}$. Larger values of this parameter cause a larger $\scal[m,SIM]{\phi}=\scal[n,sol]{K}/\scal[n,mix]{K}$ in the model, leading to an increased migration flux $\scal[sm,z]{j}$ (again based on the Eq. (\ref{eq:theory-migration-flux-definition-4})). However, the simulation results show a lower sensitivity to $\scal[n,mix]{K}$ values mainly observed in the high-concentration flows. Although higher values of this parameter lead to larger $\scal[n22,mix]{\eta}$ values, the migration flux is reduced due to the decrease in the $\scal[m,SIM]{\phi}=\scal[n,sol]{K}/\scal[n,mix]{K}$ ratio. Moreover, the model simulation results are minimally affected by the variation of the shear viscosity coefficients $\scal[s,mix]{K}$ and $\scal[s,sol]{K}$. In the studied semi-dilute flows, the relationship $\left(z/\scal[s,mix]{\eta}\right)\left(\p \scal[s,mix]{\eta}/\p z\right)=\left(\scal[s,mix]{\eta}^{'}/\scal[s,mix]{\eta}\right)\left(z \p\phis/\p z\right)\ll1$ always hold. This means that the shear rate profile remains nearly linear and unchanged regardless of the shear viscosity magnitude as $\left(z/\gammadot\right)\left(\p \gammadot/\p z\right)\simeq 1$. Consequently, the migration flux and the distribution of solids remain nearly unaffected. The sensitivity analysis also indicates that the solid and velocity distributions in the MF-MB99-B model are not highly affected by $\phim$ variations within the range of $0.585-0.68$. This is because the shear and normal viscosities in the semi-dilute regime do not significantly change with the $\phim$ value.\\
\\
The MF-roughness sensitivity results show that solid distributions of flows with $\phio<0.1$ have low sensitivity to the parameters defining the anisotropy of the PDF model, $\scal[c]{f}$ or $\scal[r]{\epsilon}$. This analysis indicates that the variation of the two closure parameters can have different impacts on the deviation of the simulation results from the experimental data depending on $\phio$ values. As shown in Fig. \ref{fig:Semwogerere_roughness_sensitivity_on_eps_r_phi_cases2356}, when increasing the $\scal[r]{\epsilon}$ value from $3.05\; \mbox{nm}$ to $100\; \mbox{nm}$ or $\scal[c]{f}$ value from $1.565$ to $3.0$, the deviation does not show a significant change in the flow cases of lower $\phio$ values (C2 and C3) but it noticeably increases in the flows with higher $\phio$ values (C5 and C6).
The different impact of the closure parameters variation on the flows of $\phio \simeq 0.1$ and $\phio\simeq 0.25$ may suggest (but not necessarily) that the width of the asymmetric region or the PDF value in the compression zones decrease with the solid fraction. According to the experimental measurements by \citet{blanc2013microstructure}, the rotation angle of the depleted region on the receding side of the particles, indicating the microstructure anisotropy, does not change for $\phis<0.13$, consistent with the results observed here. However, beyond this solid fraction, they found that the angle and, thus, the anisotropy increases with the $\phis$ values, which contradicts the sensitivity results here. Consistent with the latter experimental observations, the DNS results by \citet{morris2002microstructure} also show that the contact values of the pair correlations in the compression zone increase with the solid fraction. Notably, the two closure parameters in the MF-roughness model represent the properties of an effective anisotropic region in the PDF model and cannot be directly compared to the results found in the literature. For example, the measured PDF in \citet{blanc2013microstructure} shows that the significance of the extension of the depleted region to very high separation distances is more pronounced at lower solid fractions than at high solid fractions. The resolution of the sampled data in the literature above is also a parameter that may differ from the MF-roughness PDF model. For instance, \citet{blanc2013microstructure} sampled the PDF data using a mesh size of $\Delta \hat{r}=0.07$, larger than the outer radius of the asymmetric region in the MF-roughness model considered in the simulations.
\section{Conclusions}
This study presents the functional form of stresses for a volume-averaged MF model proposed by \citet{harvie2021tensorial} applicable for the analysis of SIM in the pressure driven suspension flows. The closure of the proposed stresses depends on some phase-specific interaction viscosities formulated based on the volume-averaged force and stresslet resulting from the particles' collision in a suspension microstructure. We numerically evaluated the interaction force and stresslets and then obtained the functional forms of the viscosities using curve fitting. For this purpose, we assumed a roughness-dependent PDF model based on the literature's direct experimental observations of solids microstructure in sheared suspensions.\\
\\
The resulting mixture shear viscosity in this MF-roughness model is larger than the solid phase and shows a good agreement with the semi-dilute experimental measurements in the literature and low sensitivity to the microstructure parameters. The solid and mixture phase normal viscosities are in the range of the DNS results and experimental correlations in the literature. These viscosities show a high increase with two microstructure parameters that define the anisotropy of the chosen microstructure model. These parameters are represented by $\scal[r]{\hat{\epsilon}}$ as the hydrodynamic roughness value defining the width asymmetric region in the PDF model and $\scal[c]{f}$ defining the PDF value in the compression quadrants of this region. Finally, the ratio of the solid to the mixture phase normal viscosity is calculated to be around $\scal[m,SIM]{\phi}\simeq0.585$ independent of the model closure parameters. The analytical analysis of the SIM resulting from the implemented MF model the current study demonstrate that the solids migration flux due to competing stress forces in the suspension flow reaches zero at a critical solid fraction equal to this viscosity ratio, having significance in concentrated flows.\\
\\
The current study also proposes two different versions of the phenomenological stress closures by \citet{morris1999curvilinear} based on the features of the MF-roughness closure. The first version, the MF-MB99-A model, is a frame-invariant, fully tensorial form of the closure that aligns with the original form in unidirectional flows. The second proposed version, the MF-MB99-B model, is consistent with averaging techniques and the MF-roughness model and partitions the mixture stress between the solid and fluid phases. This differs from the MF-MB99-A or the original \citeauthor{morris1999curvilinear}'s closure form, where the normal stress and the interaction part of the shear stress are equal for the mixture and solid phases.\\
\\
The free parameters of the three stress closures were optimised based on the simulation of SIM behaviour in channel flows. These closures were used in MF conservation equations to simulate different experimental channel flows of monodisperse $1-3\;\mu\mbox{m}$ diameter solid suspensions, as reported by \citet{semwogerere2007development} and \citet{semwogerere2008shear}. The closure parameters were globally optimised to describe the reported velocity and solid distributions across the different flows. The optimised results of the three models are very similar and show a good correspondence with the experimental data. The parameters that affect shear viscosity have very low impact on the results. However, the normal viscosity parameters, $\scal[c]{f}$ and $\scal[r]{\epsilon}$ values in the MF-roughness model and the normal viscosity coefficients in the MF-MB99-A and MF-MB99-B models, are the key parameters affecting the simulation solid distributions and their correspondence with the experimental data.\\
\\
The MF-roughness model is over-conditioned with the microstructure $\scal[c]{f}$ and $\scal[r]{\epsilon}$ values behaving as effective prefactors to the normal viscosities. Additionally, the results show that adjusting these parameters based on the particle size did not improve the agreement between the simulation results and the experimental data. Consequently, we presented different sets of optimised parameters leading to nearly normal viscosities and solid distribution results. The $\scal[c]{f}$ values corresponding to the dimensional roughness of $2-12\; \mbox{nm}$, consistent with the known physical roughness of the PMMA particles in the literature, are within the range $1-2$. Among these parameter sets, a case exists with unique dimensional roughness $\scal[r]{\epsilon}$ value and $\scal[c]{f}$ for the particles, independent of their sizes. A case also uses a unique non-dimensional roughness $\scal[r]{\hat{\epsilon}}$ value but with a $\scal[c]{f}$ value dependent on the particle size. Further SIM simulations of different suspensions and experimental observation of the microstructure of sheared suspensions in the future can help understand how these closure parameters vary with the particle size and type.\\
\\
The optimised normal viscosity of the MF-MB99-A model is significantly lower than the experimental correlations by \citet{boyer2011unifying}. However, the mixture's normal viscosity of the MF-MB99-B model due to the normal stress partitioning is found to be very close to this experimental correlation and around five times larger than that of the MF-MB99-A model. It is important to note that the solid distribution simulation results of the two models showed minimal sensitivity to the variation of the shear viscosity coefficients within the range $0.1-5$ and $\phim$ values within $0.585-0.68$.\\
\\
The sensitivity analysis indicates that solid distributions of the flows with low initial solid fraction $\phio < 0.1$ are not highly sensitive to the increase in the solid phase normal viscosities of the models from their optimum magnitudes. However, fully developed results of high concentration cases with $\phio > 0.25$ show a high sensitivity to this change, resulting in larger deviations of the simulation results from the experimental data. The assessment of this behaviour with the MF-roughness model suggests that the microstructure parameters $\scal[c]{f}$ and $\scal[r]{\epsilon}$ are constant in low solid fractions but decrease in higher solid fractions.
\bibliography{noori24a_main}
\end{document}


\maketitle
This document includes figures related to two types of sensitivity analysis in the current study. The section \ref{analysis-MF-roughness-numerical-procedures} includes some graphs related to numerical and curve-fitting procedures used to find the functional forms of the MF-roughness viscosities. The section \ref{sensitivity-analysis-MF-models-viscosities-and-results} presents the sensitivity of three MF model viscosities and simulation results to variations in mesh size, force or stress terms, and their closure parameters.
\section{The analysis of the MF-roughness numerical procedures}\label{analysis-MF-roughness-numerical-procedures}
\begin{figure}
\centering
\includegraphics[width=0.45\textwidth]{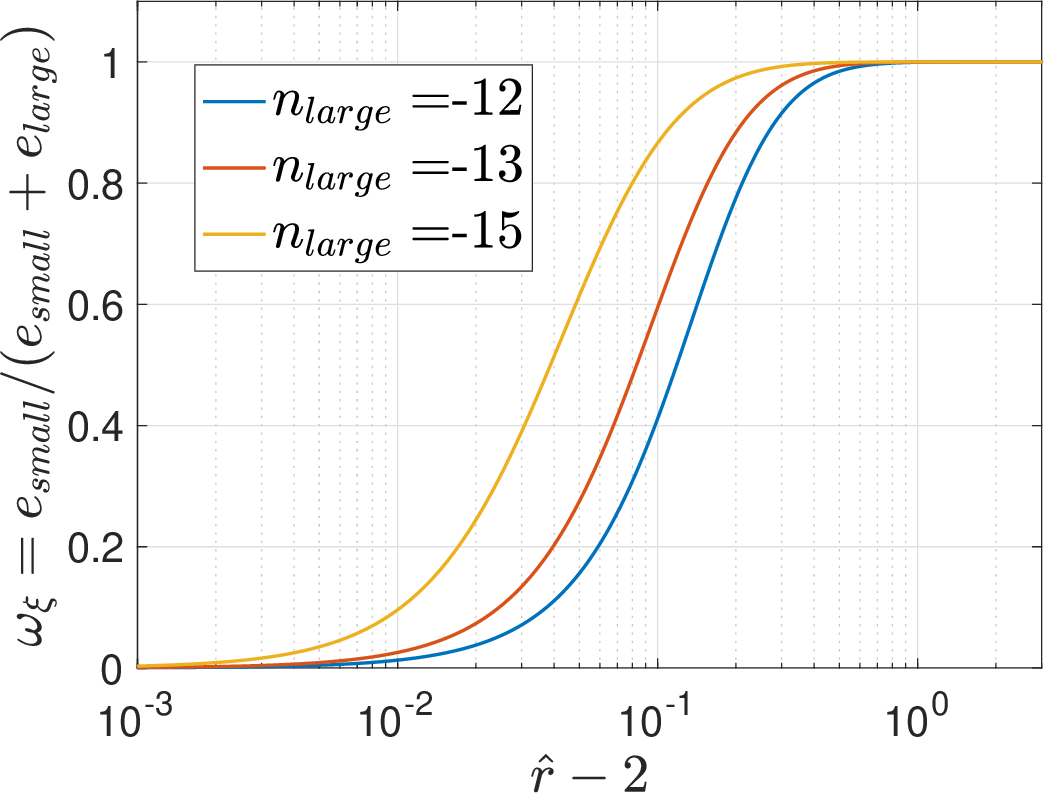}
\caption{A plot of the function used to merge the near- and far- field values of the mobility and resistance scalars.}
\label{fig:xi-scalars-interpolation-function-plot}
\end{figure}

\begin{figure}
\centering
\begin{subfigure}[b]{0.48\textwidth}
\centering
\includegraphics[width=\textwidth]{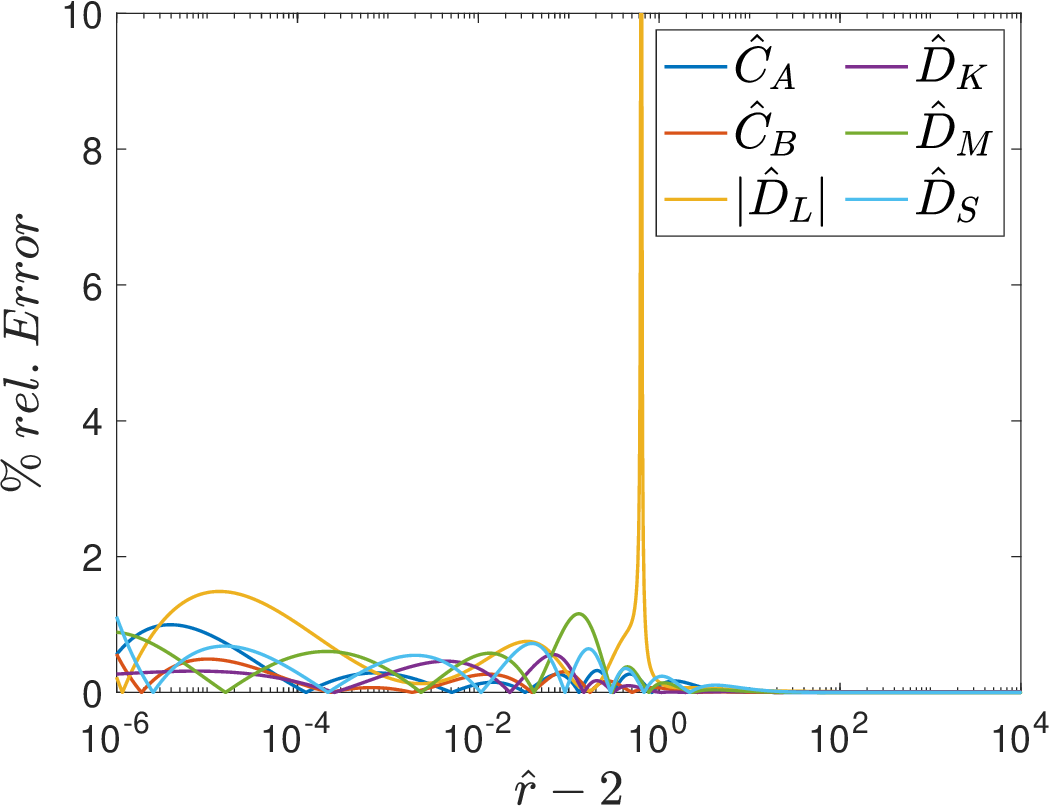}
\caption{}
\label{fig:mobility-resistance-integrals-relative-error}
\end{subfigure}
\begin{subfigure}[b]{0.48\textwidth}
\centering
\includegraphics[width=\textwidth]{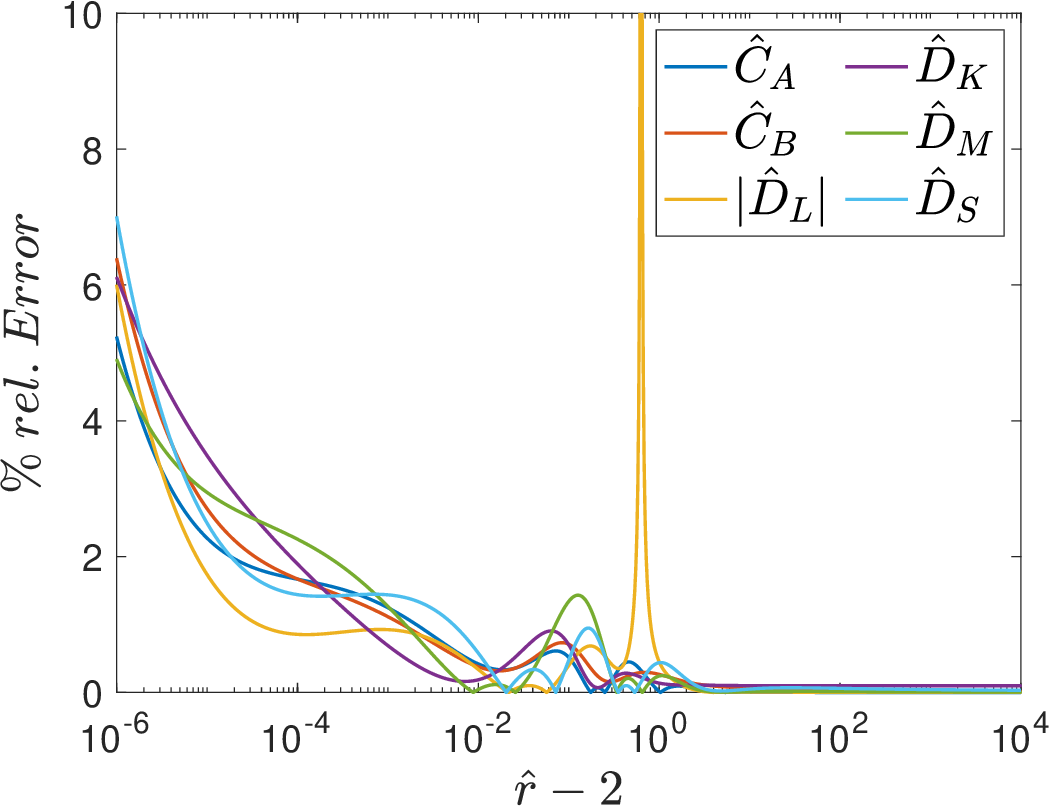}
\caption{}
\label{fig:mobility-resistance-integrals-relative-error-fine-mesh}
\end{subfigure}
\begin{subfigure}[b]{0.48\textwidth}
\centering
\includegraphics[width=\textwidth]{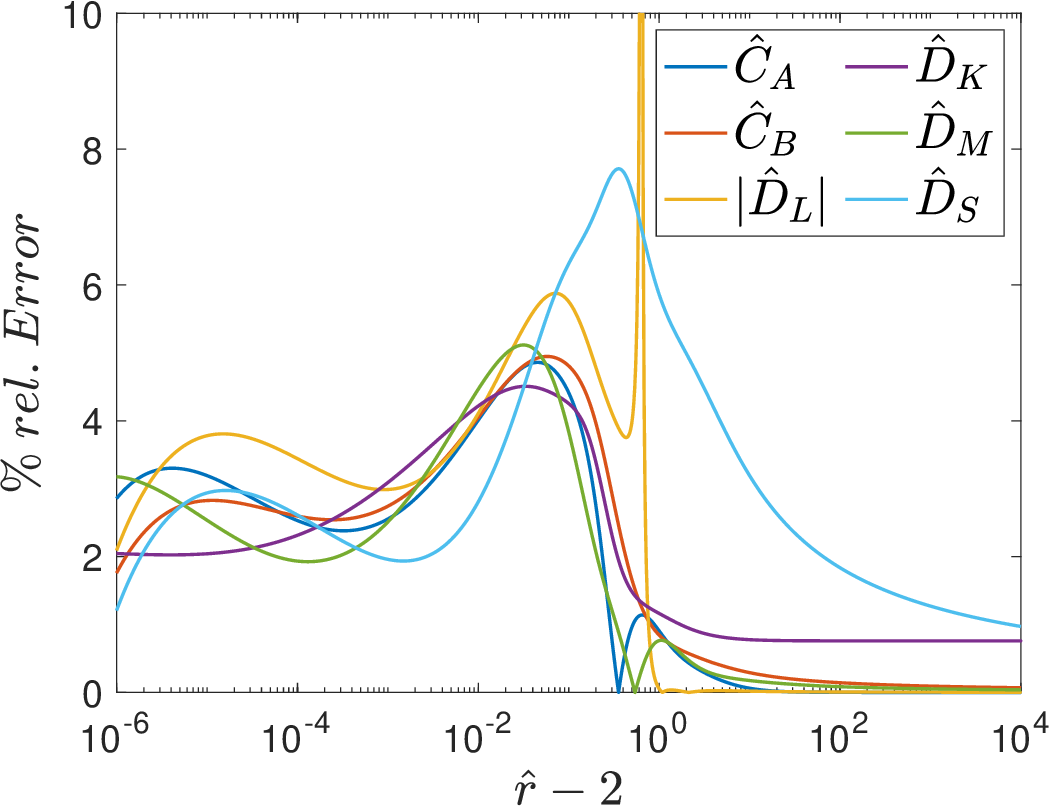}
\caption{}
\label{fig:mobility-resistance-integrals-relative-error-12}
\end{subfigure}
\begin{subfigure}[b]{0.48\textwidth}
\centering
\includegraphics[width=\textwidth]{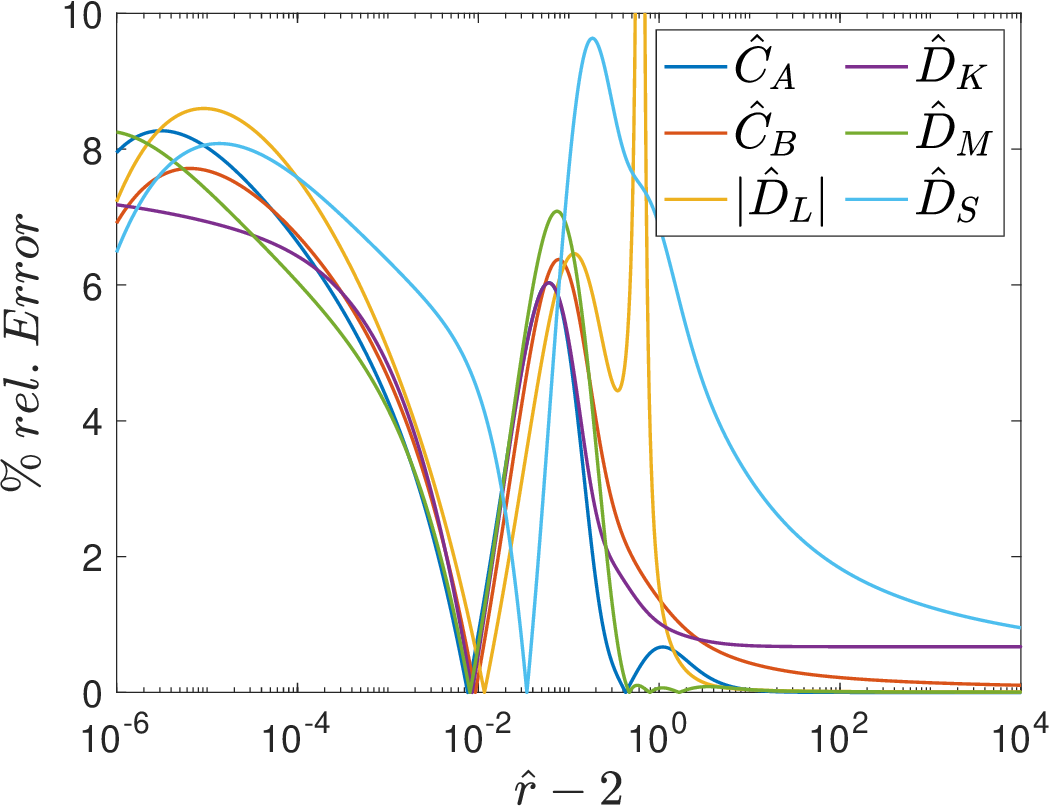}
\caption{}
\label{fig:mobility-resistance-integrals-relative-error-15}
\end{subfigure}
\caption{Percentage relative difference between the numerical and fitting function values of $\hat{C}_\Theta(\hat{r})$ and $\hat{D}_\Theta(\hat{r})$ integrals. All the four sub-plots use the same fitting functions listed in the main text. The numerical values use $\hat{r}$ growth factor and the merging parameter $\scal[large]{a}$ respectively equal to (a) $1.001$ and $-13$, (b) $1.0002$ and $-13$, (c) $1.001$ and $-12$, and (d) $1.001$ and $-15$.}
\label{fig:mobility-resistance-integrals-relative-error-fine-mesh-12-15}
\end{figure}

\begin{figure}
\centering
\includegraphics[width=0.6\textwidth]{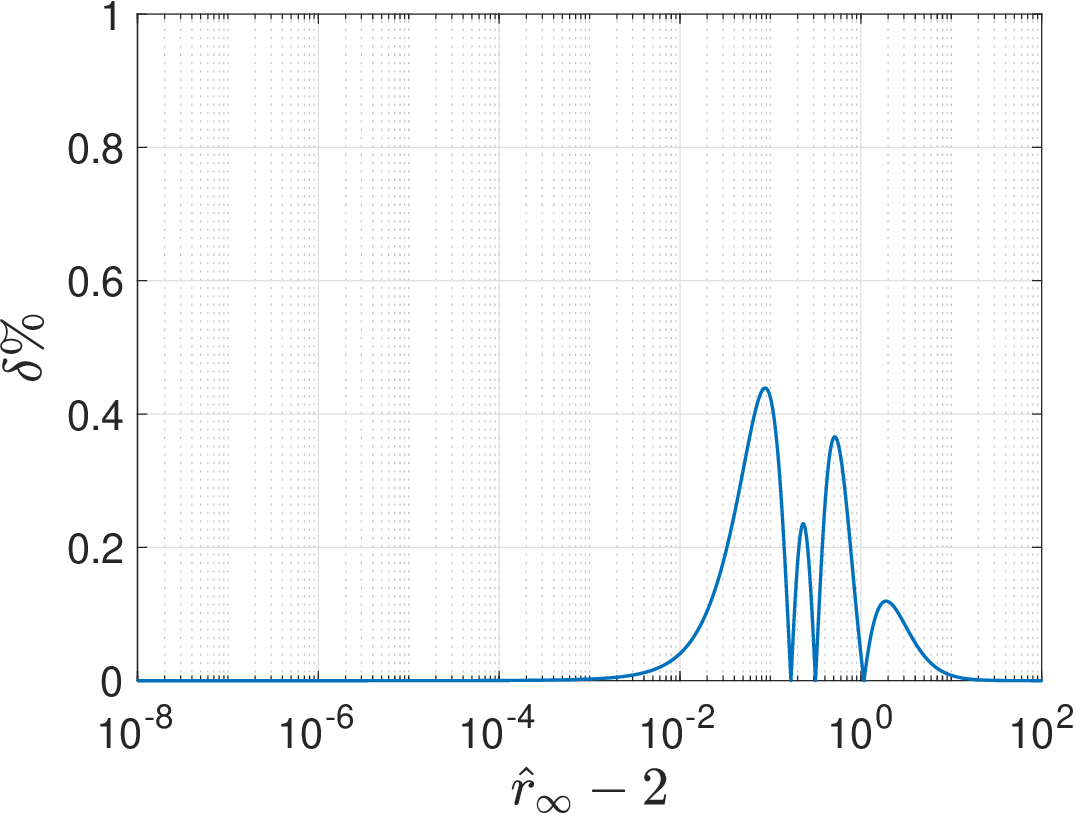}
\caption{Percentage relative difference $(\delta \%)$ between the numerical and fitting function values of $\scal[\infty]{\hat{r}}$.}
\label{fig:r_hat_infty_relative_error}
\end{figure}

\begin{figure}
\centering
\includegraphics[width=0.6\textwidth]{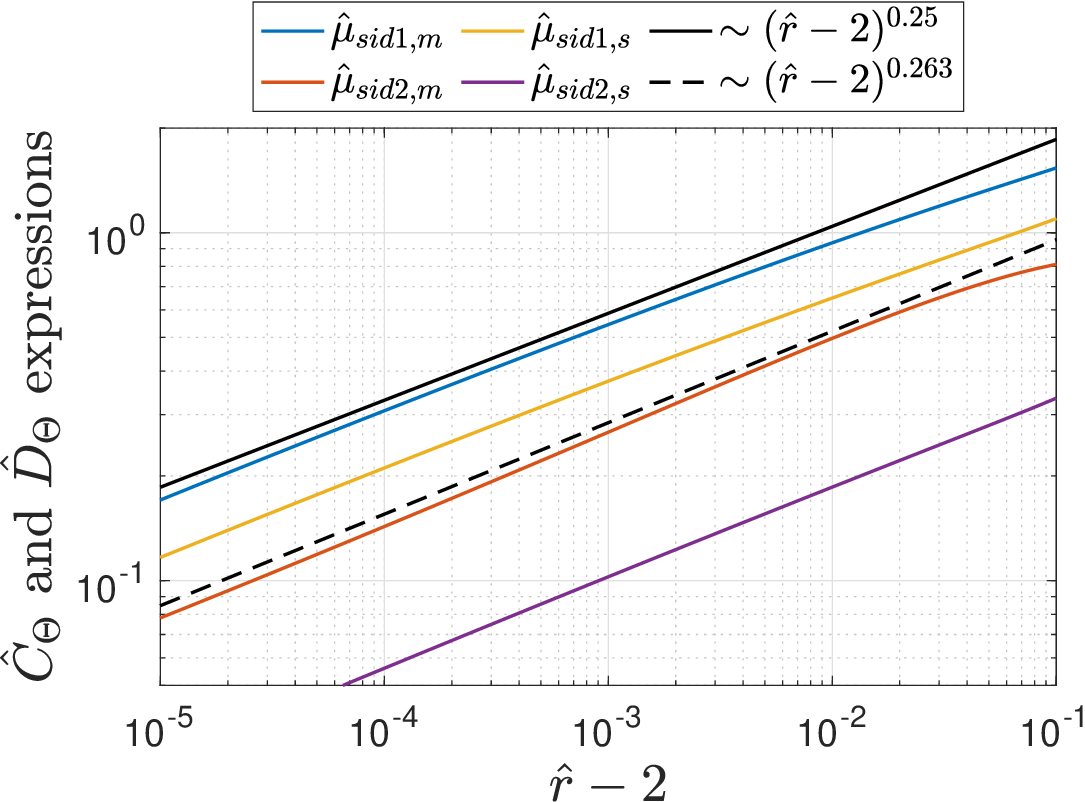}
\caption{The near-field variations of the integral expressions define the normal viscosity coefficients. The blue solid line shows the variation of $2\scal[L]{\hat{D}}+(4/7)\scal[M]{\hat{D}}$ expression (defining $\musidonem$), the red solid line shows $2\scal[S]{\hat{D}}-(4/3)\scal[L]{\hat{D}}-(8/21)\scal[M]{\hat{D}}$ expression (defining $\musidtwom$), the orange solid line shows $(9/280)(4\scal[A]{\hat{C}}+3\scal[B]{\hat{C}})$ expression (defining $\musidones$), and the purple solid line shows $(9/140)(\scal[A]{\hat{C}}-\scal[B]{\hat{C}})$ expression (defining $\musidtwos$).}
\label{fig:Chat-Dhat-integrals-summation-scaling-defining-interaction-viscosities}
\end{figure}

\begin{figure}
\centering
\includegraphics[width=0.6\textwidth]{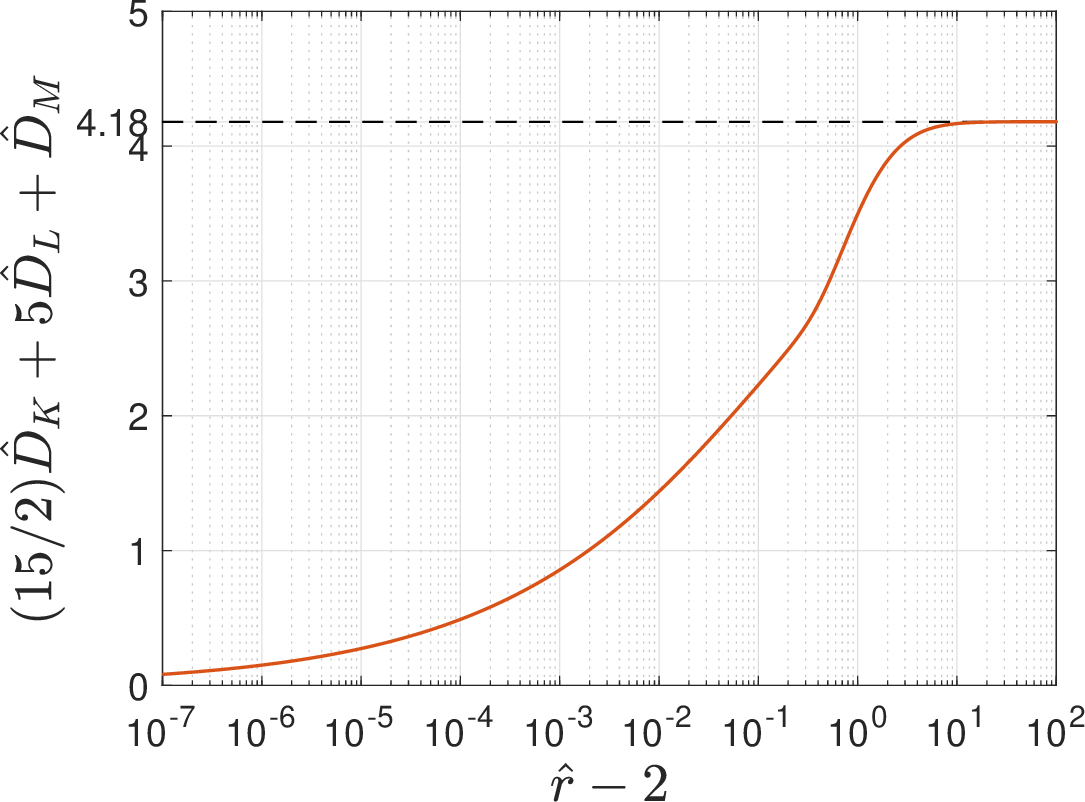}
\caption{The separation distance variation of the integral expression $(15/2)\scal[K]{\hat{D}}+5\scal[L]{\hat{D}}+\scal[M]{\hat{D}}$ that defines the mixture's interaction shear viscosity $\muintm$. The far-field limit of the expression equal to 4.18 is shown.}
\label{fig:Chat-Dhat-integrals-summation-scaling-defining-mixture-shear-viscosity}
\end{figure}

This section presents some additional figures related to the numerical analysis of the MF-roughness model interaction viscosities defined based on the $\scal[\Theta]{\hat{C}}$ and $\scal[\Theta]{\hat{D}}$ integrals. Referring to section 2.3 in the main text, we used a numerical list of $\hat{r}$ values with the specified minimum value and growth factor to calculate the numerical values of the integrals. Also, we merged the near- and far- field values of the different scalars defining these integrals as $\xi=(1-\scal[\xi]{\omega})\scal[small]{\xi}+\scal[\xi]{\omega}\scal[large]{\xi}$. The merge function is defined as $\scal[\xi]{\omega}=\scal[small]{e}/\left(\scal[small]{e}+\scal[high]{e}\right)$ with $\scal[small]{e}(\hat{r})=\scal[small]{a}\left(\hat{r}-2\right)^{\scal[small]{n}}$ and $\scal[large]{e}(\hat{r})=\scal[large]{a}\hat{r}^{\scal[large]{n}}$ (Fig. \ref{fig:xi-scalars-interpolation-function-plot}). In the section 2.3, we referred to the numerical values of the integrals calculated by using $\scal[small]{a}=0.003$, $\scal[small]{a}=3/2$, $\scal[large]{a}=1$, and $\scal[large]{a}=-13$ and based on the numerical list of $\hat{r}$ values with a minimum of $10^{-12}$ and a growth factor of $1.001$. These numerical integral values are then found to fit well with the fitting functions in the table 5. Fig. \ref{fig:mobility-resistance-integrals-relative-error-fine-mesh-12-15} compares the same fitting functions with different sets of the numerical values obtained by changing the growth factor or $\scal[large]{n}$ values. The results indicate that using a growth factor that is five times smaller or a different transition function between the near- and far-field scalars has only a limited impact on the numerical values.\\
\\
This section includes three additional figures. Fig. \ref{fig:r_hat_infty_relative_error} shows the relative difference between the numerical and fitting function values of the screen radius $\scal[\infty]{\hat{r}}$. Fig. \ref{fig:Chat-Dhat-integrals-summation-scaling-defining-interaction-viscosities} shows the separation distance variation of the $\scal[\Theta]{\hat{C}}$ and $\scal[\Theta]{\hat{D}}$ integral expressions defining the solid and mixture phase normal viscosities. In the near field, these integral expressions scale approximately as $(\hat{r}-2)^{0.25}$, indicating that the phase-specific normal viscosities also scale with non-dimensional roughness values similarly. Fig. \ref{fig:Chat-Dhat-integrals-summation-scaling-defining-mixture-shear-viscosity} shows the separation distance variation of the integral expression $(15/2)\scal[K]{\hat{D}}+5\scal[L]{\hat{D}}+\scal[M]{\hat{D}}$ with a far-field limit of $4.1815$. In an isotropic microstructure of smooth particles in extensional flows, where the PDF can be approximated as $p(\hat{r},\theta,\varphi)=q(\hat{r})$, the interaction shear viscosity of the mixture is equal to this limit of the expression $\muintm=4.1815$.
\clearpage
\section{The sensitivity analysis of the MF models viscosities and results}\label{sensitivity-analysis-MF-models-viscosities-and-results}
\begin{figure}
\centering
\begin{subfigure}{1\textwidth}
\centering
\includegraphics[width=1\textwidth]{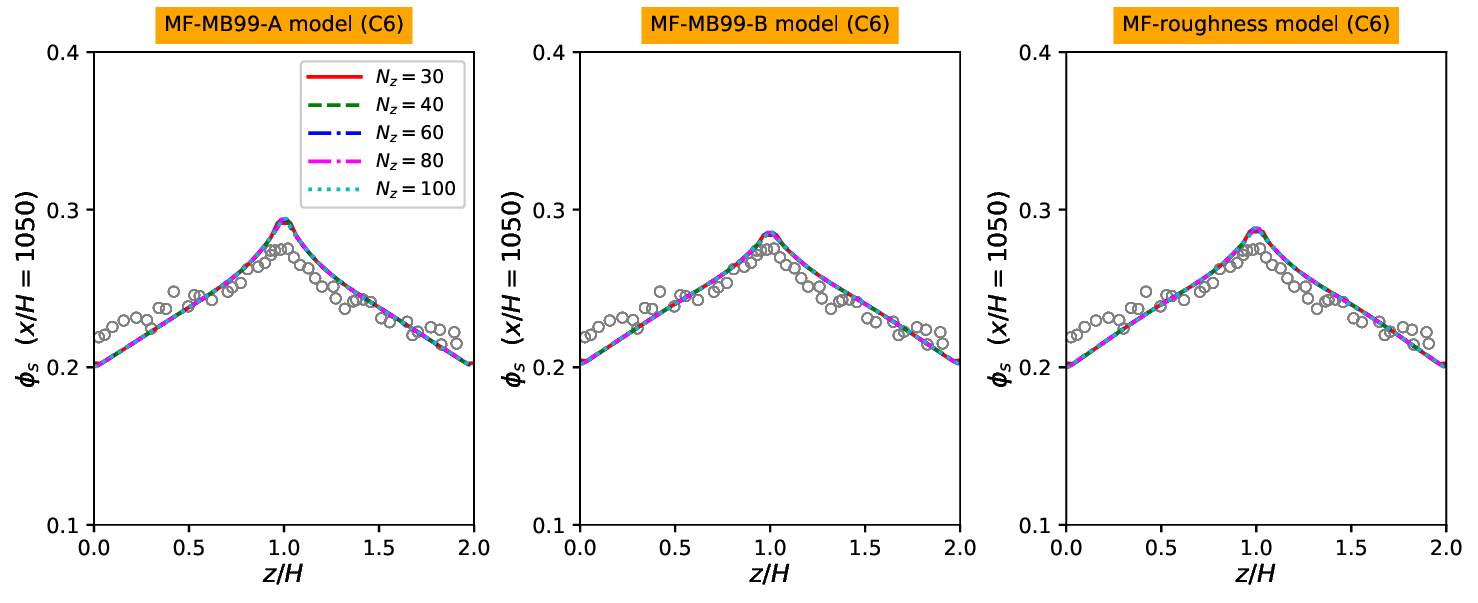}
\caption{}
\label{fig:Semwogerere_all_models_case6_mesh_dependency_with_p_osm_z_dir}
\end{subfigure}
\begin{subfigure}{1\textwidth}
\centering
\includegraphics[width=1\textwidth]{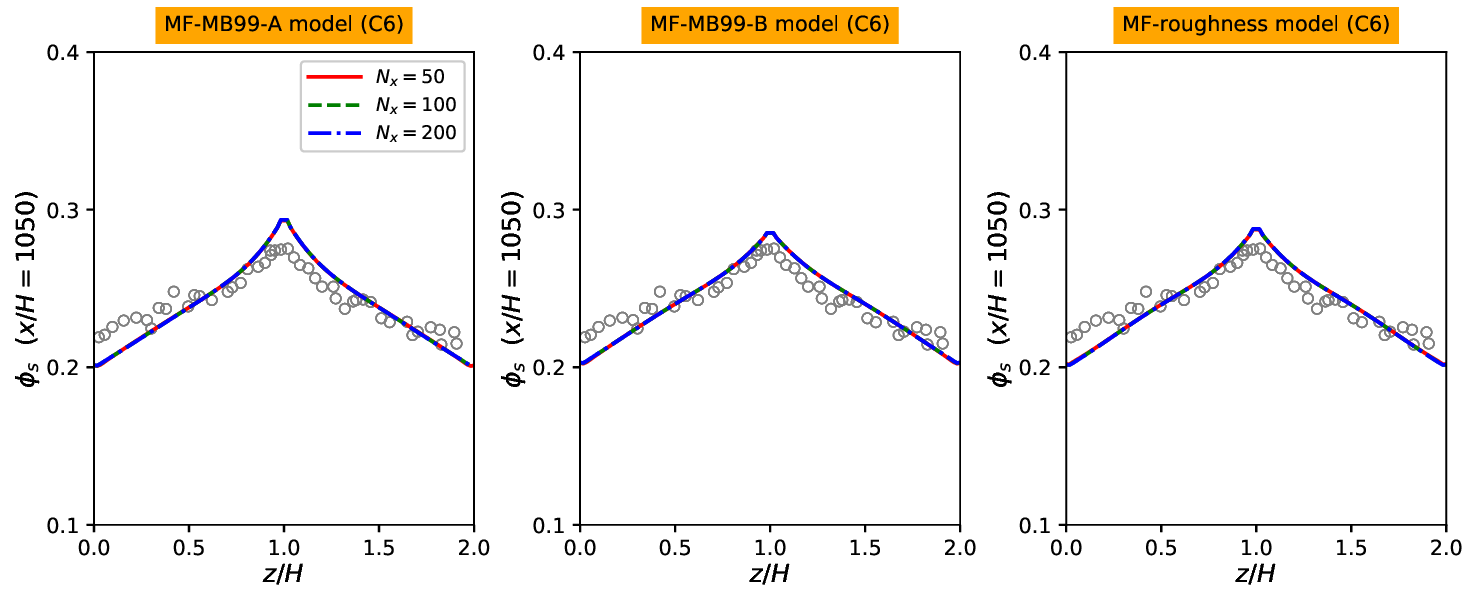}
\caption{}
\label{fig:Semwogerere_all_models_case6_mesh_dependency_with_p_osm_x_dir}
\end{subfigure}
\caption{The sensitivity of analysis of the three MF model results for the experimental case C6 on the number of mesh grids in the (a) vertical (z-axis) and (b) axial (x-axis) directions of the channel. The MF-roughness model results of the constraint type III are shown. Notably, the legends apply to all three columns in each row. The values $\scal[z]{N}$ and $\scal[x]{N}$ represent the number of cells along the z and x directions, respectively. The growth rates of the cell size along the z and x directions are fixed at $1$ and $1.02$, respectively.}
\label{fig:Semwogerere_all_models_case6_mesh_dependency_with_p_osm}
\end{figure}

\begin{figure}
\centering
\includegraphics[width=\textwidth]{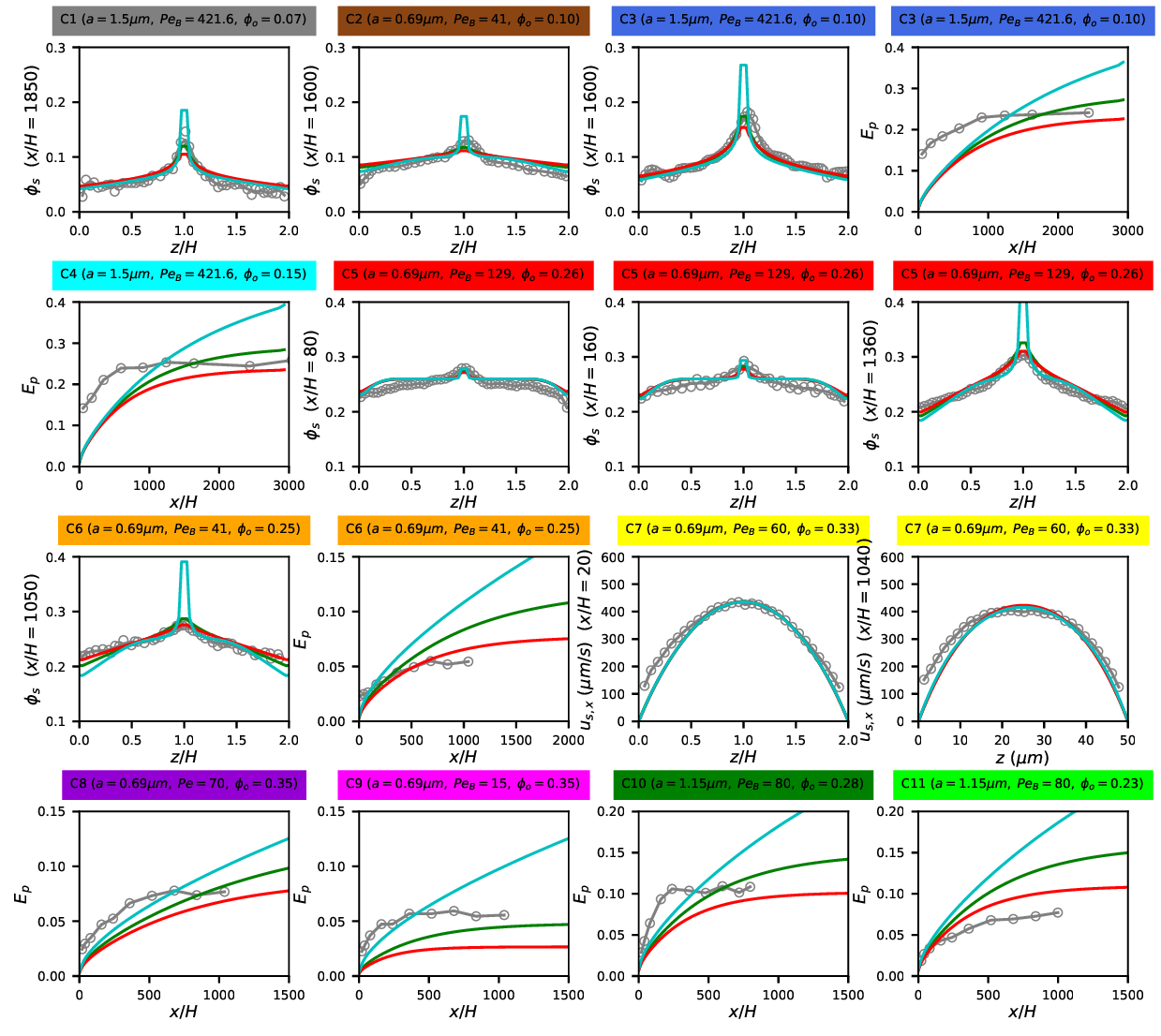}
\caption{The sensitivity of the MF-roughness model results (constraint type III) to variations in the osmotic pressure. The green solid line represents $\posm=\scal[CS]{\Pi}$, the red solid line represents $\posm=2\scal[CS]{\Pi}$, and the cyan line represents $\posm=0$.}
\label{fig:Semwogerere_roughness_sensitivity_on_p_osm}
\end{figure}

\begin{figure}
\centering
\includegraphics[width=0.95\textwidth]{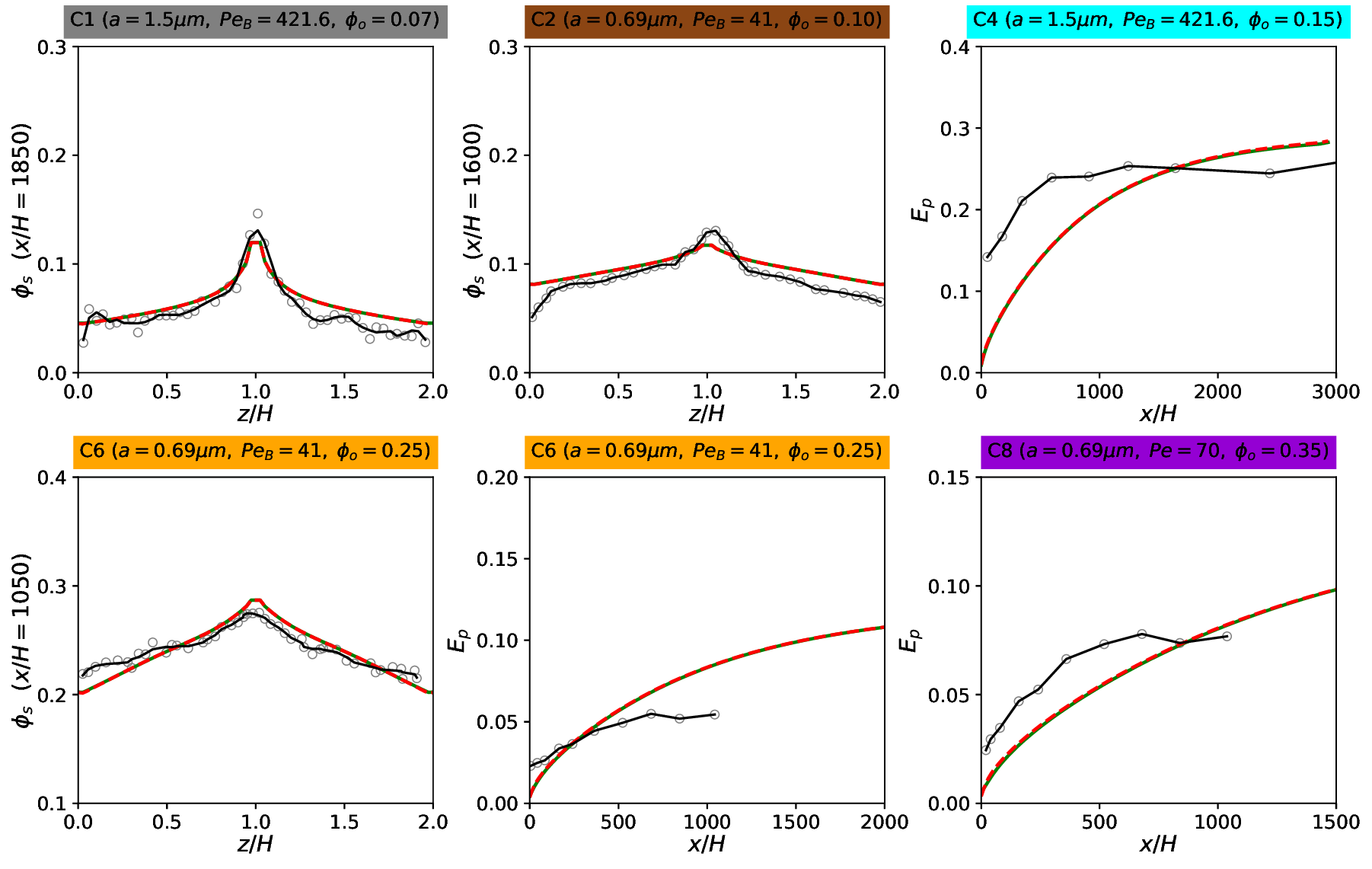}
\caption{The sensitivity of the MF-roughness model results (constraint type III) to the exclusion of the Faxen force. The green solid lines represent the results including this force, and the red dashed lines represent the results excluding this force.}
\label{fig:Semwogerere_roughness_sensitivity_on_Faxen}
\end{figure}

\begin{figure}
\centering
\includegraphics[width=0.95\textwidth]{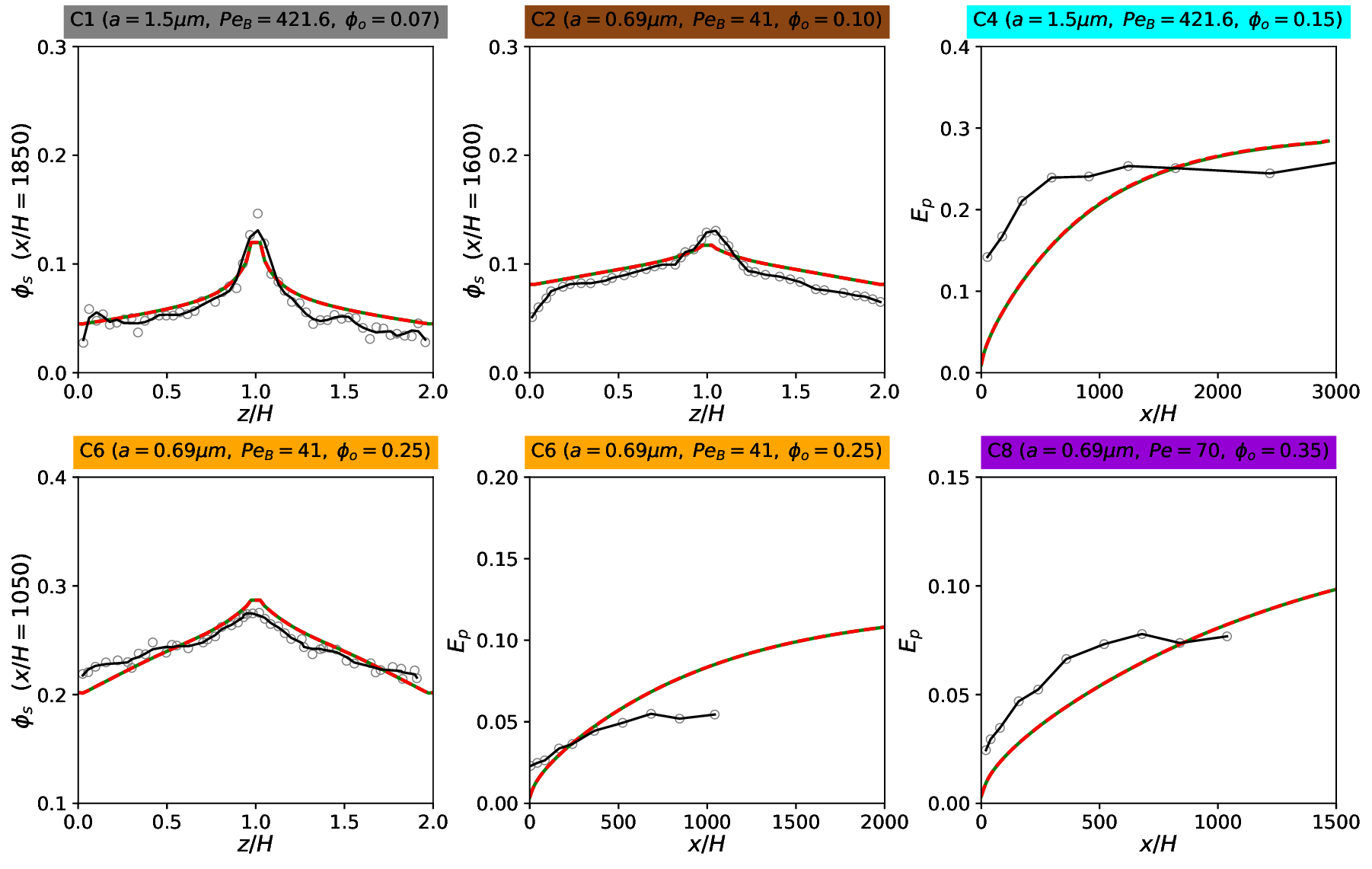}
\caption{The sensitivity of the MF-roughness model results (constraint type III) to the exclusion of the dilation viscosity terms $\mubulkones$ and $\mubulkonem$. The green solid lines represent the results including these terms, and the red dashed lines represent the results excluding these terms.}
\label{fig:Semwogerere_roughness_sensitivity_on_bulk1}
\end{figure}

\begin{figure}
\centering
\includegraphics[width=\textwidth]{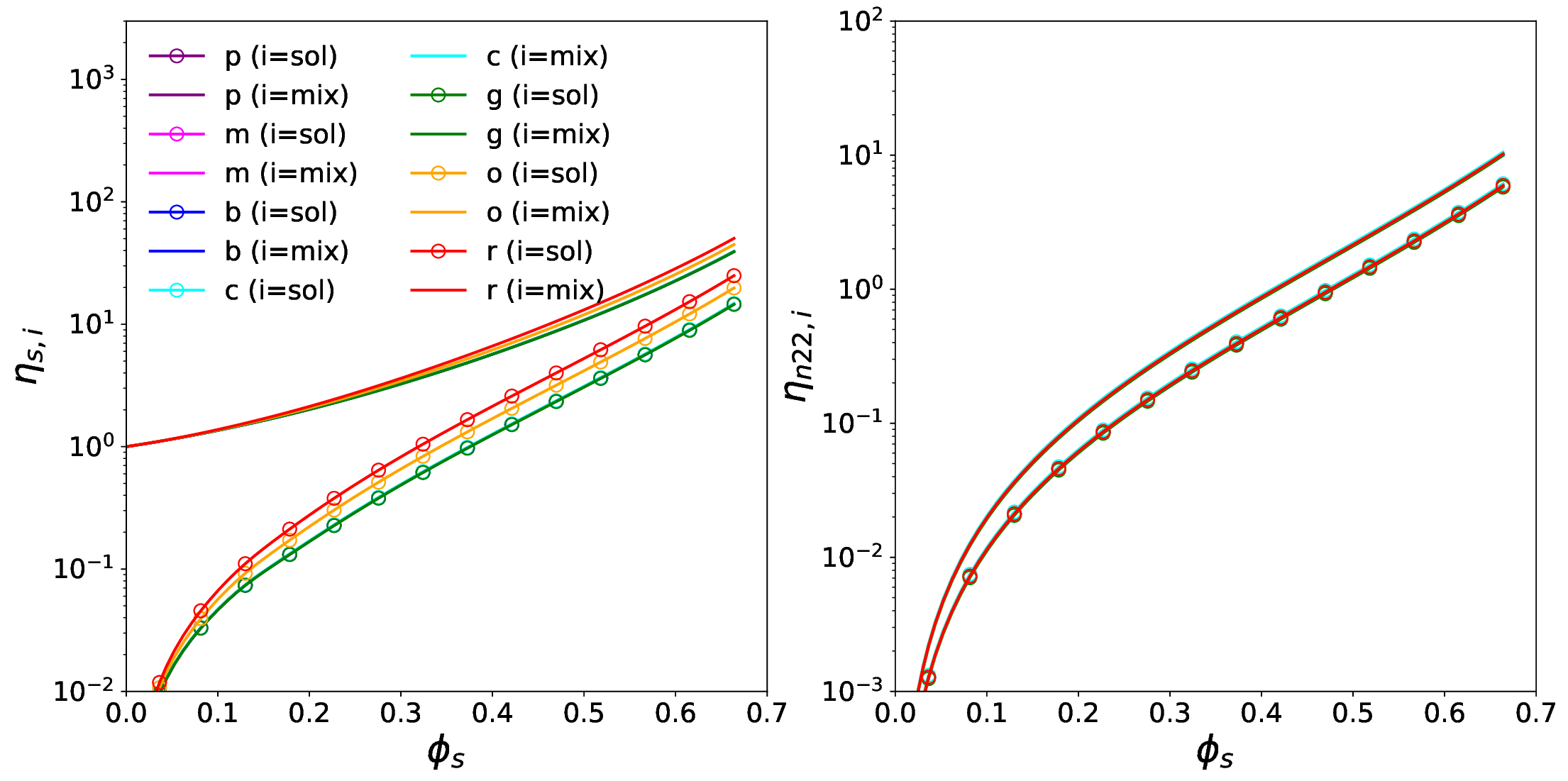}
\begin{tabular}{|cccccccc|}
\hline
 &p&  m&  b& c& g& o& r\\
 \hline
 $\scal[{2-\scal[\infty]{\hat{r}}}]{N}$ &1&0.5&10&1&1&1&1\\
$\scal[o]{\hat{\hat{r}}}\times 10^{5}$&4&4&4&3&5&4&4\\
$\scal[e]{f}$ &0&0&0&0&0&1&2\\
$\scal[c]{f}$ &1.565&1.565&1.565&1.565&1.565&2.565&3.565\\
\hline
\end{tabular}
\caption{The impact of the variation of the MF-roughness parameters on the phase-specific shear and normal viscosities. The legends on the left subplot also apply to the right subplot. The table summarises the closure parameter values for each legend colour code. The viscosity graphs use a fixed dimensional roughness value of $\scal[r]{\epsilon}=3.09\;\mbox{nm}$ and a fixed particle radius of $a=0.69\;\mu\mbox{m}$. Notably, the viscosity graphs in purple, magenta, blue, cyan, and green colours show insignificant differences, with only the green graphs visible in the figure. This indicates the minimal impact of varying $\scal[{2-\scal[\infty]{\hat{r}}}]{N}$ and $\scal[o]{\hat{\hat{r}}}$ in the model. The small variation between the green, orange, and red graphs shows how the model viscosities change with $\scal[c]{f}+\scal[e]{f}$ at a fixed $\scal[c]{f}-\scal[e]{f}=1.565$ value.}
\label{fig:Semwogerere_roughness_sensitivity_on_N_ro_fe_viscosities}
\end{figure}

\begin{figure}
\centering
\includegraphics[width=0.8\textwidth]{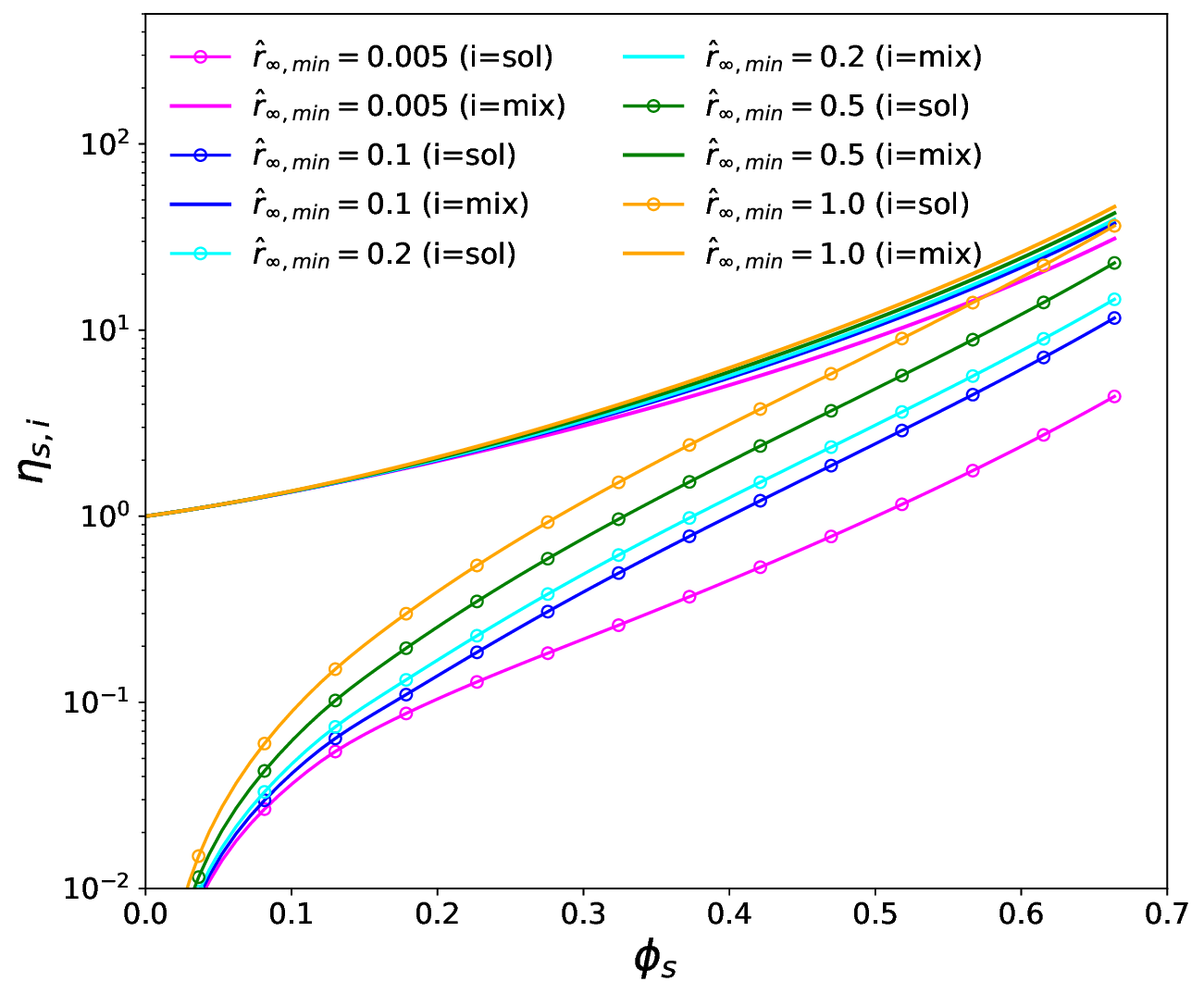}
\caption{The sensitivity of the phase-specific shear viscosities in the MF-roughness model (constraint type III) to the variation of $\scal[\infty,min]{\hat{r}}$ values. Notably, the normal viscosities of the model are independent of this parameter. For the simulations and the viscosity plots in the current study, we have chosen to implement $\scal[\infty]{\hat{r}}+\scal[\infty,min]{\hat{r}}$ with $\scal[\infty,min]{\hat{r}}=0.2$ as the outer radius of the symmetric region of the PDF model instead of using $\scal[\infty]{\hat{r}}$.}
\label{fig:Semwogerere_roughness_sensitivity_on_rhat_infinity_min}
\end{figure}

\begin{figure}
\centering
\begin{subfigure}{1\textwidth}
\centering
\includegraphics[width=1\textwidth]{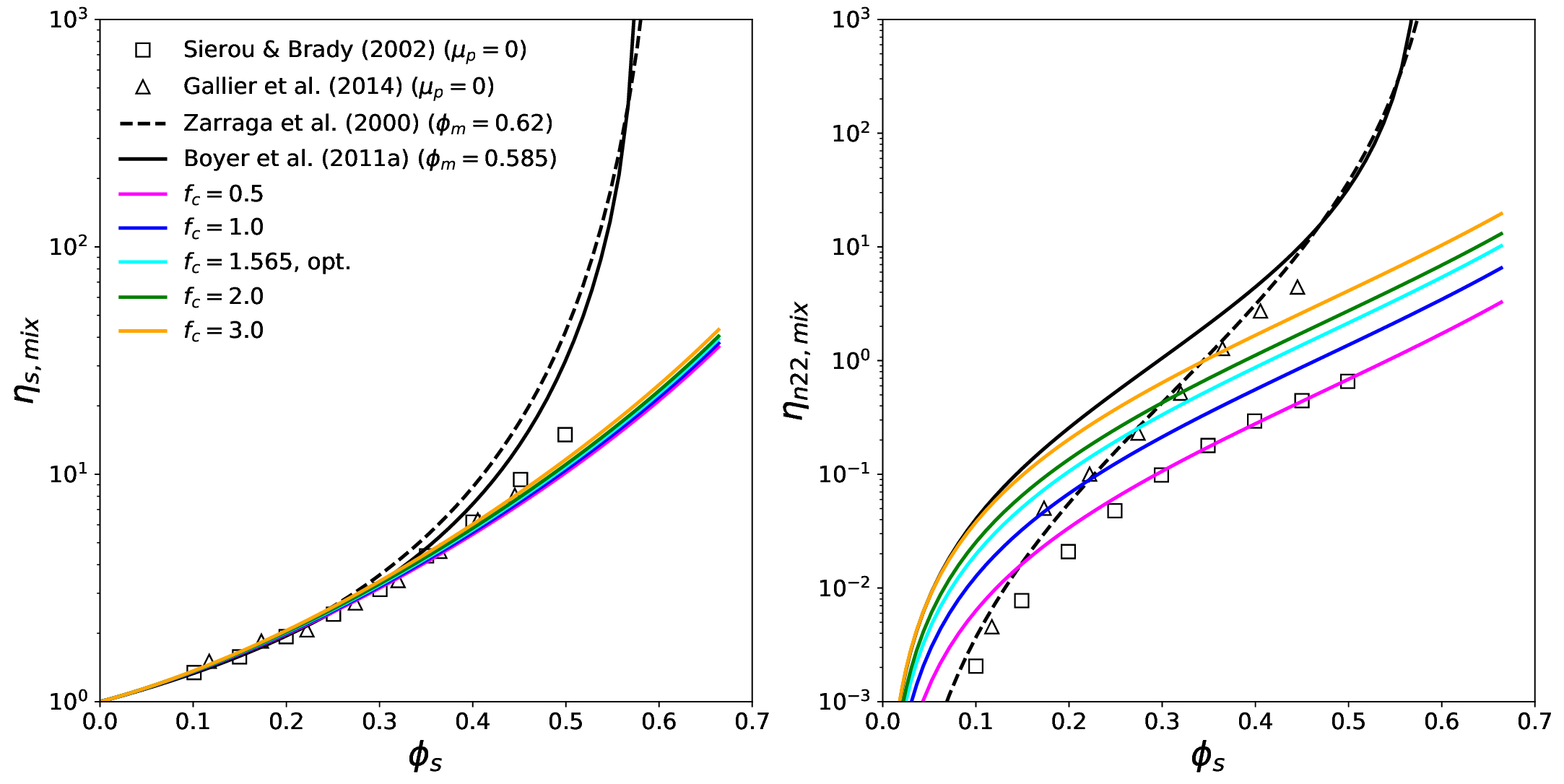}
\caption{}
\label{fig:Semwogerere_roughness_sensitivity_on_f_c_viscosities}
\end{subfigure}
\begin{subfigure}{1\textwidth}
\centering
\includegraphics[width=1\textwidth]{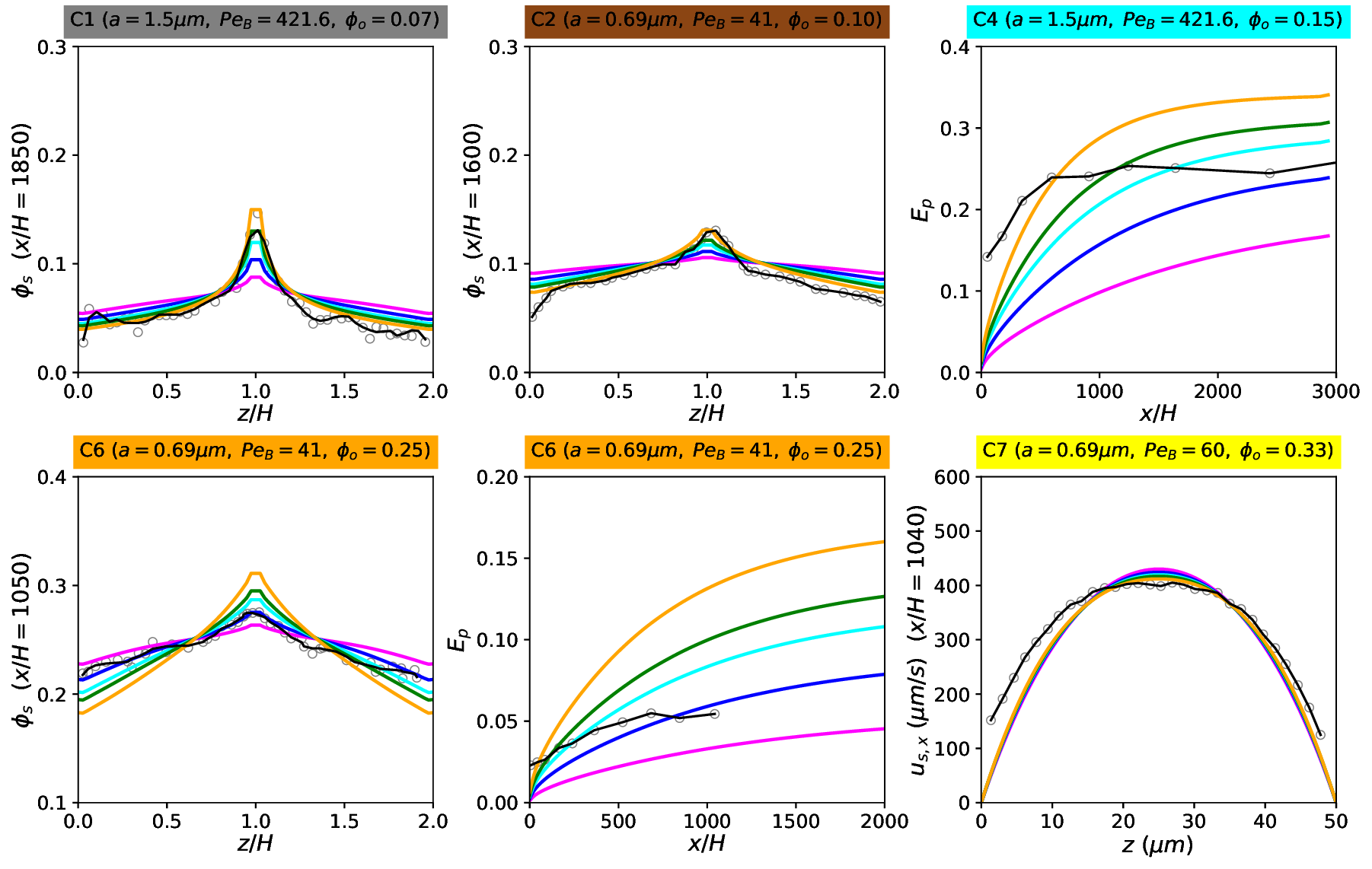}
\caption{}
\label{fig:Semwogerere_roughness_sensitivity_on_f_c_phi}
\end{subfigure}
\caption{The sensitivity of the MF-roughness (constraint type III) viscosities of the small particle (a) and simulation results (b) to the variation of $\scal{f}{c}$. The legends on the shear viscosity graphs apply to all the sub-figures in parts (a) and (b).}
\label{fig:Semwogerere_roughness_sensitivity_on_f_c_phi_viscosities}
\end{figure}

\begin{figure}
\centering
\begin{subfigure}{1\textwidth}
\centering
\includegraphics[width=0.95\textwidth]{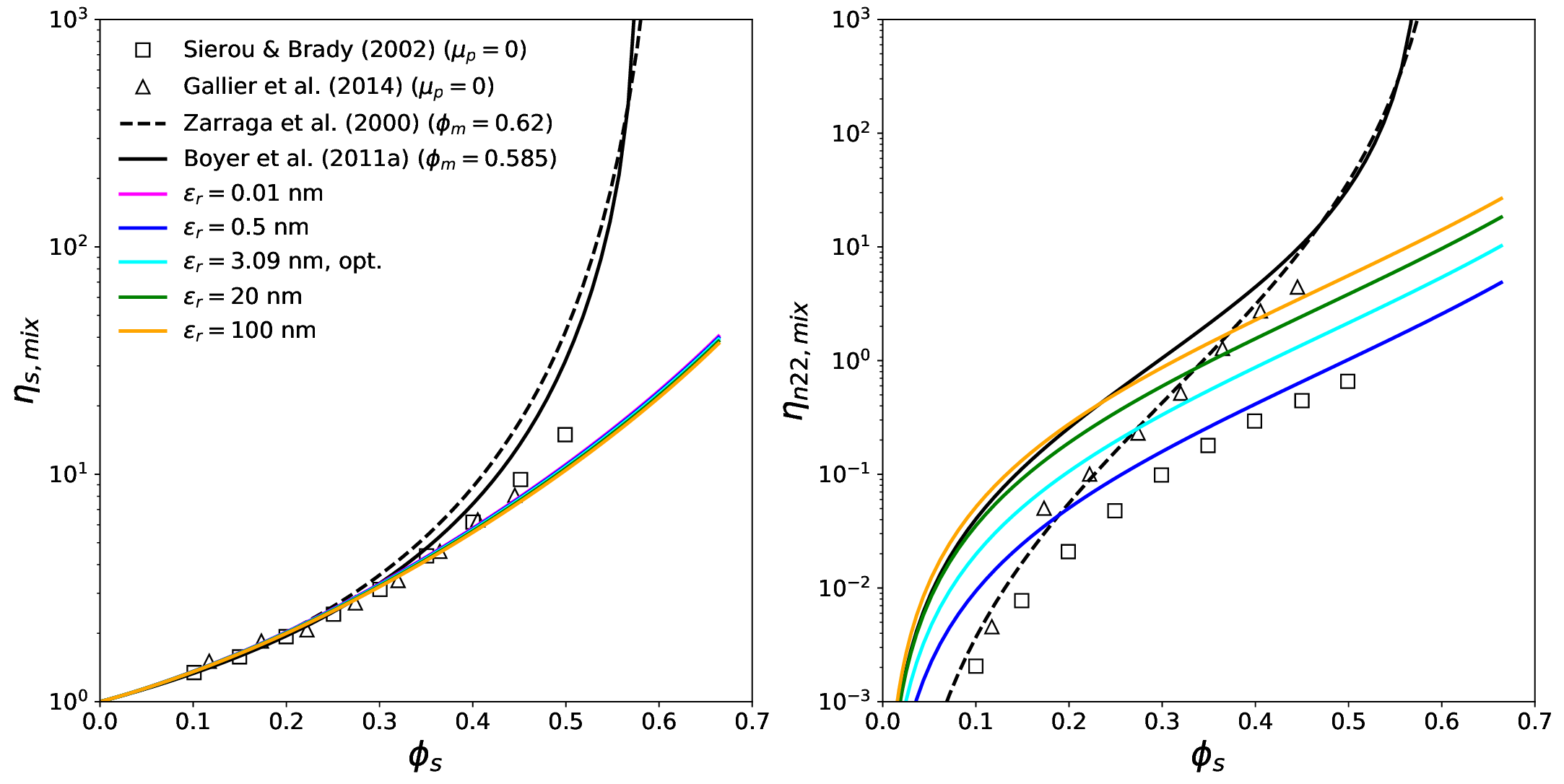}
\caption{}
\label{fig:Semwogerere_roughness_sensitivity_on_eps_r_viscosities}
\end{subfigure}
\begin{subfigure}{1\textwidth}
\centering
\includegraphics[width=1\textwidth]{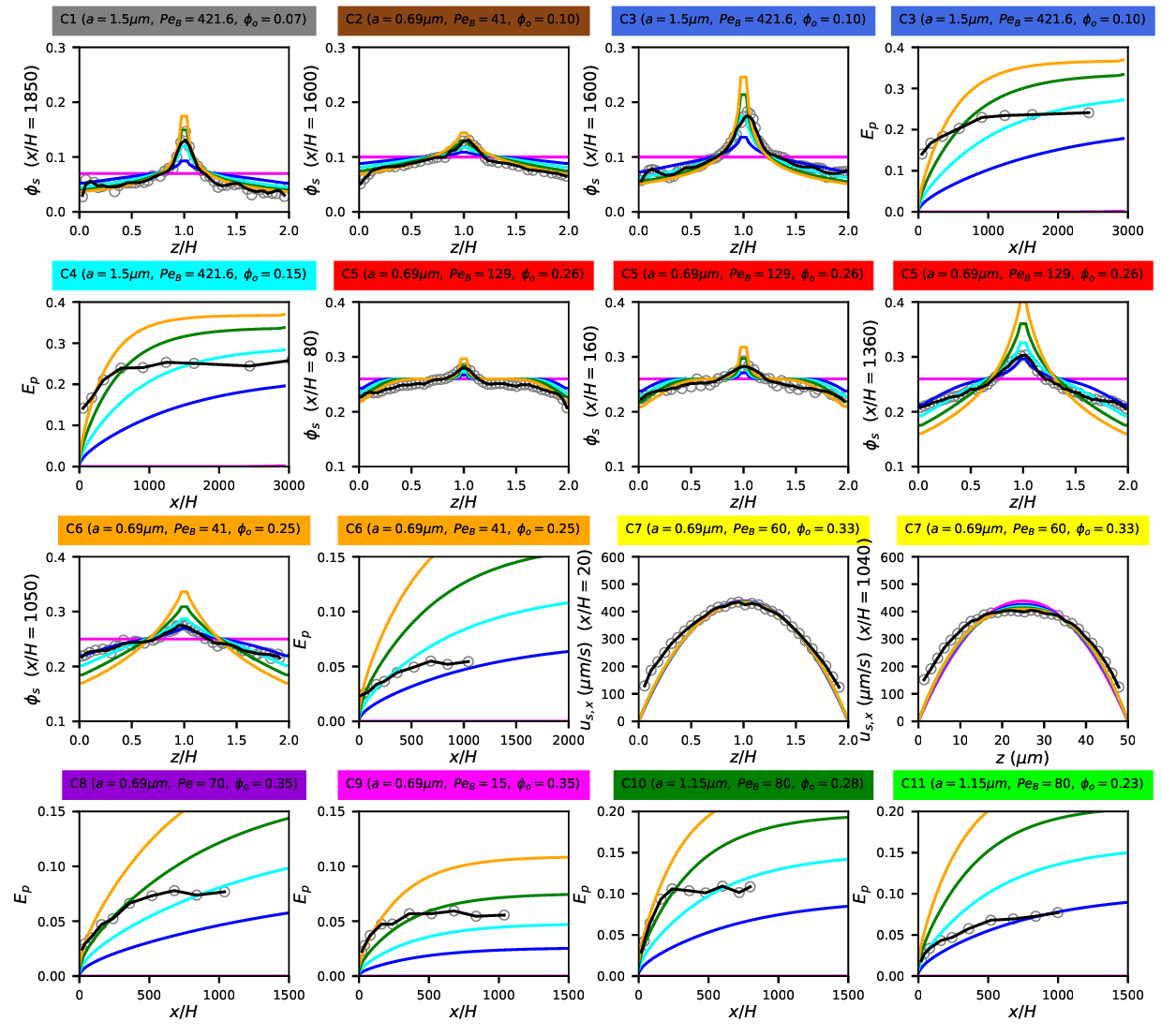}
\caption{}
\label{fig:Semwogerere_roughness_sensitivity_on_eps_r_phi}
\end{subfigure}
\caption{The sensitivity of the MF-roughness (constraint type III) viscosities of the small particle (a) and simulation results (b) to the variation of dimensional roughness value $\scal[r]{\epsilon}\; (\mbox{nm})$. The legends on the shear viscosity graphs apply to all the sub-figures in parts (a) and (b).}
\label{fig:Semwogerere_roughness_sensitivity_on_eps_r_phi_viscosities}
\end{figure}

\begin{figure}
\centering
\begin{subfigure}{1\textwidth}
\centering
\includegraphics[width=0.95\textwidth]{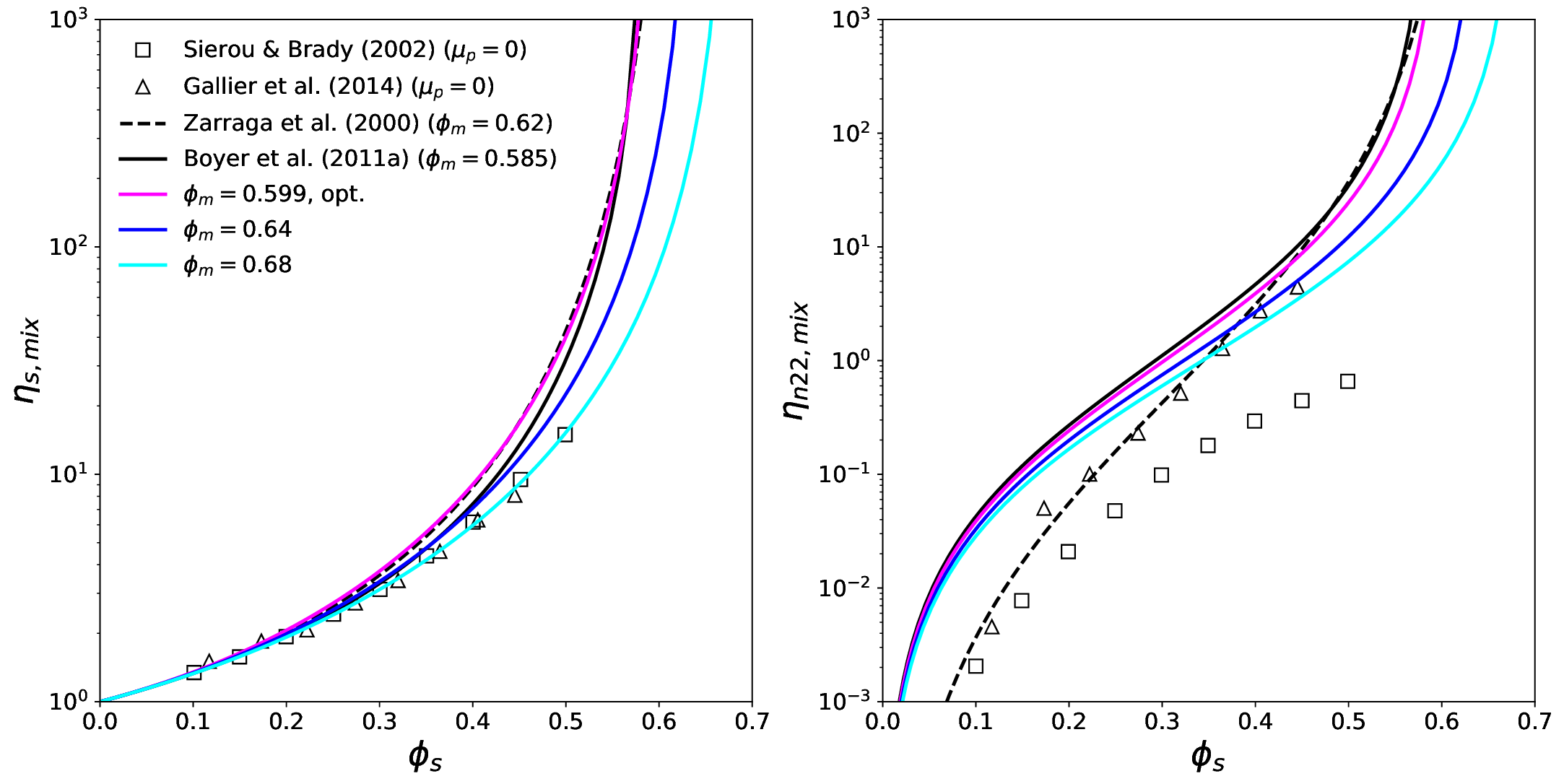}
\caption{}
\label{fig:Semwogerere_MB99B_sensitivity_on_phim_viscosities}
\end{subfigure}
\begin{subfigure}{1\textwidth}
\centering
\includegraphics[width=1\textwidth]{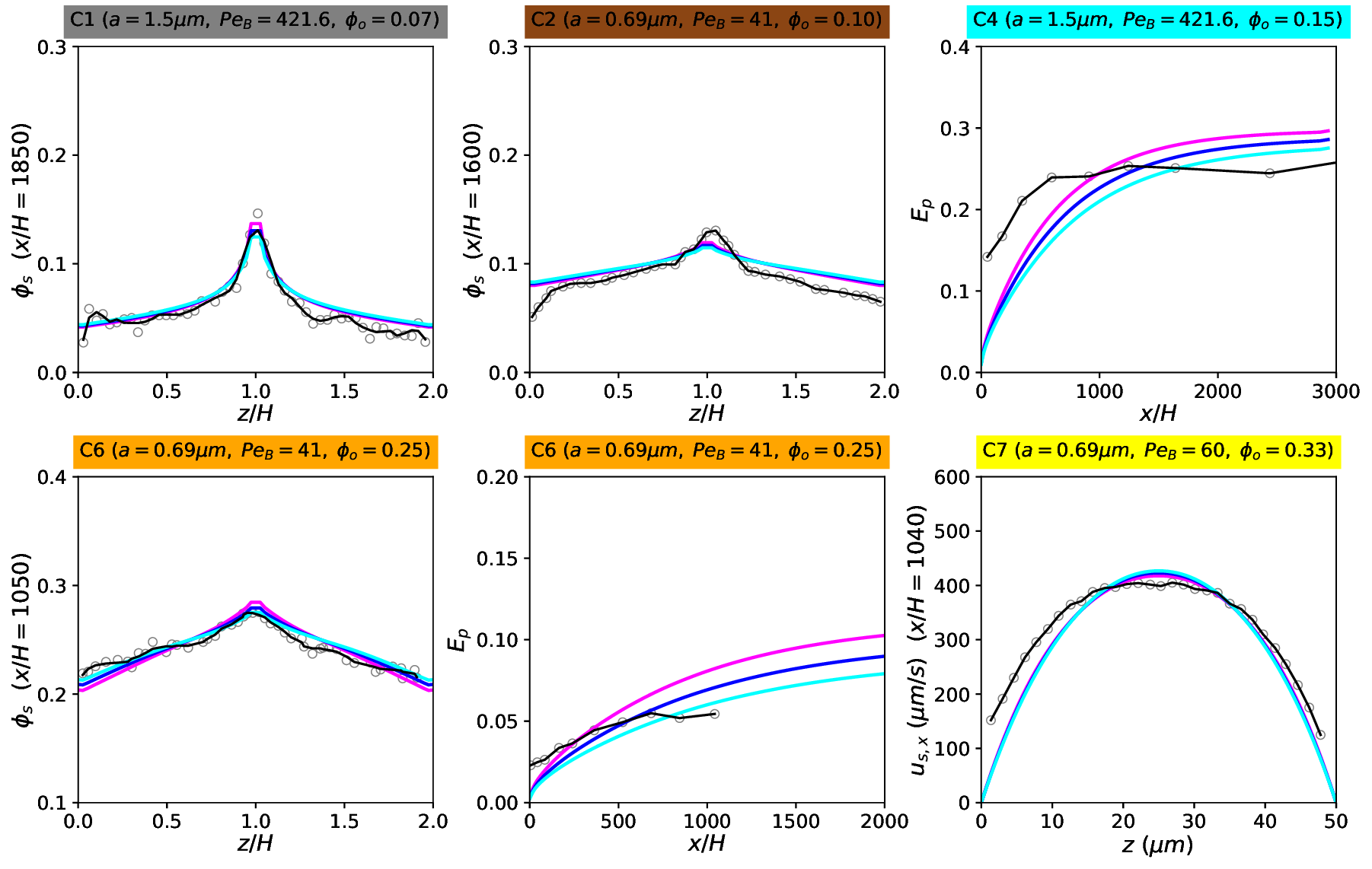}
\caption{}
\label{fig:Semwogerere_MB99B_sensitivity_on_phim_phi}
\end{subfigure}
\caption{The sensitivity of the MF-MB99-B model viscosities (a) and simulation results (b) to the variation of $\phim$ value. The legends on the shear viscosity graphs apply to all the sub-figures in parts (a) and (b). }
\label{fig:Semwogerere_MB99B_sensitivity_on_phim_phi_viscosities}
\end{figure}

\begin{figure}
\centering
\begin{subfigure}{1\textwidth}
\centering
\includegraphics[width=1\textwidth]{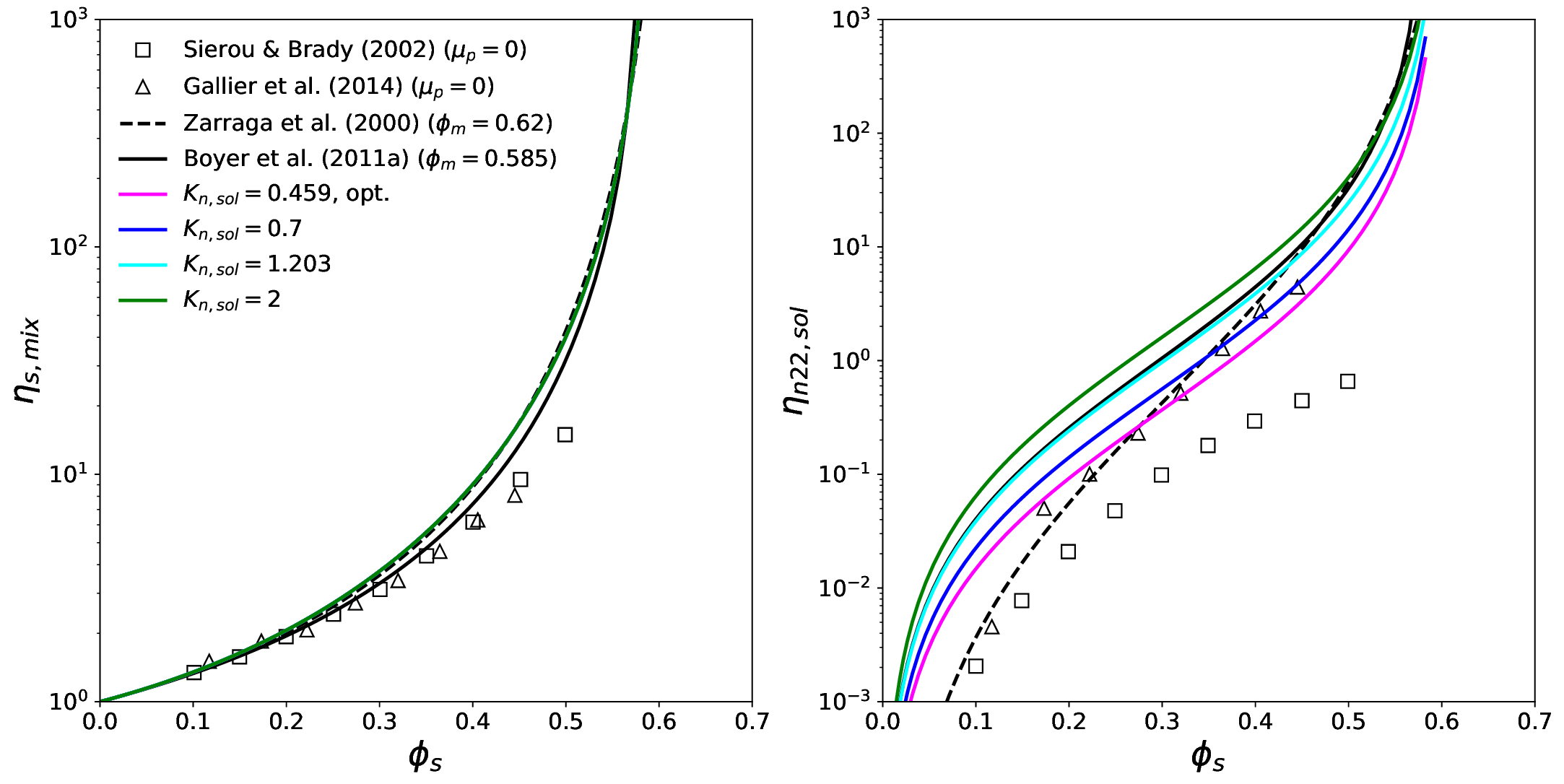}
\caption{}
\label{fig:Semwogerere_MB99B_sensitivity_on_Kn_sol_viscosities}
\end{subfigure}
\begin{subfigure}{1\textwidth}
\centering
\includegraphics[width=1\textwidth]{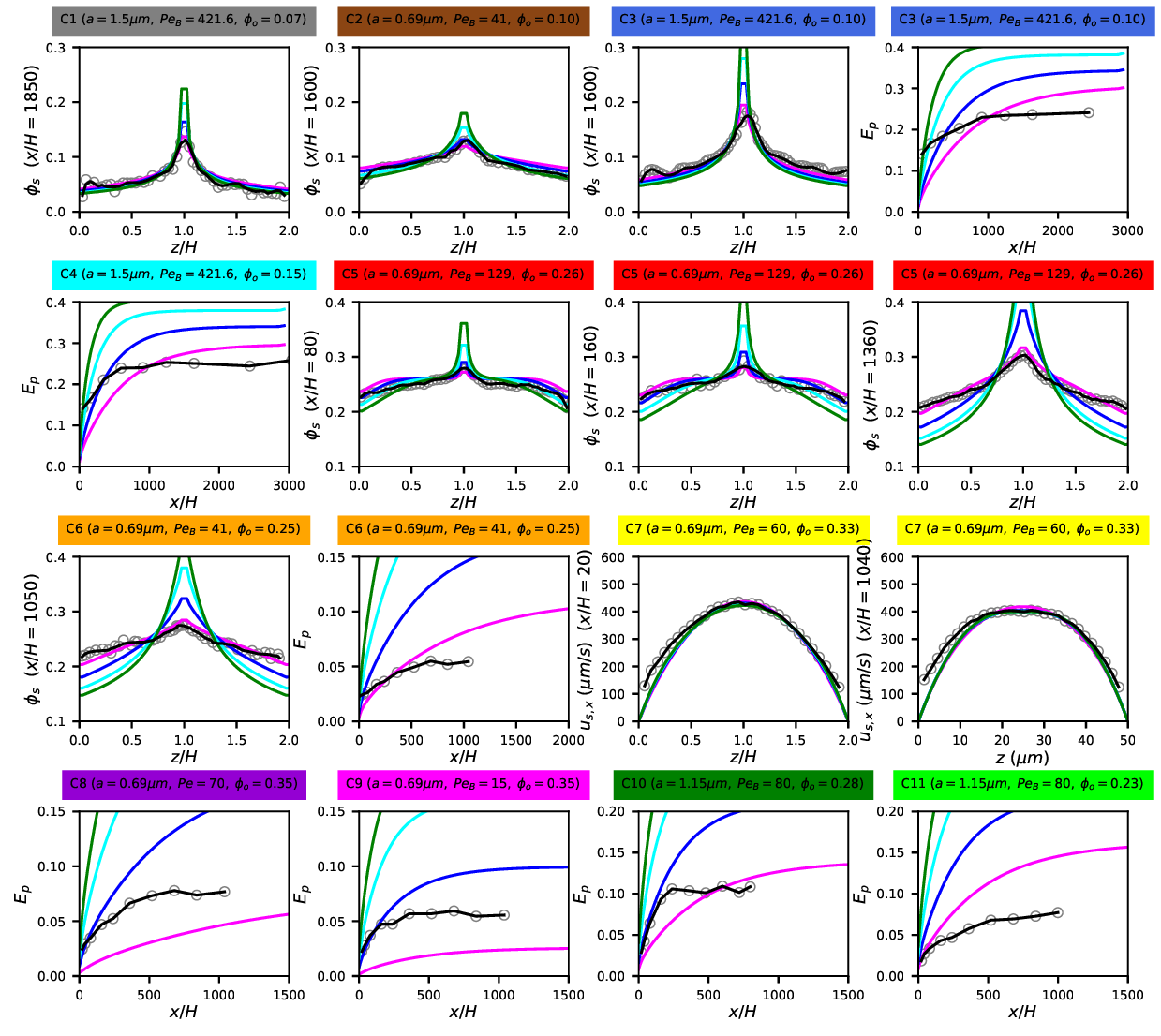}
\caption{}
\label{fig:Semwogerere_MB99B_sensitivity_on_Kn_sol_phi}
\end{subfigure}
\caption{The sensitivity of the MF-MB99-B model viscosities (a) and simulation results (b) to the variation of $\scal[n,sol]{K}$ value. The legends on the shear viscosity graphs apply to all the sub-figures in parts (a) and (b). }
\label{fig:Semwogerere_MB99B_sensitivity_on_Kn_sol_phi_viscosities}
\end{figure}

\begin{figure}
\centering
\begin{subfigure}{1\textwidth}
\centering
\includegraphics[width=1\textwidth]{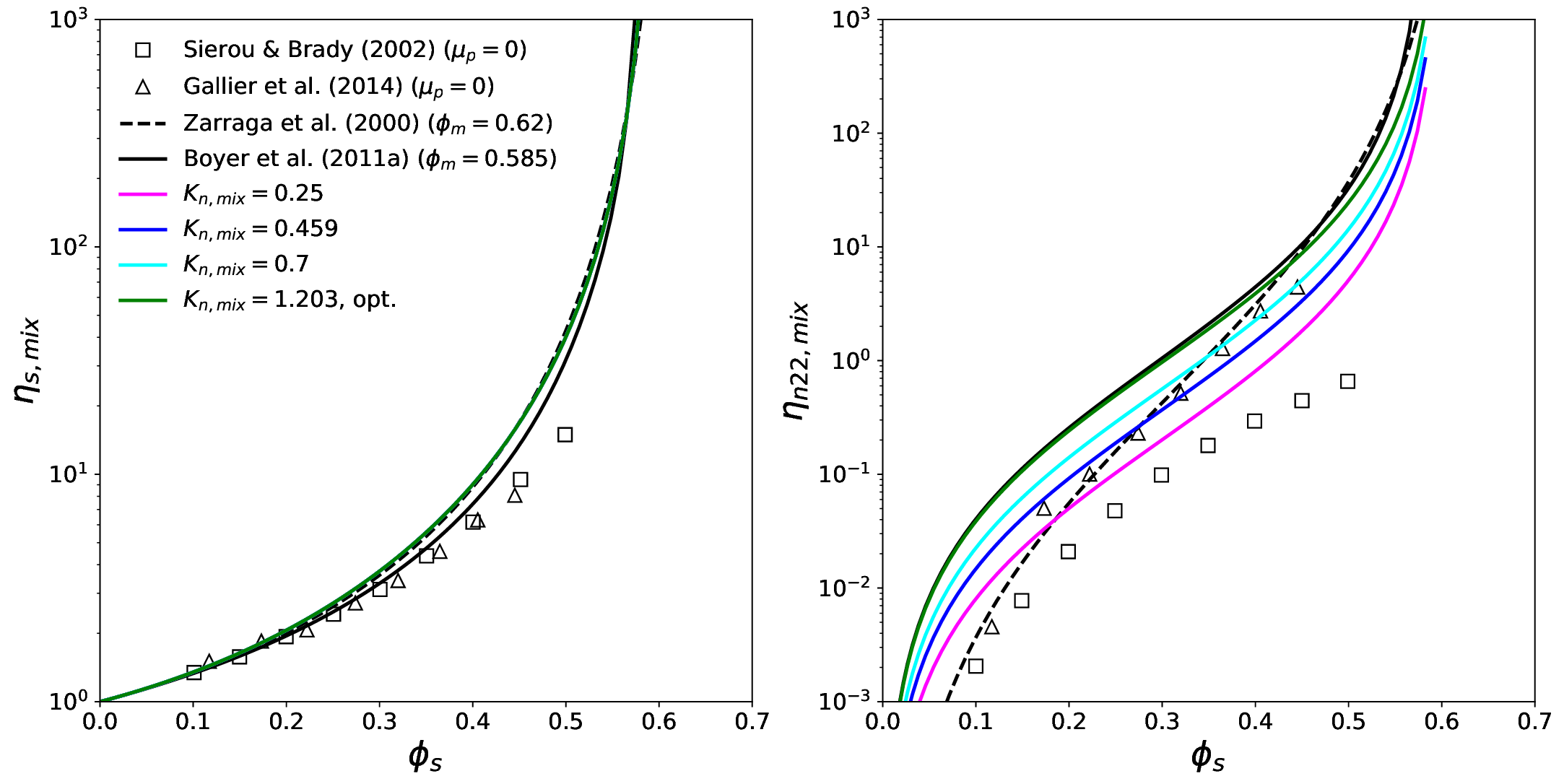}
\caption{}
\label{fig:Semwogerere_MB99B_sensitivity_on_Kn_mix_viscosities}
\end{subfigure}
\begin{subfigure}{1\textwidth}
\centering
\includegraphics[width=1\textwidth]{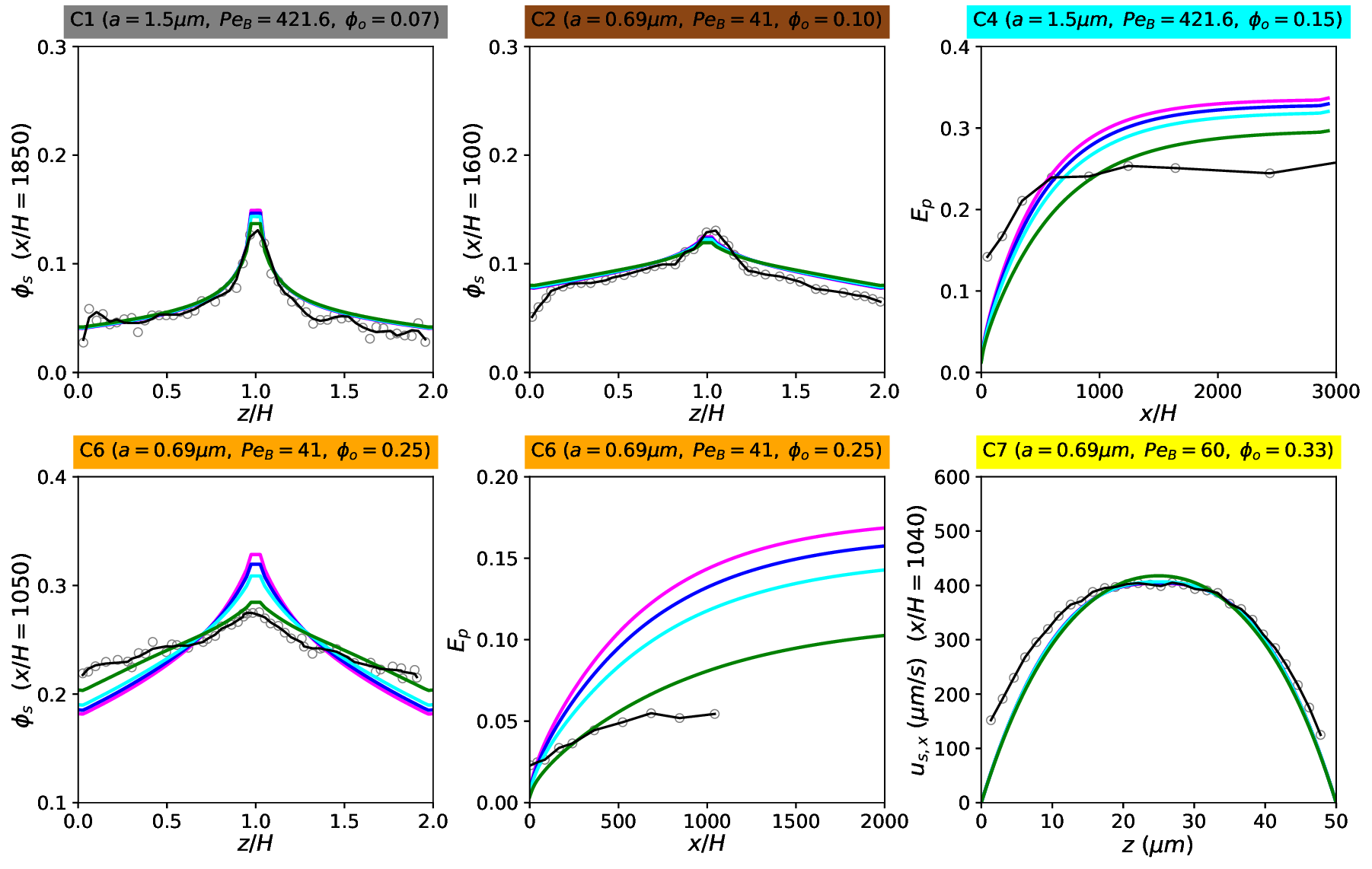}
\caption{}
\label{fig:Semwogerere_MB99B_sensitivity_on_Kn_mix_phi}
\end{subfigure}
\caption{The sensitivity of the MF-MB99-B model viscosities (a) and simulation results (b) to the variation of $\scal[n,mix]{K}$ values. The legends on the shear viscosity graphs apply to all the sub-figures in parts (a) and (b). }
\label{fig:Semwogerere_MB99B_sensitivity_on_Kn_mix_phi_viscosities}
\end{figure}

\begin{figure}
\centering
\begin{subfigure}{1\textwidth}
\centering
\includegraphics[width=1\textwidth]{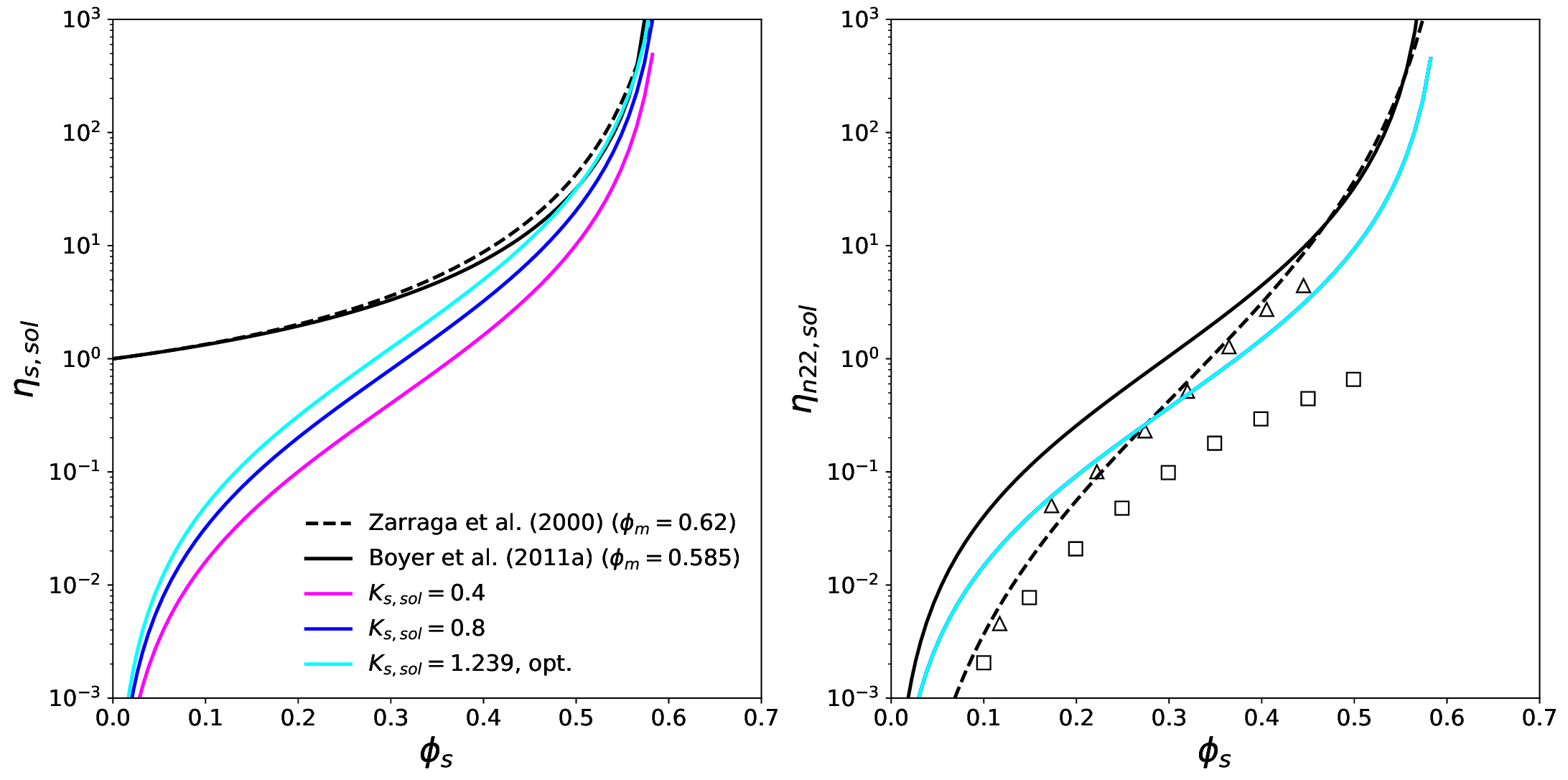}
\caption{}
\label{fig:Semwogerere_MB99B_sensitivity_on_Ks_sol_viscosities}
\end{subfigure}
\begin{subfigure}{1\textwidth}
\centering
\includegraphics[width=1\textwidth]{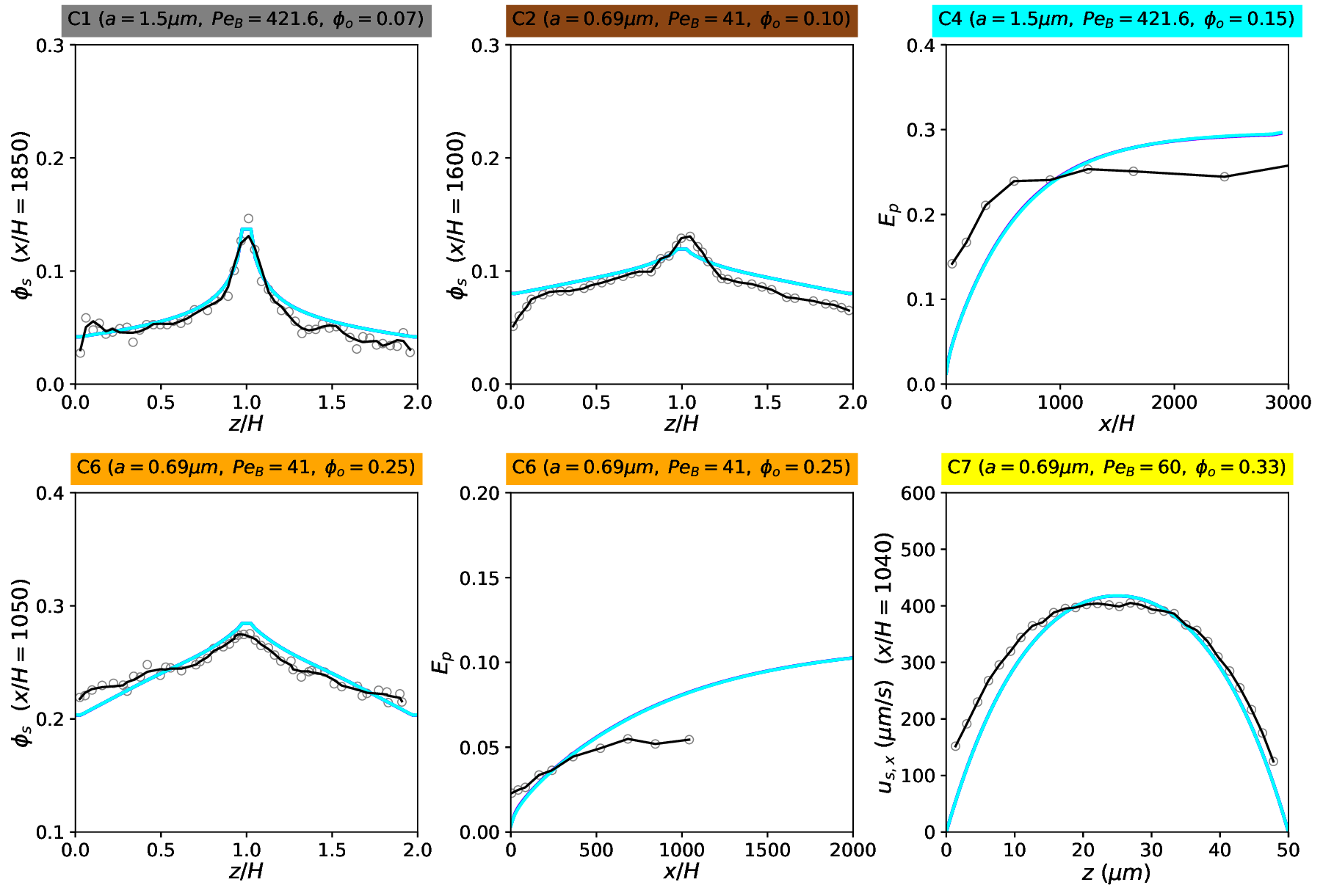}
\caption{}
\label{fig:Semwogerere_MB99B_sensitivity_on_Ks_sol_phi}
\end{subfigure}
\caption{The sensitivity of the MF-MB99-B model viscosities (a) and simulation results (b) to the variation of $\scal[s,sol]{K}$ values. The legends on the shear viscosity graphs apply to all the sub-figures in parts (a) and (b). }
\label{fig:Semwogerere_MB99B_sensitivity_on_Ks_sol_phi_viscosities}
\end{figure}

\begin{figure}
\centering
\begin{subfigure}{1\textwidth}
\centering
\includegraphics[width=1\textwidth]{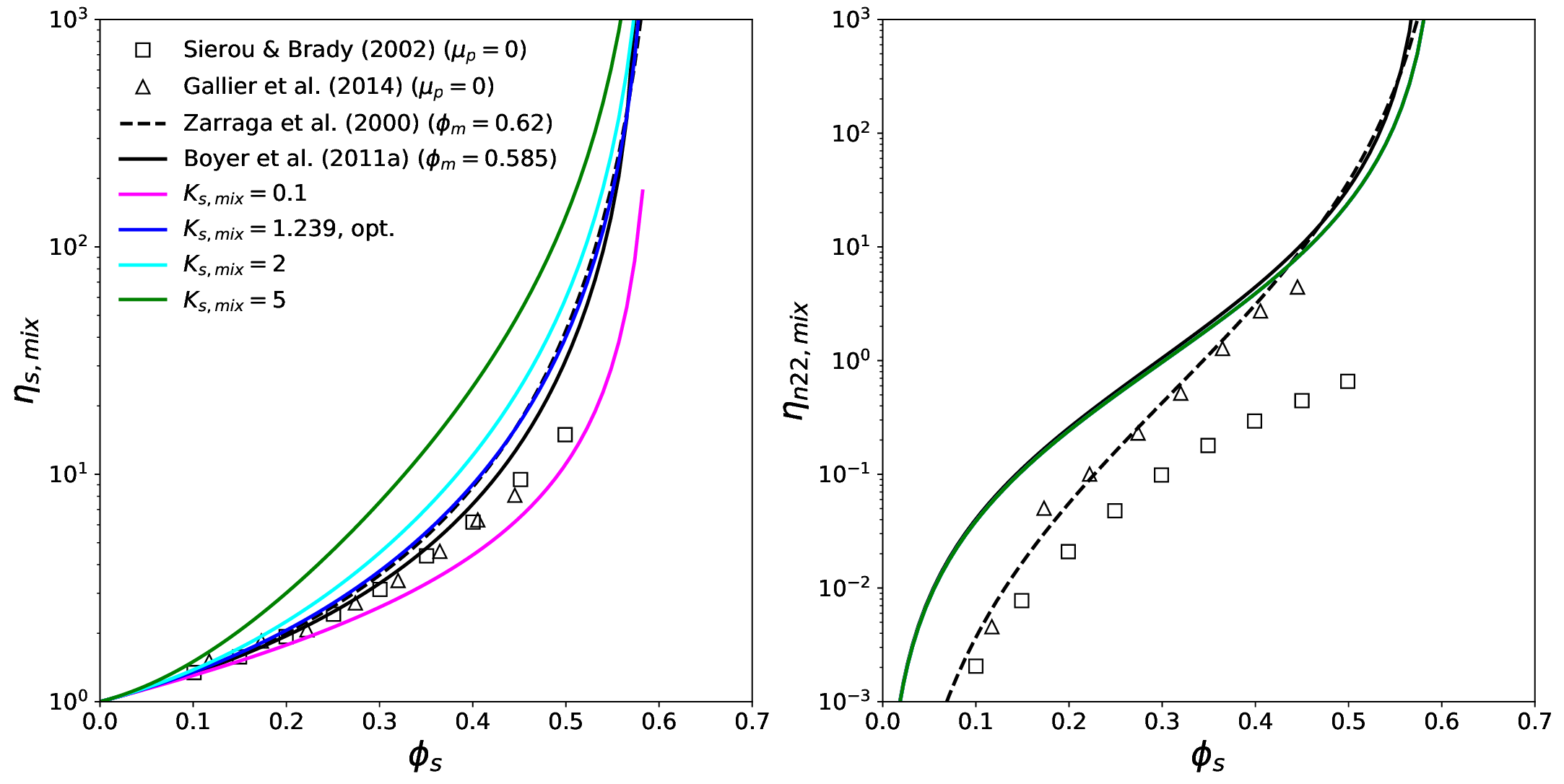}
\caption{}
\label{fig:Semwogerere_MB99B_sensitivity_on_Ks_mix_viscosities}
\end{subfigure}
\begin{subfigure}{1\textwidth}
\centering
\includegraphics[width=1\textwidth]{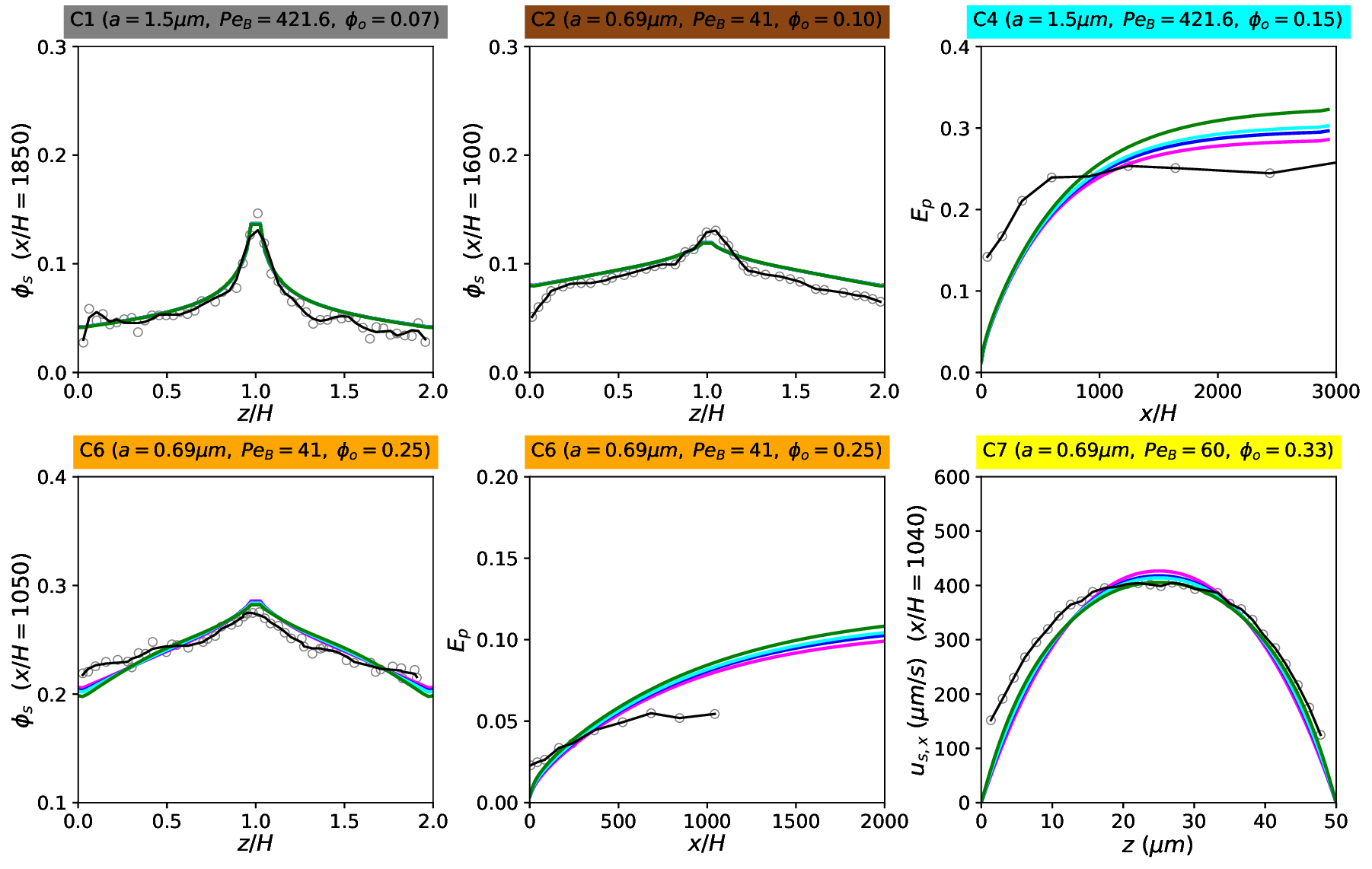}
\caption{}
\label{fig:Semwogerere_MB99B_sensitivity_on_Ks_mix_phi}
\end{subfigure}
\caption{The sensitivity of the MF-roughness model viscosities (a) and simulation results (b) to the variation of $\scal[s,mix]{K}$ values. The legends on the shear viscosity graphs apply to all the sub-figures in parts (a) and (b). }
\label{fig:Semwogerere_MB99B_sensitivity_on_Ks_mix_phi_viscosities}
\end{figure}

This section presents the supplementary figures related to the sensitivity of the models' viscosities or simulation results to changes in mesh size, force or stress terms, and their closure parameters. The figures are listed as follows:\\
\begin{itemize}
\item Fig. \ref{fig:Semwogerere_all_models_case6_mesh_dependency_with_p_osm} shows the sensitivity of the three model results to the mesh size.\\
    \item Fig. \ref{fig:Semwogerere_roughness_sensitivity_on_p_osm} shows the sensitivity of the MF-roughness model results to doubling or excluding the osmotic pressure. \\
    \item Fig. \ref{fig:Semwogerere_roughness_sensitivity_on_Faxen} shows the sensitivity of the MF-roughness model results to the exclusion of the Faxen force.\\
    \item Fig. \ref{fig:Semwogerere_roughness_sensitivity_on_bulk1} shows the sensitivity of the MF-roughness results to the exclusion of the dilation viscosity terms $\mubulkones$ and $\mubulkonem$.\\
    \item Figs. \ref{fig:Semwogerere_roughness_sensitivity_on_N_ro_fe_viscosities}, \ref{fig:Semwogerere_roughness_sensitivity_on_rhat_infinity_min}, \ref{fig:Semwogerere_roughness_sensitivity_on_f_c_phi_viscosities}, and \ref{fig:Semwogerere_roughness_sensitivity_on_eps_r_phi_viscosities} show the sensitivity of the MF-roughness viscosities to the variation of the closure parameters $\scal[{2-\scal[\infty]{\hat{r}}}]{N}$, $\scal[o]{\hat{\hat{r}}}$, $\scal[c]{f}+\scal[e]{f}$, $\scal[\infty,min]{\hat{r}}$, $\scal[c]{f}$, and $\scal[r]{\epsilon}$ values.\\
    \item Figs. \ref{fig:Semwogerere_MB99B_sensitivity_on_phim_phi_viscosities}, \ref{fig:Semwogerere_MB99B_sensitivity_on_Kn_sol_phi_viscosities}, \ref{fig:Semwogerere_MB99B_sensitivity_on_Kn_mix_phi_viscosities}, \ref{fig:Semwogerere_MB99B_sensitivity_on_Ks_sol_phi_viscosities}, and \ref{fig:Semwogerere_MB99B_sensitivity_on_Ks_mix_phi_viscosities} respectively show the sensitivity of the MF-MB99-B model viscosities and simulation results to the variation of the $\phim$, $\scal[n,sol]{K}$, $\scal[n,mix]{K}$, $\scal[s,sol]{K}$, and $\scal[s,mix]{K}$ values.\\
\end{itemize}


